\documentclass[twocolumn]{aastex62}

\usepackage{natbib}
\usepackage{ulem}
\hypersetup{linkcolor=red,citecolor=blue,filecolor=cyan,urlcolor=magenta}
\usepackage{amsmath}
\usepackage{bbding}
\usepackage{graphicx}

\newcommand{\MO}{$M_\odot$}
\newcommand{\ZO}{$Z_\odot$}
\newcommand{\Chandra}{\textit{Chandra}}
\newcommand{\XMM}{\textit{XMM-Newton}}
\newcommand{\HST}{\textit{HST}}

\received{}
\revised
\accepted{}
\submitjournal{ApJ}

\shorttitle{Gas density perturbations in cool cores}
\shortauthors{Ueda et al.}

\begin{document}

\title{Gas Density Perturbations in Cool Cores of CLASH Galaxy Clusters}


\correspondingauthor{Shutaro Ueda}
\email{sueda@asiaa.sinica.edu.tw}
\author[0000-0001-6252-7922]{Shutaro Ueda}
\affiliation{Academia Sinica Institute of Astronomy and Astrophysics (ASIAA), No. 1, Section 4, Roosevelt Road, Taipei 10617, Taiwan}

\author{Yuto Ichinohe}
\affiliation{Department of Physics, Rikkyo University, 3-34-1 Nishi-Ikebukuro, Toshima-ku, Tokyo 171-8501, Japan}

\author{Sandor M. Molnar}
\affiliation{Academia Sinica Institute of Astronomy and Astrophysics (ASIAA), No. 1, Section 4, Roosevelt Road, Taipei 10617, Taiwan}

\author{Keiichi Umetsu}
\affiliation{Academia Sinica Institute of Astronomy and Astrophysics (ASIAA), No. 1, Section 4, Roosevelt Road, Taipei 10617, Taiwan}

\author{Tetsu Kitayama}
\affiliation{Department of Physics, Toho University, 2-2-1 Miyama, Funabashi, Chiba 274-8510, Japan}

%



\begin{abstract}
We present a systematic study of gas density perturbations in cool cores of high-mass galaxy clusters. We select 12 relaxed clusters from the Cluster Lensing And Supernova survey with {\it Hubble} (CLASH) sample and analyze their cool core features observed with the {\it Chandra X-ray Observatory}. We focus on the X-ray residual image characteristics after subtracting their global profile of the X-ray surface brightness distribution. We find that all the galaxy clusters in our sample have, at least, both one positive and one negative excess regions in the X-ray residual image, indicating the presence of gas density perturbations. We identify and characterize the locally perturbed regions using our detection algorithm, and extract X-ray spectra of the intracluster medium (ICM). The ICM temperature in the positive excess region is lower than that in the negative excess region, whereas the ICM in both regions is in pressure equilibrium in a systematic manner. These results indicate that gas sloshing in cool cores takes place in more than $80\,\%$ of relaxed clusters (95\,\% CL). We confirm this physical picture by analyzing synthetic X-ray observations of a cool-core cluster from a hydrodynamic simulation, finding that our detection algorithm can accurately extract both the positive and negative excess regions and can reproduce the temperature difference between them. Our findings support the picture that the gas density perturbations are induced by gas sloshing, and a large fraction of cool-core clusters have undergone gas sloshing, indicating that gas sloshing may be capable of suppressing runaway cooling of the ICM.   
\end{abstract}


\section{Introduction}

Galaxy clusters are still growing in mass through mergers and continuous accretion of material from their surrounding large-scale environments. Galaxy clusters contain a large amount of X-ray emitting gas, i.e., the intracluster medium (ICM), which is thermalized in the gravitational potential well dominated by dark matter. The thermal evolution of the ICM is thus coupled with the formation and evolution of galaxy clusters.  

Most of galaxy clusters host cool cores in their center, in the form of the dense, relatively cool, and metal-enriched ICM. Since cool cores are considered to be formed by radiatively cooling gas, it is expected that runaway cooling occurs in the cool cores because of the fact that their cooling time estimated by electron density is much shorter than the age of galaxy clusters \citep[e.g.,][for a review]{Peterson06}. This expectation is, however, inconsistent with observational evidence  \citep[e.g.,][]{Tamura01, Peterson01, ODea08, McDonald11, McDonald18}, the temperature of the central part of the ICM remains hot (i.e. several keV). No evidence is found of clusters whose ICM temperature is lower than 1\,keV. Such inconsistency indicates that a heating source is required to suppress runaway cooling. In addition, a balance between cooling and heating must be kept during a lifetime of galaxy clusters. 

The origin of heating sources and heating mechanisms is a long-standing problem in cluster astrophysics. Feedback from active galactic nuclei (AGN) in the brightest cluster galaxies (BCG) is considered to be a plausible mechanism \citep[e.g.,][for reviews]{Fabian12, McNamara12}. Mechanical feedback (e.g., jets and sound waves) from AGN interacts with the ICM and induces a characteristic feature in the X-ray surface brightness \citep[e.g.,][]{McNamara05, Hlavacek-Larrondo12, Hlavacek-Larrondo13b, Hlavacek-Larrondo15, Shin16}. In fact, X-ray cavities are found in a large sample of relaxed galaxy clusters over a wide redshift range. Some of them are considered to be most likely due to AGN radio jets \citep[e.g.,][for a review]{McNamara12}. The mechanical power estimated by the size of X-ray cavities is almost sufficient to balance between the cooling and heating rates \citep[e.g.,][for a review]{McNamara07}. On the other hand, gas sloshing in the core of galaxy clusters has been proposed as an alternative heating source \citep[e.g.,][]{Fujita04b, ZuHone10}. Gas sloshing is induced by (minor) mergers with a nonzero impact parameter \citep[see][for a review]{Markevitch07}. Bulk motion in a cool core, generated by gas sloshing, was recently found \citep{Hitomi18d, Ueda19} and its kinetic energy is considered to be converted into heat by dissipating turbulence produced by e.g., the Kelvin-Helmholtz instability \citep[KHI; e.g.,][]{ZuHone10, Roediger12, ZuHone16b}. Evidence of gas sloshing is frequently found in cool cores \citep[e.g.,][]{Churazov03, Clarke04, Blanton11, Owers11, OSullivan12, Canning13, Rossetti13, Ghizzardi14, Sanders14, Ichinohe15, Ueda17, Su17, Liu18, Ueda18, Calzadilla19, Ueda19}. The well-known characteristic features of gas sloshing are that (1) the ICM in the cool core exhibits a characteristic pair of positive and negative density perturbations in the X-ray brightness distribution, (2) the temperature (abundance) of the ICM in the positive excess region is lower (higher) than that in the negative excess area, and (3) the ICM in both regions is in pressure equilibrium. We note that \cite{Khochfar08} have presented detailed studies of gas physics within semi-analytic models of galaxy formation by focusing on the role played by environmental effects. They have studied gravitational heating of the ICM by gravitational potential energy released from stripped gas from infalling substructures and pointed out that gravitational heating appears to be an efficient heating source able to prevent cooling in environments. The context of gravitational heating in \cite{Khochfar08} appears to be associated with gas sloshing.

In the context of cosmic structure formation, mergers play a key role in the formation and evolution of galaxy clusters. Cluster mergers have a large impact on the thermal properties of the ICM for a sufficiently long time scale (say, several Gyrs). Hence, revealing their influence on the ICM properties provides us with an important clue to understanding the role of structure formation on the evolution of galaxy clusters \citep[e.g.,][]{Fujita18}. In principle, the thermodynamic properties of the ICM in cool cores can be perturbed and modified by gas sloshing motions associated with infalling substructures \citep[e.g.,][]{Fujita04b, ZuHone10}. However, thus far, the population of sloshing cool cores has been poorly studied both observationally and theoretically. Therefore, systematic observational studies of sloshing features in cool-core clusters are critically important for understanding the role of gas sloshing in cluster evolution over cosmic time. 

In this paper, we present a systematic study of gas sloshing features by analyzing thermodynamic properties of the ICM in cool cores of relaxed galaxy clusters. To this end, we develop a novel method to identify characteristic features of gas sloshing in X-ray residual images, with an aid of strong-lensing mass models. In addition, we test and validate our method by analyzing synthetic X-ray observations of a cool-core cluster with a merging substructure using a high-resolution hydrodynamic simulation. 

Throughout the paper, we assume $\Omega_{\rm m}=0.27$, $\Omega_{\rm \Lambda}=0.73$, and the Hubble constant of $H_{0} = 70\,$km\,s$^{-1}$\,Mpc$^{-1}$. Unless stated otherwise, quoted errors correspond to $1\sigma$ uncertainties.

\section{Sample}

In this study, we focus on a sample of high-mass galaxy clusters targeted by the Cluster Lensing And Supernova survey with {\it Hubble} \citep[CLASH;][]{Postman12}. The CLASH sample of 25 galaxy clusters has been studied extensively in multiwavelength observations \citep[e.g.,][{}]{Biviano13, Donahue14, Meneghetti14, Umetsu14, Merten15, Zitrin15, Donahue16, Umetsu16, Molino17, Umetsu17, Chiu18, Sereno18, Umetsu18, Yu18}. Detailed X-ray studies of the CLASH sample were presented in \cite{Donahue14} and \cite{Donahue16}, which focused on the global ICM properties of the CLASH sample using \Chandra ~and \XMM ~observations. In this paper, we used the data taken with \Chandra ~X-ray observations.

Here, we select a subsample of 12 CLASH clusters that were identified as cool-core systems on the basis of their X-ray concentration \citep{Donahue16}. For these selected clusters, a clear temperature drop towards the center is observed \citep{Donahue14}, and the cooling time within 100\,kpc is much shorter than 10\,Gyrs in each case \citep{Hlavacek-Larrondo12}. In addition, the central mass distribution from CLASH strong-lens (SL) modeling is publicly available for all of the clusters  \citep{Zitrin15}. Hence, these observational datasets allow us to systematically study and characterize gas density perturbations in the X-ray brightness distribution for a sample of 12 cool-core CLASH clusters. Our cluster sample is summarized in Table~\ref{tab:list}. We note that a detailed study of RXJ\,1347.5-1145 was conducted by \cite{Ueda18} from a combined analysis of X-ray, Sunyaev--Zel'dovich effect (SZE), and \HST~SL observations. Here we will analyze RXJ\,1347.5-1145 using the same methodology developed in this work.

\section{Observation and data reduction}

We used archival X-ray data of each cluster taken with the Advanced CCD Imaging Spectrometer \citep[ACIS;][]{Garmire03} on board the {\it Chandra X-ray Observatory}. The ObsID of all datasets analyzed is summarized in Table~\ref{tab:list}. Since some of the datasets were used in \cite{Donahue16}, the newly added data in this paper are highlighted in boldface in Table~\ref{tab:list}. We used the versions of 4.10 and 4.7.9 for \Chandra ~Interactive Analysis of Observations \cite[CIAO;][]{Fruscione06} and the calibration database (CALDB), respectively. We checked the light curve of each cluster using the {\tt lc\_clean} task in CIAO, filtering flare data. The blanksky data included in the CALDB were adopted as background data. We extracted X-ray spectra from each dataset with {\tt specexctract} in CIAO and combined them after making individual spectrum, response, and ancillary response files for the spectral fitting. We used {\tt XSPEC} version 12.10.0e \citep{Arnaud96} and the atomic database for plasma emission modeling version 3.0.9 in the X-ray spectral analysis, assuming that the ICM is in collisional ionization equilibrium \citep{Smith01}. The abundance table of \cite{Lodders09} is used. The Galactic absorption (i.e., $N_{\rm H}$) for each cluster was estimated using \cite{Kalberla05} and fixed to the estimated value in the X-ray spectral analysis.

\begin{table*}[ht]
\begin{center}
\caption{
Summary of our galaxy cluster our sample: sky coordinates, redshift, net exposure time, and datasets taken with \Chandra.
}\label{tab:list}
\begin{tabular}{lccccl}
\hline\hline	
Sample				& RA			& Dec			& Redshift		& Expo. time (ksec)	& ObsID\tablenotemark{a}							\\ \hline
A383					& 02:48:03.40	& -03:31:44.9		& 0.187		& 48.8				& {\bf 524}, {\bf 2320}, 2321					\\
MACSJ\,0329.6-0211	& 03:29:41.56	& -02:11:46.1		& 0.450		& 76.5				& {\bf 3257}, 3582, 6108, {\bf 7719}			\\
MACSJ\,0429.6-0253	& 04:29:36.05	& -02:53:06.1		& 0.399		& 23.2				& 3271							\\
MACSJ\,1115.8+0129	& 11:15:51.90	& 01:29:55.1		& 0.352		& 55.5				& {\bf 3275}, 9375						\\
MACSJ\,1311.0-0310	& 13:11:01.80	& -03:10:39.8		& 0.494		& 114.9				& {\bf 3258}, 6110, {\bf 7721}, 9381				\\
RXJ\,1347.5-1145		& 13:47:30.62	& -11:45:09.4		& 0.451		& 233.8				& {\bf 506}, {\bf 507}, 3592, {\bf 13516}, {\bf 13999}, {\bf 14407}	\\
MACSJ\,1423.8+2404	& 14:23:47.88	& 24:04:42.5		& 0.545		& 134.1				& {\bf 1657}, 4195						\\
RXJ\,1532.9+3021		& 15:32:53.78	& 30:20:59.4		& 0.363		& 108.2				& 1649, 1665, {\bf 14009}					\\
MACS\,J1720.2+3536	& 17:20:16.78	& 35:36:26.5		& 0.391		& 63.7				& {\bf 3280}, 6107, {\bf 7225}, {\bf 7718}			\\
MACS\,J1931.8-2634	& 19:31:49.62	& -26:34:32.9		& 0.352		& 112.5				& 3282, 9382						\\
RXJ\,2129.6+0005		& 21:29:39.96	& 00:05:21.2		& 0.234		& 39.6				& 552, 9370						\\
MS\,2137.3-2353		& 21:40:15.17	& -23:39:40.2		& 0.313		& 141.5				& {\bf 928}, 4974, 5250					\\
\hline
\end{tabular}
\end{center}
\tablenotetext{a}{
The ObsID in boldface indicates newly added data, which are not included in \cite{Donahue16}.
}
\end{table*}

\section{Analyses and Results}

\subsection{X-ray imaging analysis}
\label{sec:img}

In order to detect gas density perturbations in the cool core of each cluster, we analyzed their X-ray surface brightness in the $0.5 - 7.0$\,keV band. The left side of Figure~\ref{fig:image} shows the X-ray surface brightness of our sample taken with \Chandra. Following \cite{Ueda17}, we first modeled a mean surface brightness using the concentric ellipse fitting algorithm, by minimizing the variance of the X-ray surface brightness relative to the ellipse model. In this analysis, we fixed the center of the ellipse model to the peak position of the SL mass map\footnote{\url{https://archive.stsci.edu/prepds/clash/}} that assumed an elliptical Navarro-Frenk-White \citep[NFW;][]{Navarro97} profile by \cite{Zitrin15}. The location of the center, position angle (PA)\footnote{The position angle is measured for the major axis of an ellipse from north ($0^{\circ}$) to east ($90^{\circ}$).}, and axis ratio (AR) of each cluster are listed in Table~\ref{tab:sky}. After modeling, we subtracted the obtained model from the original X-ray surface brightness. The right side of Figure~\ref{fig:image} shows the X-ray residual image of each cluster. For a comparison, we also listed the PA and AR of the dark matter halo in each cluster summarized in \cite{Donahue16} in Table~\ref{tab:sky}. As already reported by \cite{Donahue16}, the PA and AR of the X-ray surface brightness are in good agreement with those in the central mass distribution of the dark matter halo, respectively.

We found at least one positive and one negative excess regions in the X-ray residual images of all 12 galaxy clusters in our sample, which indicate the presence of gas density perturbations in their cool cores. Locally perturbed regions are defined as those where the amplitude of a fluctuation exceeds 20\,\% of the extreme values found in the X-ray residual images around the position of the peak. If another perturbed region is found (e.g., the cases of A383 and RXJ\,1532.9+3021), we apply this procedure again after masking the previous region. In this case, the maximum value of the fluctuation is the same as the previous one but the peak position is shifted to the second local peak. This procedure is repeated until no additional fluctuations are found. For RXJ\,1347.5-1145, the positive excess region is found in the southeast region. However, this excess has been recognized as a stripped gas originally in an infalling subcluster \citep{Kreisch16, Ueda18}. We, therefore, avoided defining this excess as a perturbed region in the cool core. In addition, a dipolar pattern in the X-ray residual image is found. The shape of this pattern is consistent with that found in \cite{Ueda18}. The number of the perturbed regions identified along with the net photon counts are summarized in Table~\ref{tab:net}. The perturbed regions are also shown in the right side of Figure~\ref{fig:image}.

We note here that if we define a threshold to identify perturbed regions to be 10\,\% of the extreme values in the X-ray residual images instead of 20\,\%, we find that most regions are so large that a large amount of ambient ICM is included in these regions. Hence, a threshold of 10\,\% is not suitable for studying thermodynamic properties of the ICM in cool cores. On the other hand, if we increase the threshold from 20\% to 30\,\%, the area of perturbed regions is reduced considerably. Since the net photon counts in each region decrease significantly, such a high threshold also prevents us from accurately measuring thermodynamic properties of the ICM. Therefore, we define perturbed regions using a threshold of 20\,\%.

\begin{table*}[ht]
\begin{center}
\caption{
Location of the peak of the best-fit NFW profile, the position angle (PA) and axis ratio (AR) of the X-ray surface brightness. We include those derived by the previous lensing analysis for reference.
}\label{tab:sky}
\begin{tabular}{lcccccc}
\hline\hline	
Sample				& RA			& Dec		& PA (X-ray) ($^{\circ}$)\tablenotemark{a}	& AR (X-ray)\tablenotemark{b}	& PA (lensing) ($^{\circ}$)\tablenotemark{c}	& AR (lensing)\tablenotemark{c}	\\ \hline
A383					& 42.014094	& -3.529372	& 16								& 0.89					& $13 \pm 15$								& $0.91 \pm 0.08$				\\
MACSJ\,0329.6-0211	& 52.423220	& -2.196229	& 152							& 0.83					& $144 \pm 9$								& $0.84 \pm 0.07$				\\
MACSJ\,0429.6-0253	& 67.400045	& -2.885190	& 162							& 0.73					& $172 \pm 3$								& $0.78 \pm 0.15$				\\
MACSJ\,1115.8+0129	& 168.96625	& 1.4986388	& 140							& 0.64					& $139 \pm 5$								& $0.87 \pm 0.07$				\\
MACSJ\,1311.0-0310	& 197.75751	& -3.177704	& 135							& 0.81					& $1 \pm 35$								& $0.86 \pm 0.14$				\\
RXJ\,1347.5-1145		& 206.87754	& -11.752634	& 167							& 0.68					& $26 \pm 4$								& $0.81 \pm 0.10$				\\
MACSJ\,1423.8+2404	& 215.94948	& 24.078454	& 19								& 0.85					& $26 \pm 2$								& $0.79 \pm 0.14$				\\
RXJ\,1532.9+3021		& 233.22410	& 30.349815	& 52								& 0.73					& $39 \pm 18$								& $0.84 \pm 0.11$				\\
MACS\,J1720.2+3536	& 260.06980	& 35.607306	& 40								& 0.90					& $11 \pm 3$								& $0.74 \pm 0.21$				\\
MACS\,J1931.8-2634	& 292.95677	& -26.575715	& 173							& 0.62					& $177 \pm 4$								& $0.70 \pm 0.18$				\\
RXJ\,2129.6+0005		& 322.41648	& 0.089240	& 64								& 0.64					& $68 \pm 2$								& $0.68 \pm 0.23$				\\
MS\,2137.3-2353		& 325.06314	& -23.661148	& 48								& 0.90					& $59 \pm 8$								& $0.88 \pm 0.08$				\\
\hline
\end{tabular}
\end{center}
\tablenotetext{a}{
Uncertainty of this parameter is $\pm 1^{\circ}$.
}
\tablenotetext{b}{
Uncertainty of this parameter is $\pm 0.01$.
}
\tablenotetext{c}{
These data are taken from Table~5 in \cite{Donahue16}.
}
\end{table*}

\begin{table*}[ht]
\begin{center}
\caption{
Net photon count in the positive and negative excess regions in the $0.4 - 7.0$\,keV band.
}\label{tab:net}
\begin{tabular}{lccc}
\hline\hline	
Sample				& ID	& Positive excess	& Negative excess	\\ \hline
A383					& 1	& 785			& 850			\\
					& 2	& 2637			& 2482			\\
MACSJ\,0329.6-0211	& 1	& 2243			& 2602			\\
MACSJ\,0429.6-0253	& 1	& 539			& 514			\\
MACSJ\,1115.8+0129	& 1	& 2693			& 2639			\\
MACSJ\,1311.0-0310	& 1	& 2527			& 1921			\\
RXJ\,1347.5-1145		& 1	& 41773			& 31393			\\
MACSJ\,1423.8+2404	& 1	& 2038			& 5067			\\
					& 2	& 2657			& None			\\
RXJ\,1532.9+3021		& 1	& 4272			& 2711			\\
					& 2	& 10807			& 1010			\\
					& 3	& 2965			& 760			\\
					& 4	& 693			& 352			\\
MACS\,J1720.2+3536	& 1	& 1322			& 1148			\\
					& 2	& 616			& 407			\\
MACS\,J1931.8-2634	& 1	& 5514			& 3162			\\
					& 2	& None			& 3357			\\
RXJ\,2129.6+0005		& 1	& 2318			& 1473			\\
MS\,2137.3-2353		& 1	& 8994			& 12065			\\
					& 2	& 10199			& 3193			\\
\hline
\end{tabular}
\end{center}
\end{table*}

\subsection{X-ray spectral analysis}
\label{sec:spec_ana}

We extracted an X-ray spectrum of each perturbed region identified in Section~\ref{sec:img}. The X-ray spectra in the $0.4 - 7.0$\,keV band are analyzed using the model of {\tt phabs * apec} in {\tt XSPEC}. The cluster redshift is fixed to the value in Table~\ref{tab:list}. The blanksky data provided by CALDB are adopted as the background data for the spectral analysis. For MACS\,J1931.8-2634, we excluded the luminous AGN \citep{Ehlert11} with $r < 1''$ to reduce contamination of the AGN emission. The best-fit parameters of the ICM temperature and abundance are shown in Table~\ref{tab:fit}. Most of them in the perturbed regions were measured well, except several regions with poor statistics (net counts $< 1000$, see Table~\ref{tab:net}). Hence, we obtained the upper limits of the abundance in the region \#2 of RXJ\,1532.9+3021 and the regions \#1 and \#2 of MACS\,J1720.2+3536.

We focus here on the temperature and abundance differences between the positive and negative excess regions, respectively, to reveal the origin of gas density perturbations in the cool core. Gas sloshing is expected to create a positive excess with lower temperature and higher abundance, and a negative excess with higher temperature and lower abundance in the cool core. In Table~\ref{tab:fit}, we list the properties of the gas sloshing regions, where a clear feature expected by gas sloshing is found, by taking the statistical uncertainty into account. We also include in the table additional cases that could potentially be identified as gas sloshing regions, but that cannot be confirmed due to the statistical uncertainty

We also calculated the ICM electron number density ($n_{\rm e}$), pressure ($p_{\rm e} = kT \times n_{\rm e}$), and entropy ($K_{\rm e} = kT \times n_{\rm e}^{-2/3}$), assuming that a length of the line-of-sight is $L$/1\,Mpc. Table~\ref{tab:npe} shows these parameters in the positive and negative excess regions, respectively.

\begin{table*}[ht]
\begin{center}
\caption{
Best-fit parameters of the X-ray spectral analyses performed in the positive and negative excess regions in each cluster.
}\label{tab:fit}
\begin{tabular}{lcccccc}
\hline\hline	
Sample				& ID	& \multicolumn2c{Positive excess}				& \multicolumn2c{Negative excess}				& Sloshing?\tablenotemark{a}	\\
					&	& $kT$	(keV)		& $Z$ (\ZO)			& $kT$ (keV)  			& $Z$(\ZO)			&			\\ \hline
A383					& 1	& $2.07^{+0.30}_{-0.11}$	& $0.85^{+0.28}_{-0.21}$	& $2.58^{+0.32}_{-0.20}$	& $0.82^{+0.34}_{-0.25}$	& ?			\\
					& 2	& $3.64^{+0.21}_{-0.20}$	& $0.73^{+0.18}_{-0.16}$	& $3.71^{+0.20}_{-0.20}$	& $0.93^{+0.21}_{-0.18}$	&			\\
MACSJ\,0329.6-0211	& 1	& $4.05^{+0.19}_{-0.19}$	& $1.19^{+0.26}_{-0.22}$	& $4.26^{+0.25}_{-0.17}$	& $0.99^{+0.21}_{-0.17}$	&			\\
MACSJ\,0429.6-0253	& 1	& $3.95^{+0.57}_{-0.44}$	& $0.57^{+0.55}_{-0.42}$	& $4.08^{+0.54}_{-0.40}$	& $1.21^{+0.77}_{-0.54}$	&			\\
MACSJ\,1115.8+0129	& 1	& $4.43^{+0.27}_{-0.22}$	& $0.78^{+0.19}_{-0.17}$	& $4.99^{+0.30}_{-0.29}$	& $0.74^{+0.18}_{-0.16}$	& ?			\\
MACSJ\,1311.0-0310	& 1	& $4.69^{+0.31}_{-0.29}$	& $0.41^{+0.14}_{-0.12}$	& $6.46^{+0.56}_{-0.52}$	& $0.66^{+0.21}_{-0.18}$	& ?			\\
RXJ\,1347.5-1145		& 1	& $8.50^{+0.17}_{-0.16}$	& $0.47^{+0.03}_{-0.03}$	& $13.50^{+0.79}_{-0.47}$& $0.33^{+0.05}_{-0.05}$	& \Checkmark	\\
MACSJ\,1423.8+2404	& 1	& $3.90^{+0.20}_{-0.19}$	& $1.05^{+0.28}_{-0.23}$	& $5.14^{+0.20}_{-0.19}$	& $0.93^{+0.16}_{-0.14}$	& ?			\\
					& 2	& $5.10^{+0.28}_{-0.27}$	& $0.79^{+0.21}_{-0.18}$	& $-$				& $-$				&			\\
RXJ\,1532.9+3021		& 1	& $4.08^{+0.17}_{-0.16}$	& $0.52^{+0.12}_{-0.13}$	& $4.46^{+0.28}_{-0.25}$	& $0.63^{+0.18}_{-0.16}$	&			\\
					& 2	& $4.50^{+0.13}_{-0.13}$	& $0.81^{+0.09}_{-0.09}$	& $7.68^{+1.49}_{-1.13}$	& $<0.23$				& \Checkmark	\\
					& 3	& $6.03^{+0.44}_{-0.42}$	& $0.34^{+0.15}_{-0.13}$	& $5.70^{+0.67}_{-0.56}$	& $1.77^{+0.85}_{-0.61}$	&			\\
					& 4	& $5.86^{+0.87}_{-0.69}$	& $0.90^{+0.54}_{-0.43}$	& $4.63^{+1.01}_{-0.87}$	& $2.81^{+5.50}_{-1.82}$	&			\\
MACS\,J1720.2+3536	& 1	& $4.22^{+0.40}_{-0.26}$	& $1.28^{+0.17}_{-0.15}$	& $4.22^{+0.56}_{-0.37}$	& $<0.15$				&			\\
					& 2	& $5.00^{+0.60}_{-0.49}$	& $1.73^{+0.30}_{-0.29}$	& $5.78^{+1.39}_{-1.05}$	& $<0.63$				& ?			\\
MACS\,J1931.8-2634	& 1	& $4.88^{+0.19}_{-0.19}$	& $0.66^{+0.11}_{-0.10}$	& $6.20^{+0.40}_{-0.38}$	& $0.42^{+0.14}_{-0.13}$	& \Checkmark	\\
					& 2	& $-$				&$-$					& $5.12^{+0.27}_{-0.25}$	& $0.55^{+0.14}_{-0.13}$	&			\\
RXJ\,2129.6+0005		& 1	& $3.84^{+0.20}_{-0.20}$	& $0.77^{+0.22}_{-0.19}$	& $4.95^{+0.43}_{-0.41}$	& $0.47^{+0.27}_{-0.24}$	& ?			\\
MS\,2137.3-2353		& 1	& $4.02^{+0.11}_{-0.11}$	& $0.61^{+0.09}_{-0.08}$	& $4.87^{+0.14}_{-0.14}$	& $0.59^{+0.09}_{-0.08}$	& ?			\\
					& 2	& $4.70^{+0.15}_{-0.15}$	& $0.59^{+0.08}_{-0.08}$	& $4.96^{+0.28}_{-0.27}$	& $0.80^{+0.20}_{-0.17}$	&			\\
\hline
\end{tabular}
\end{center}
\tablenotetext{a}{
If a clear sloshing feature is found (i.e., the ICM temperature and abundance in the positive excess region are lower and higher than those in the negative excess region, respectively), a check mark (i.e., \Checkmark) is assigned. In the case of a possible candidate, ? is entered.
}
\end{table*}

\begin{table*}[ht]
\begin{center}
\caption{
Electron number density ($n_{\rm e}$), pressure ($p_{\rm e}$), and entropy ($K_{\rm e}$) in the positive and negative excess regions in each cluster.
}\label{tab:npe}
\begin{tabular}{lccccccc}
\hline\hline	
Sample				& ID	& \multicolumn3c{Positive excess}				& \multicolumn3c{Negative excess}				\\
					&	& $n_{\rm e}$\tablenotemark{a}		& $p_{\rm e}$\tablenotemark{b}		& $K_{\rm e}$\tablenotemark{c}		& $n_{\rm e}$\tablenotemark{a}	& $p_{\rm e}$\tablenotemark{b}		& $K_{\rm e}$\tablenotemark{c}		\\ \hline
A383					& 1	& $1.46_{-0.08}^{+0.08}$	& $3.02^{+0.47}_{-0.23}$	& $34.65^{+5.18}_{-2.23}$	& $1.25_{-0.06}^{+0.07}$	& $3.23^{+0.44}_{-0.29}$	& $47.90^{+6.20}_{-4.02}$	\\
					& 2	& $0.94_{-0.02}^{+0.02}$	& $3.42^{+0.21}_{-0.20}$	& $81.72^{+4.86}_{-4.64}$	& $0.75_{-0.02}^{+0.02}$	& $2.78^{+0.17}_{-0.17}$	& $96.83^{+5.50}_{-5.50}$	\\
MACSJ\,0329.6-0211	& 1	& $1.50_{-0.04}^{+0.04}$	& $6.08^{+0.33}_{-0.33}$	& $66.59^{+3.34}_{-3.34}$	& $1.08_{-0.03}^{+0.03}$	& $4.60^{+0.30}_{-0.22}$	& $87.19^{+5.37}_{-3.84}$	\\
MACSJ\,0429.6-0253	& 1	& $2.07_{-0.14}^{+0.13}$	& $8.18^{+1.29}_{-1.07}$	& $52.39^{+7.87}_{-6.30}$	& $1.32_{-0.10}^{+0.10}$	& $5.39^{+0.82}_{-0.67}$	& $73.05^{+10.35}_{-8.06}$	\\
MACSJ\,1115.8+0129	& 1	& $1.50_{-0.03}^{+0.03}$	& $6.64^{+0.43}_{-0.36}$	& $72.84^{+4.54}_{-3.75}$	& $1.17_{-0.02}^{+0.02}$	& $5.84^{+0.36}_{-0.35}$	& $96.82^{+5.92}_{-5.73}$	\\
MACSJ\,1311.0-0310	& 1	& $1.01_{-0.02}^{+0.02}$	& $4.74^{+0.33}_{-0.31}$	& $100.37^{+6.77}_{-6.35}$	& $0.67_{-0.01}^{+0.01}$	& $4.33^{+0.38}_{-0.35}$	& $181.77^{+15.86}_{-14.74}$	\\
RXJ\,1347.5-1145		& 1	& $2.49_{-0.01}^{+0.01}$	& $21.16^{+0.43}_{-0.41}$& $99.68^{+2.01}_{-1.90}$	& $1.62_{-0.01}^{+0.01}$	&$21.87^{+1.27}_{-0.77}$	& $210.86^{+12.21}_{-7.39}$	\\
MACSJ\,1423.8+2404	& 1	& $2.67_{-0.09}^{+0.09}$	& $10.41^{+0.64}_{-0.62}$& $43.66^{+2.44}_{-2.34}$	& $1.52_{-0.02}^{+0.02}$	& $7.81^{+0.32}_{-0.31}$	& $83.77^{+3.34}_{-3.18}$	\\
					& 2	& $1.62_{-0.04}^{+0.03}$	& $8.26^{+0.48}_{-0.48}$	& $79.66^{+4.48}_{-4.42}$	& $-$				& $-$						& $-$					\\
RXJ\,1532.9+3021		& 1	& $2.06_{-0.04}^{+0.03}$	& $8.40^{+0.37}_{-0.37}$	& $54.29^{+2.32}_{-2.24}$	& $1.72_{-0.04}^{+0.04}$	& $7.67^{+0.51}_{-0.47}$	& $66.93^{+4.33}_{-3.89}$	\\
					& 2	& $1.70_{-0.02}^{+0.02}$	& $7.65^{+0.24}_{-0.24}$	& $68.06^{+2.04}_{-2.04}$	& $1.48_{-0.03}^{+0.02}$	& $11.37^{+2.21}_{-1.69}$	& $127.41^{+24.74}_{-18.82}$	\\
					& 3	& $0.91_{-0.02}^{+0.01}$	& $5.49^{+0.40}_{-0.40}$	& $138.34^{+10.15}_{-9.85}$	& $0.91_{-0.05}^{+0.05}$	& $5.19^{+0.67}_{-0.58}$	& $130.77^{+16.10}_{-13.71}$	\\
					& 4	& $1.30_{-0.06}^{+0.06}$	& $7.62^{+1.18}_{-0.96}$	& $105.99^{+16.07}_{-12.90}$	& $2.11_{-0.28}^{+0.39}$	& $9.77^{+2.79}_{-2.25}$	& $60.64^{+15.19}_{-12.59}$	\\
MACS\,J1720.2+3536	& 1	& $1.36_{-0.06}^{+0.06}$	& $5.74^{+0.60}_{-0.43}$	& $74.07^{+7.35}_{-5.06}$	& $1.32_{-0.04}^{+0.03}$	& $5.57^{+0.75}_{-0.52}$	& $75.55^{+10.09}_{-6.80}$	\\
					& 2	& $0.71_{-0.05}^{+0.05}$	& $3.55^{+0.49}_{-0.43}$	& $135.35^{+17.44}_{-14.71}$	& $0.46_{-0.03}^{+0.02}$	& $2.66^{+0.65}_{-0.51}$	& $208.97^{+50.62}_{-39.03}$	\\
MACS\,J1931.8-2634	& 1	& $2.40_{-0.03}^{+0.03}$	& $11.71^{+0.48}_{-0.48}$	& $58.65^{+2.34}_{-2.34}$	& $2.36_{-0.04}^{+0.04}$	& $14.63^{+0.98}_{-0.93}$	& $75.36^{+4.94}_{-4.70}$	\\
					& 2	& $-$					& $-$					& $-$					& $1.15_{-0.02}^{+0.02}$	& $5.89^{+0.33}_{-0.31}$	& $100.49^{+5.43}_{-5.04}$	\\
RXJ\,2129.6+0005		& 1	& $1.29_{-0.04}^{+0.04}$	& $4.95^{+0.30}_{-0.30}$	& $69.81^{+3.91}_{-3.91}$	& $1.08_{-0.03}^{+0.03}$	& $5.35^{+0.49}_{-0.47}$	& $101.31^{+9.00}_{-8.60}$	\\
MS\,2137.3-2353		& 1	& $2.04_{-0.02}^{+0.02}$	& $8.20^{+0.24}_{-0.24}$	& $53.84^{+1.51}_{-1.51}$	& $1.52_{-0.02}^{+0.01}$	& $7.40^{+0.22}_{-0.23}$	& $79.37^{+2.31}_{-2.39}$	\\
					& 2	& $1.28_{-0.01}^{+0.01}$	& $6.02^{+0.20}_{-0.20}$	& $85.89^{+2.78}_{-2.78}$	& $1.15_{-0.02}^{+0.02}$	& $5.70^{+0.34}_{-0.33}$	& $97.35^{+5.61}_{-5.42}$	\\
\hline
\end{tabular}
\end{center}
\tablenotetext{a}{
Electron number density in units of $10^{-2}$\,cm\,$^{-3}$\,($L$/1\,{\rm Mpc})$^{-1/2}$.
}
\tablenotetext{b}{
Electron pressure in units of $10^{-2}$\,keV\,cm$^{-3}$\,($L$/1\,{\rm Mpc})$^{-1/2}$.
}
\tablenotetext{c}{
Electron entropy in units of keV\,cm$^2$\,($L$/1\,{\rm Mpc})$^{1/3}$.
}
\end{table*}

\subsection{Quantification of the contrast of gas density perturbations}
\label{sec:mdIx}

To quantify the density perturbations found in the X-ray residual images, we use the contrast, typically defined as $| \Delta I_{\rm X}| / \langle I_{\rm X} \rangle$, where $ \langle I_{\rm X} \rangle$ is the mean X-ray surface brightness, following \cite{Ueda18}. We calculated $| \Delta I_{\rm X}| / \langle I_{\rm X} \rangle$ for each cluster and summarized the quantified value in Table~\ref{tab:dIx}. Note that the contrast found in all cases stays well below unity, which means that the amplitude of the fluctuations in the cool core is modest regardless of mechanisms to create perturbations.

\begin{table}[ht]
\begin{center}
\caption{
Level of gas density perturbations in the cool core of each cluster.
}\label{tab:dIx}
\begin{tabular}{lc}
\hline\hline	
Sample				& $|\Delta I_{\rm X}| / \langle I_{\rm X} \rangle$	\\ \hline
A383					& $0.380 \pm 0.016$		\\
MACSJ\,0329.6-0211	& $0.357 \pm 0.021$		\\
MACSJ\,0429.6-0253	& $0.488 \pm 0.035$		\\
MACSJ\,1115.8+0129	& $0.392 \pm 0.019$		\\
MACSJ\,1311.0-0310	& $0.528 \pm 0.020$		\\
RXJ\,1347.5-1145		& $0.260 \pm 0.013$		\\
MACSJ\,1423.8+2404	& $0.246 \pm 0.026$		\\
RXJ\,1532.9+3021		& $0.259 \pm 0.014$		\\
MACS\,J1720.2+3536	& $0.509 \pm 0.020$		\\
MACS\,J1931.8-2634	& $0.357 \pm 0.015$		\\
RXJ\,2129.6+0005		& $0.415 \pm 0.022$		\\
MS\,2137.3-2353		& $0.216 \pm 0.022$		\\
\hline
\end{tabular}
\end{center}
\end{table}

\begin{figure*}
 \begin{center}
  \includegraphics[width=7.5cm]{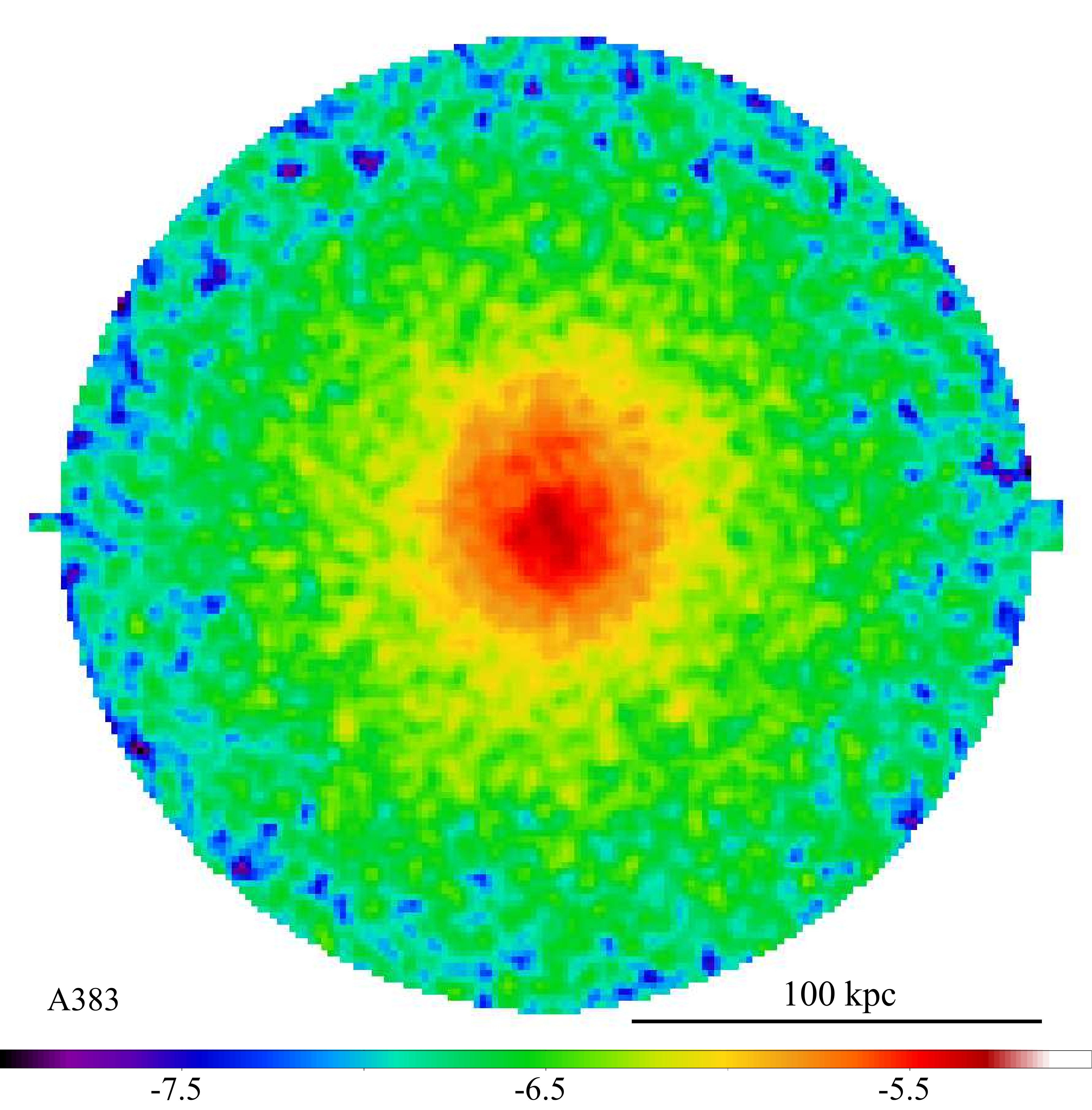}
  \includegraphics[width=7.5cm]{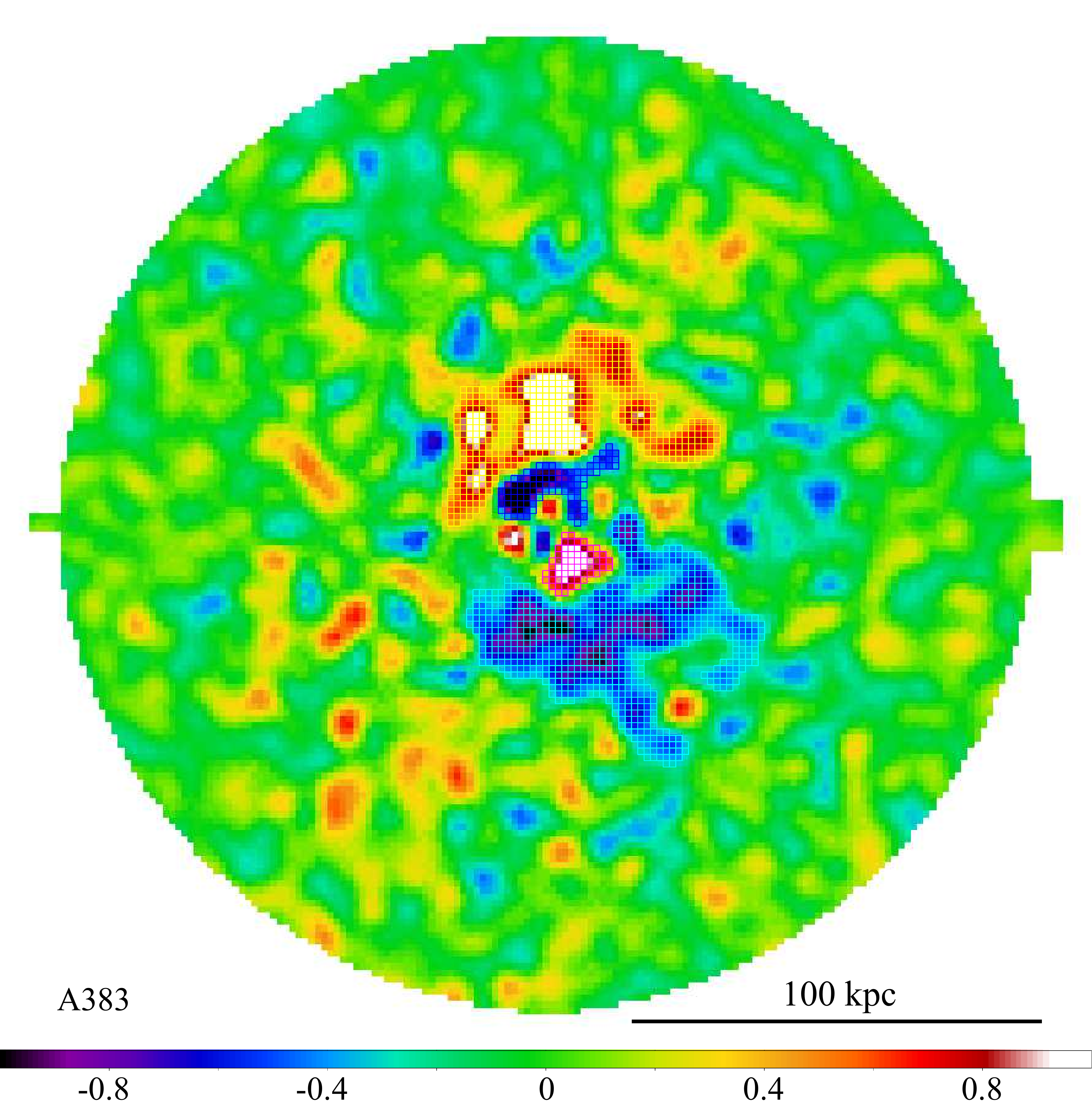} 
  \includegraphics[width=7.5cm]{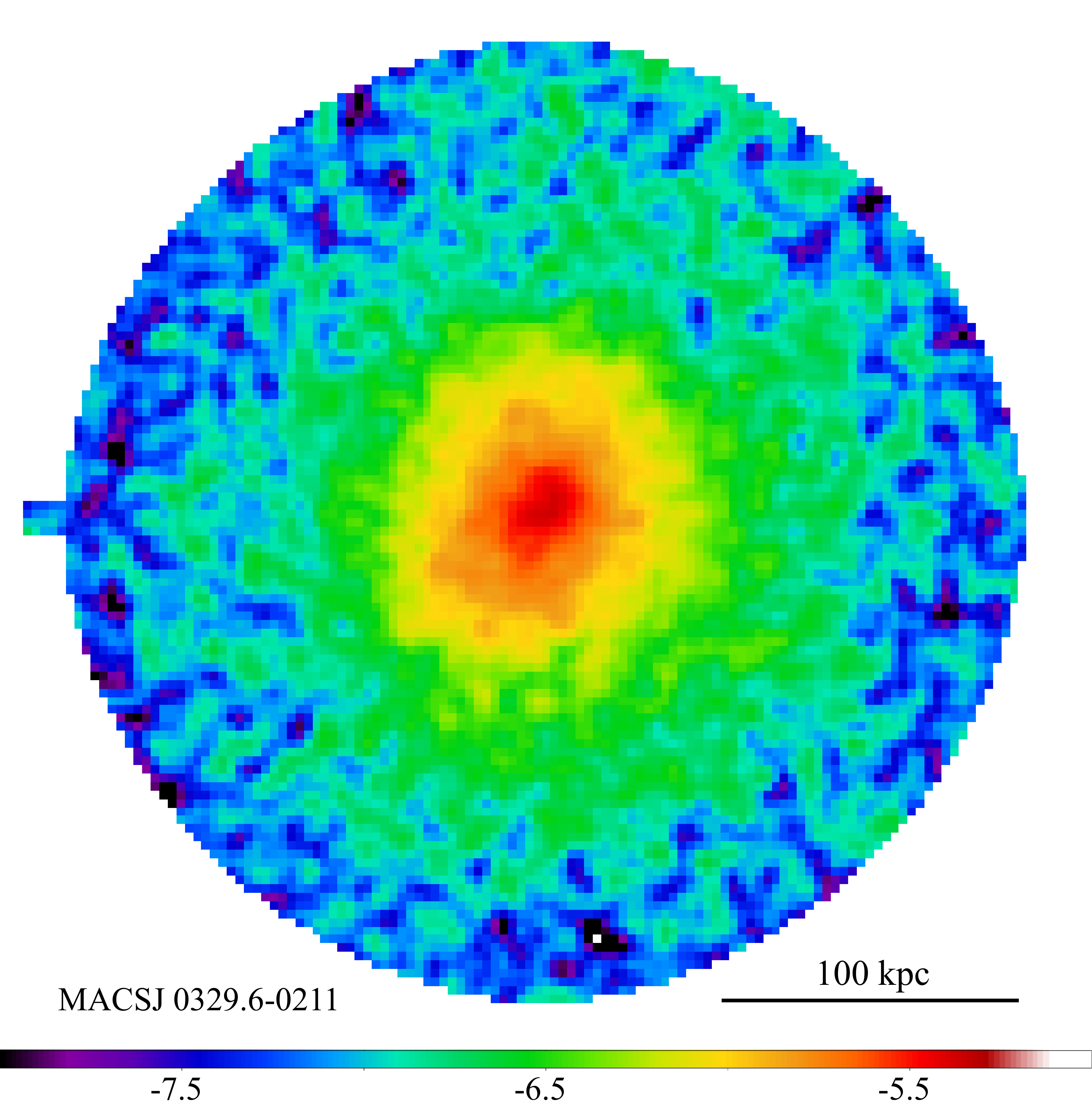}
  \includegraphics[width=7.5cm]{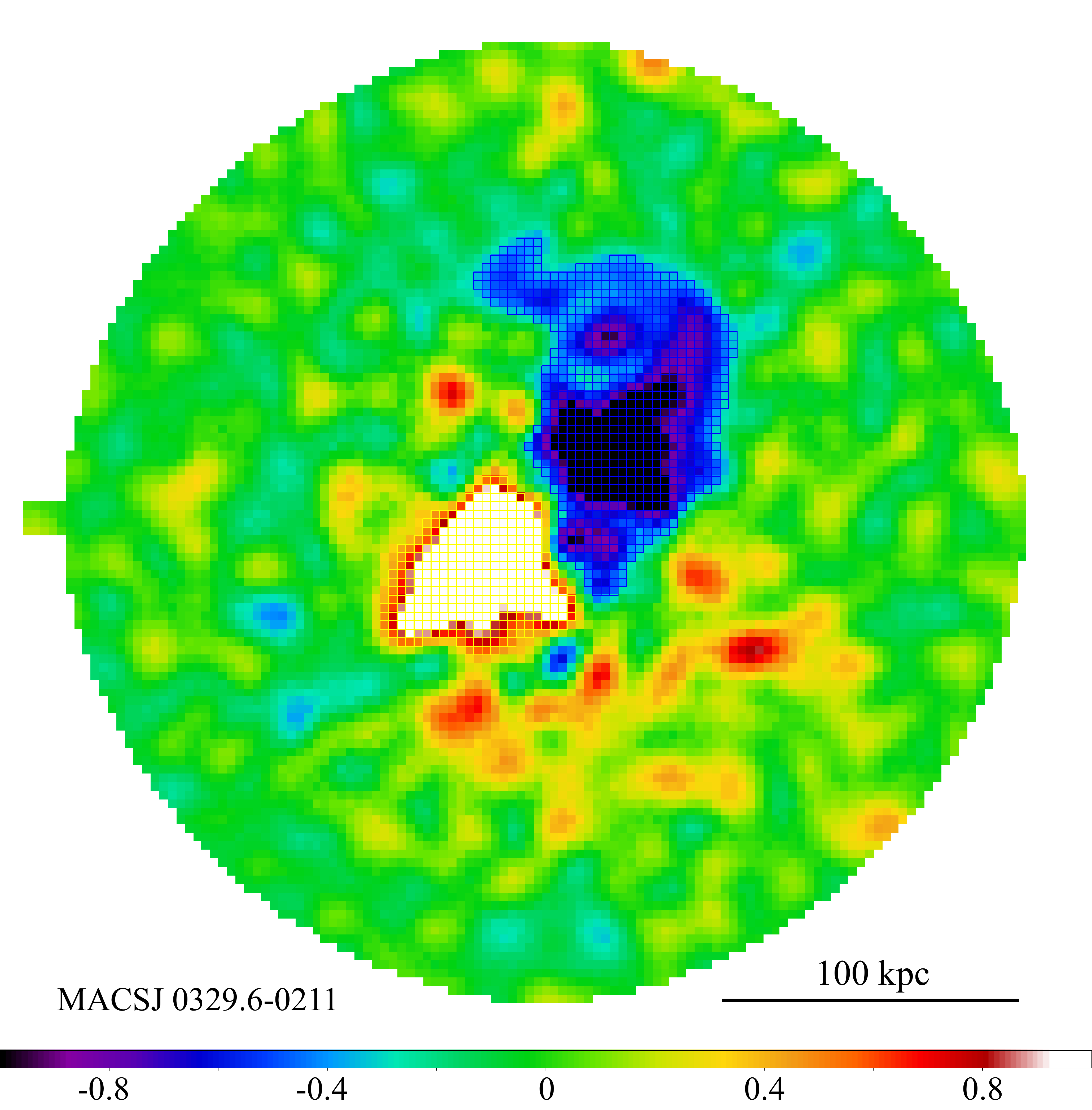} 
 \end{center}
\caption{X-ray surface brightness (left) and its residual image after removing the mean profile (right). 
Left: The X-ray surface brightness in the $0.5 - 7.0$\,keV band is shown on a logarithm scale in units of photon\,sec$^{-1}$\,arcsec$^{-2}$\,cm$^{-2}$. This image is smoothed with a Gaussian kernel with $2.3''$ FWHM.
Right: The X-ray residual image is smoothed with a Gaussian kernel with $4.6''$ FWHM.
The yellow, magenta, green, and red boxes correspond to each positive excess region, respectively. On the other hand, the blue, cyan, white, and black boxes represent each negative excess region, respectively. 
The panels from top to bottom show A383 and MACSJ\,0329.6-0211, respectively.
}
\label{fig:image}
\end{figure*}

\begin{figure*}
\addtocounter{figure}{-1}
 \begin{center}
  \includegraphics[width=7.5cm]{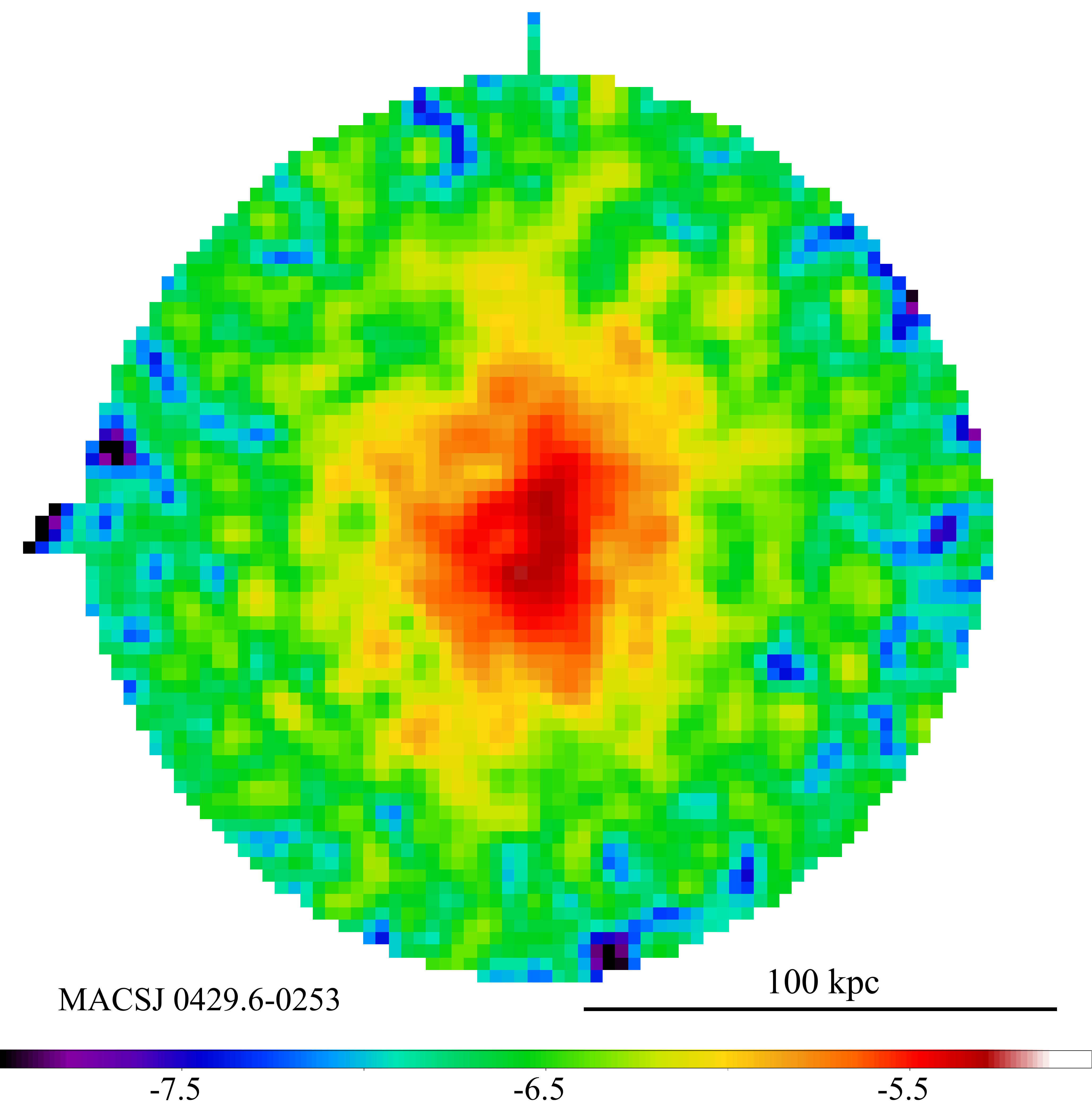}
  \includegraphics[width=7.5cm]{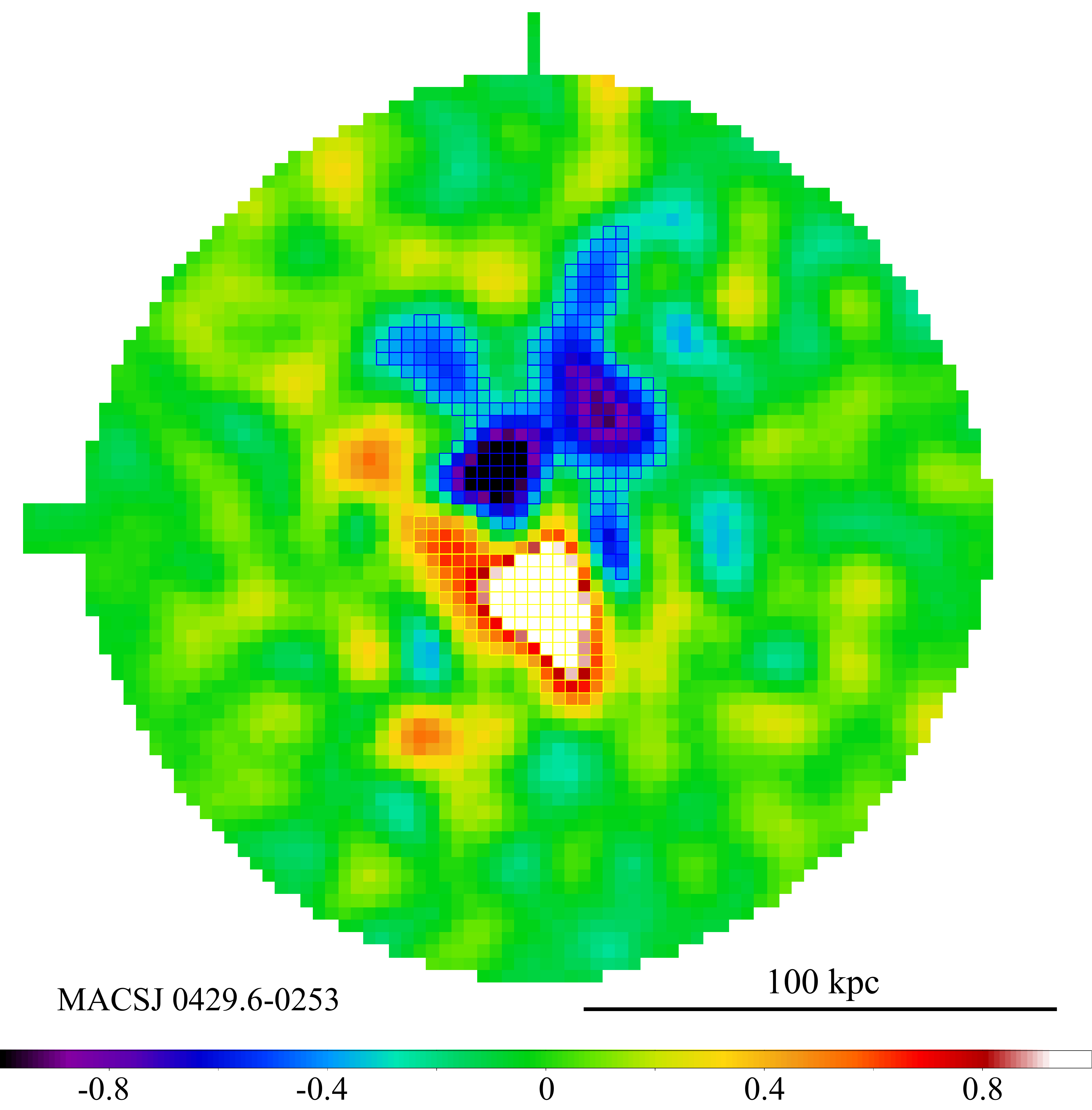} 
  \includegraphics[width=7.5cm]{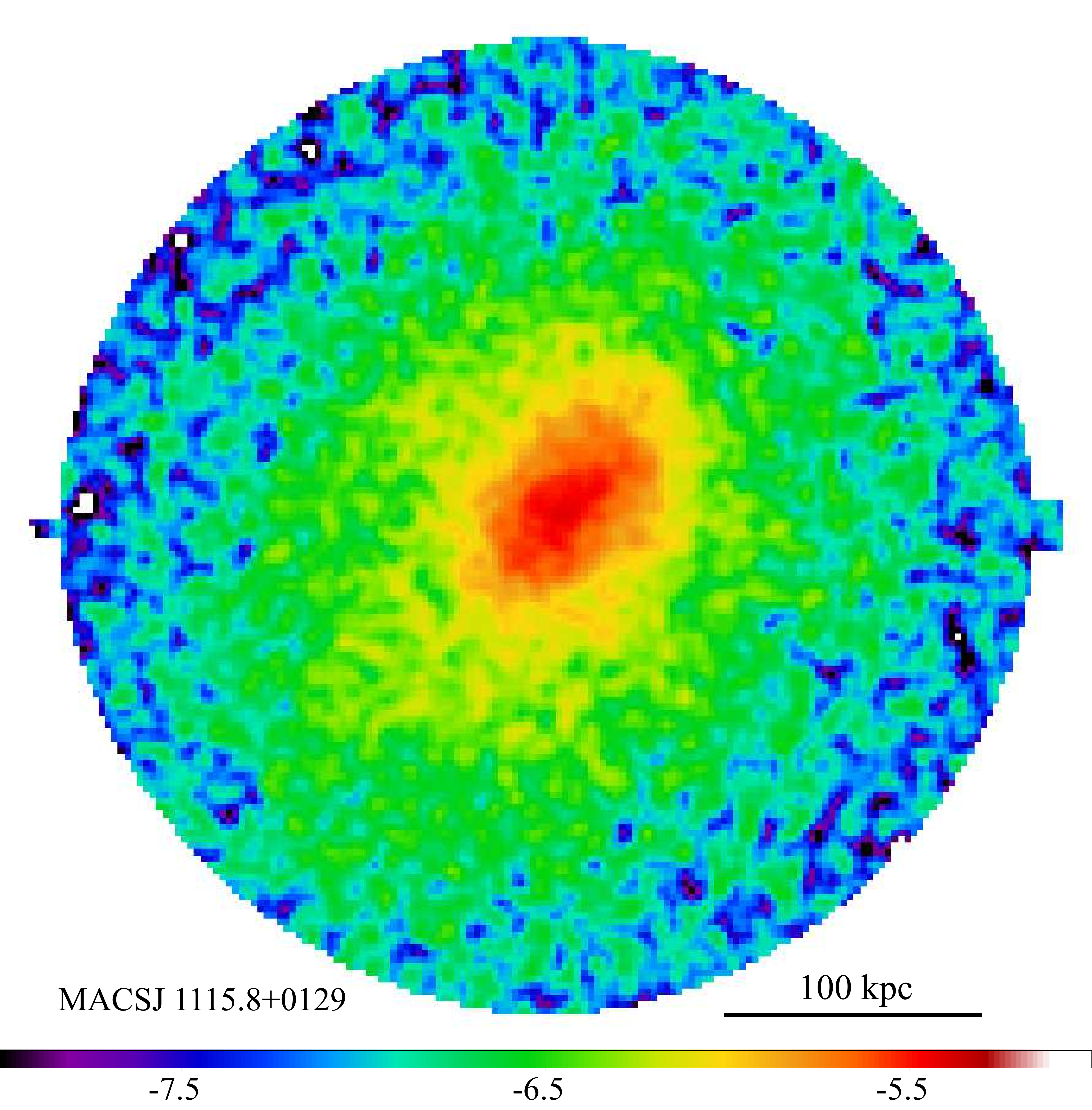}
  \includegraphics[width=7.5cm]{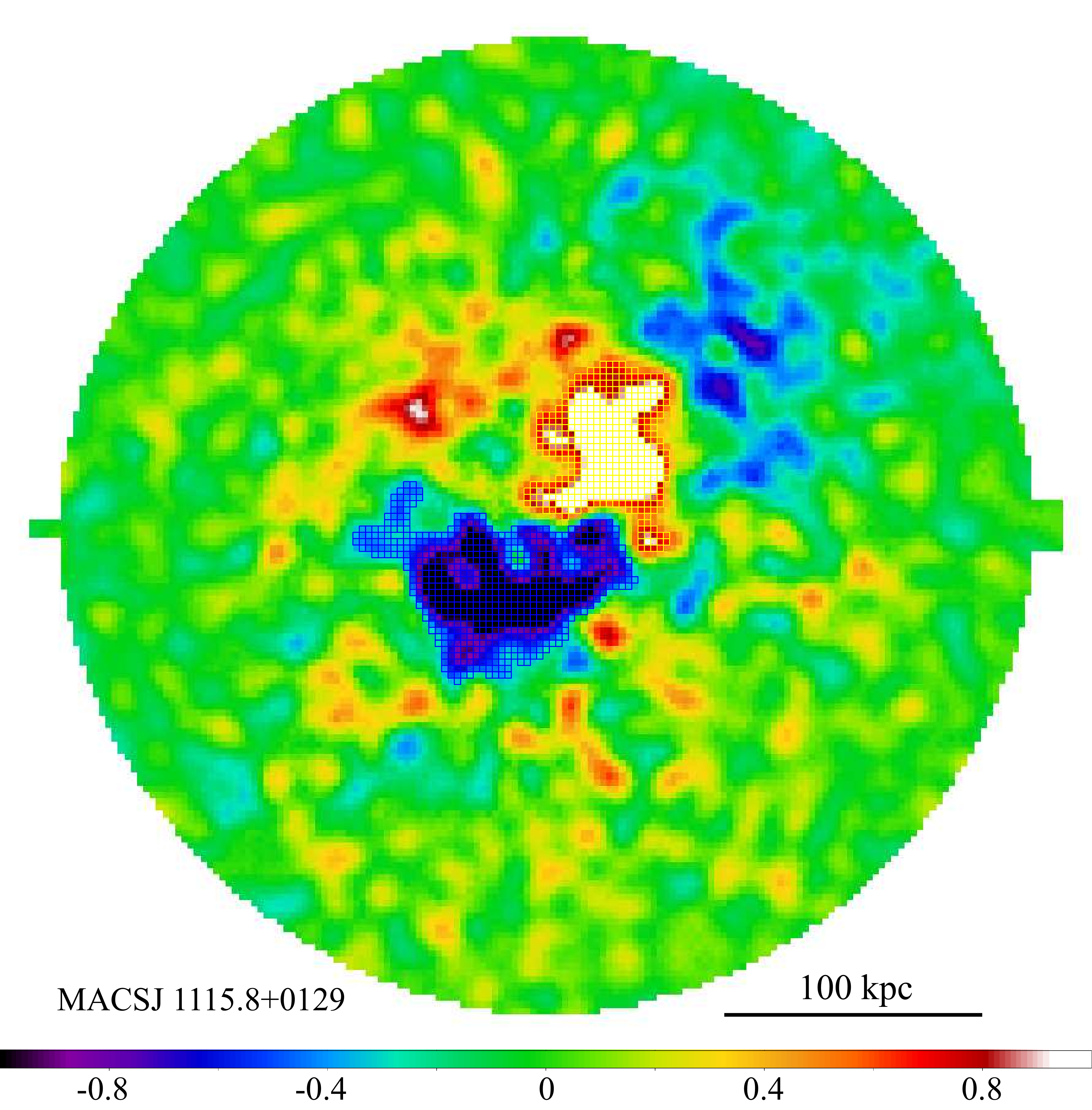} 
  \includegraphics[width=7.5cm]{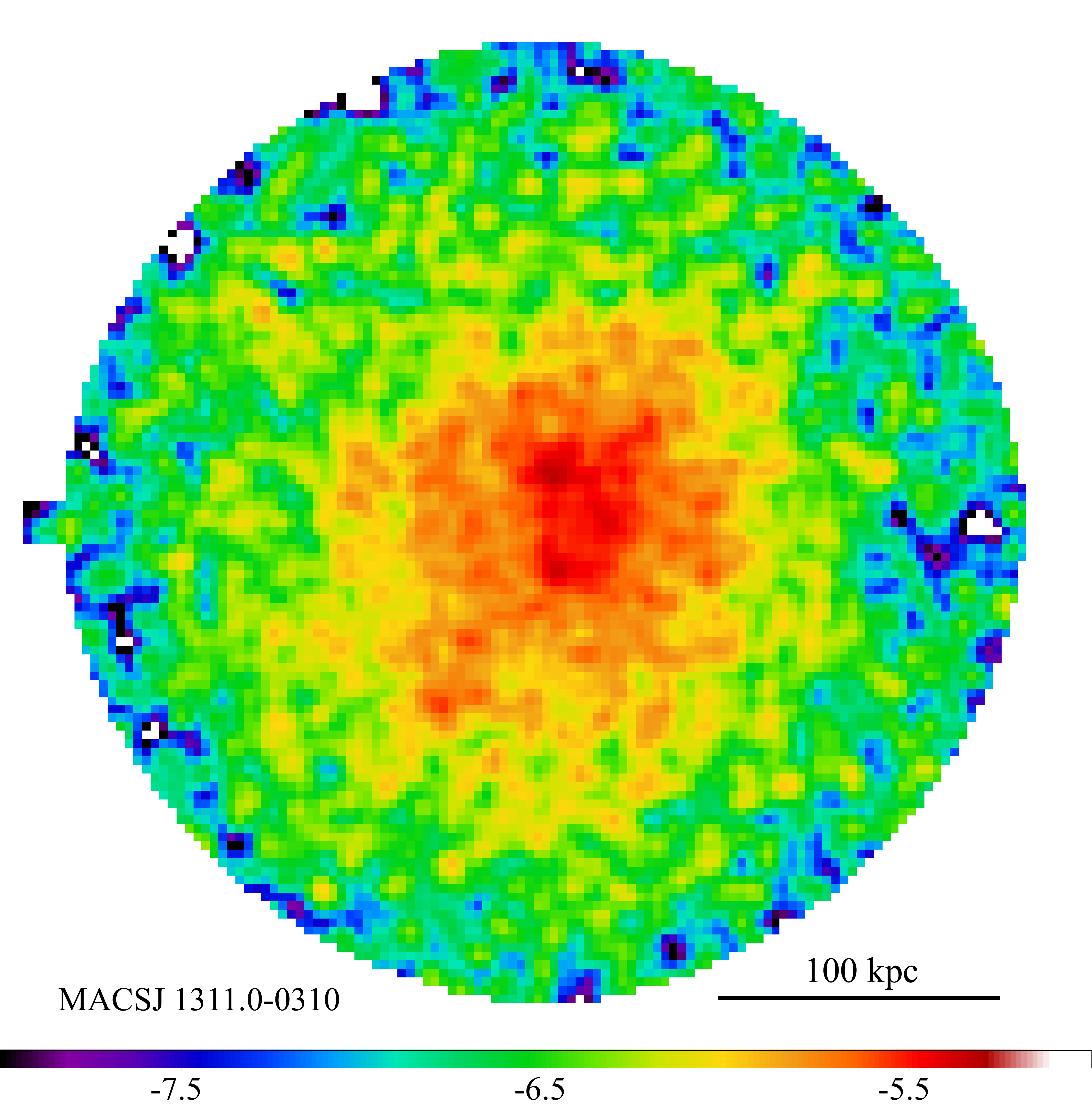}
  \includegraphics[width=7.5cm]{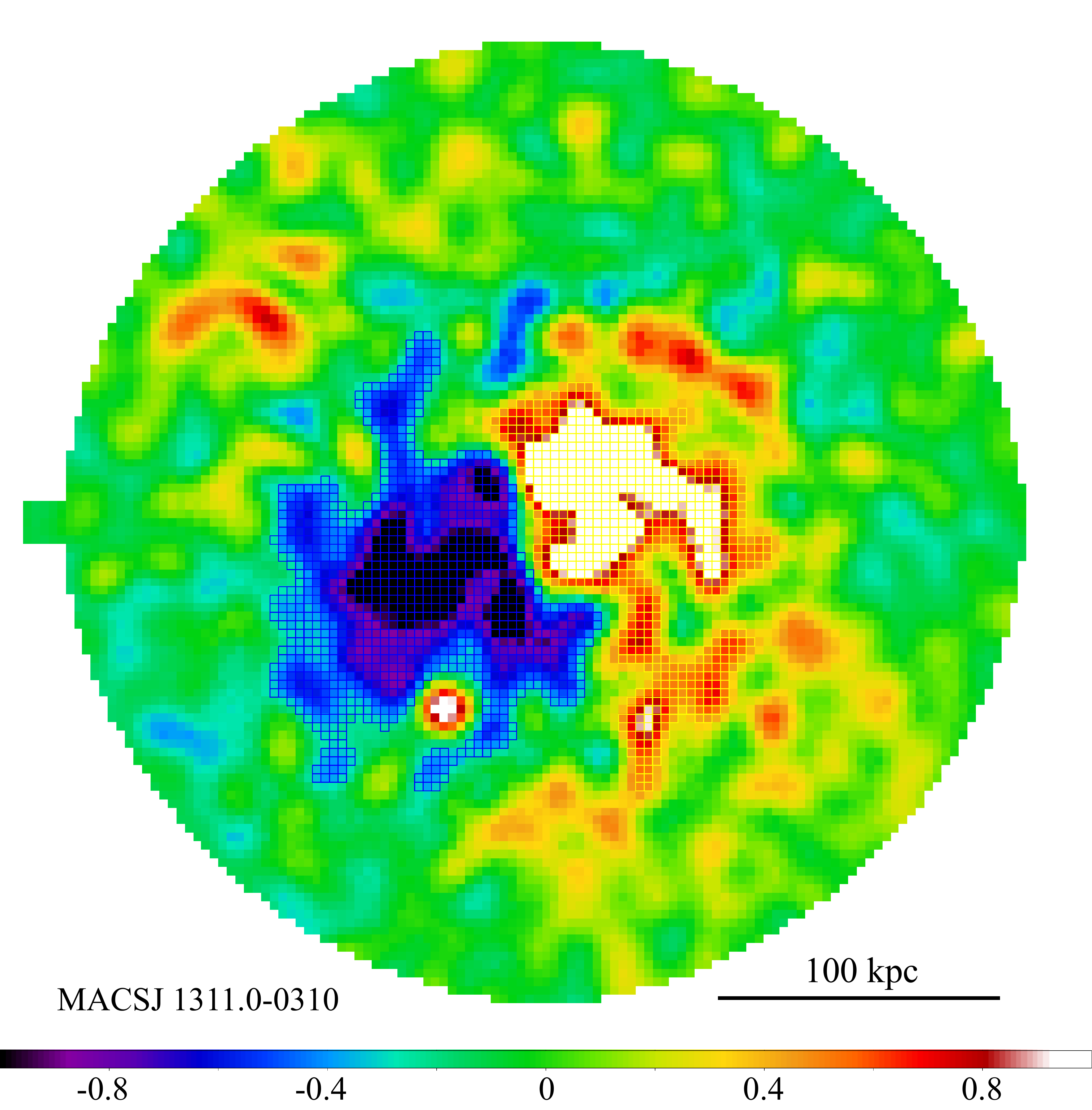} 
 \end{center}
\caption{Continued. MACSJ\,0429.6-0253 (top), MACSJ\,1115.8+0129 (middle), and MACSJ\,1311.0-0310 (bottom).
}
\end{figure*}

\begin{figure*}
\addtocounter{figure}{-1}
 \begin{center}
  \includegraphics[width=7.5cm]{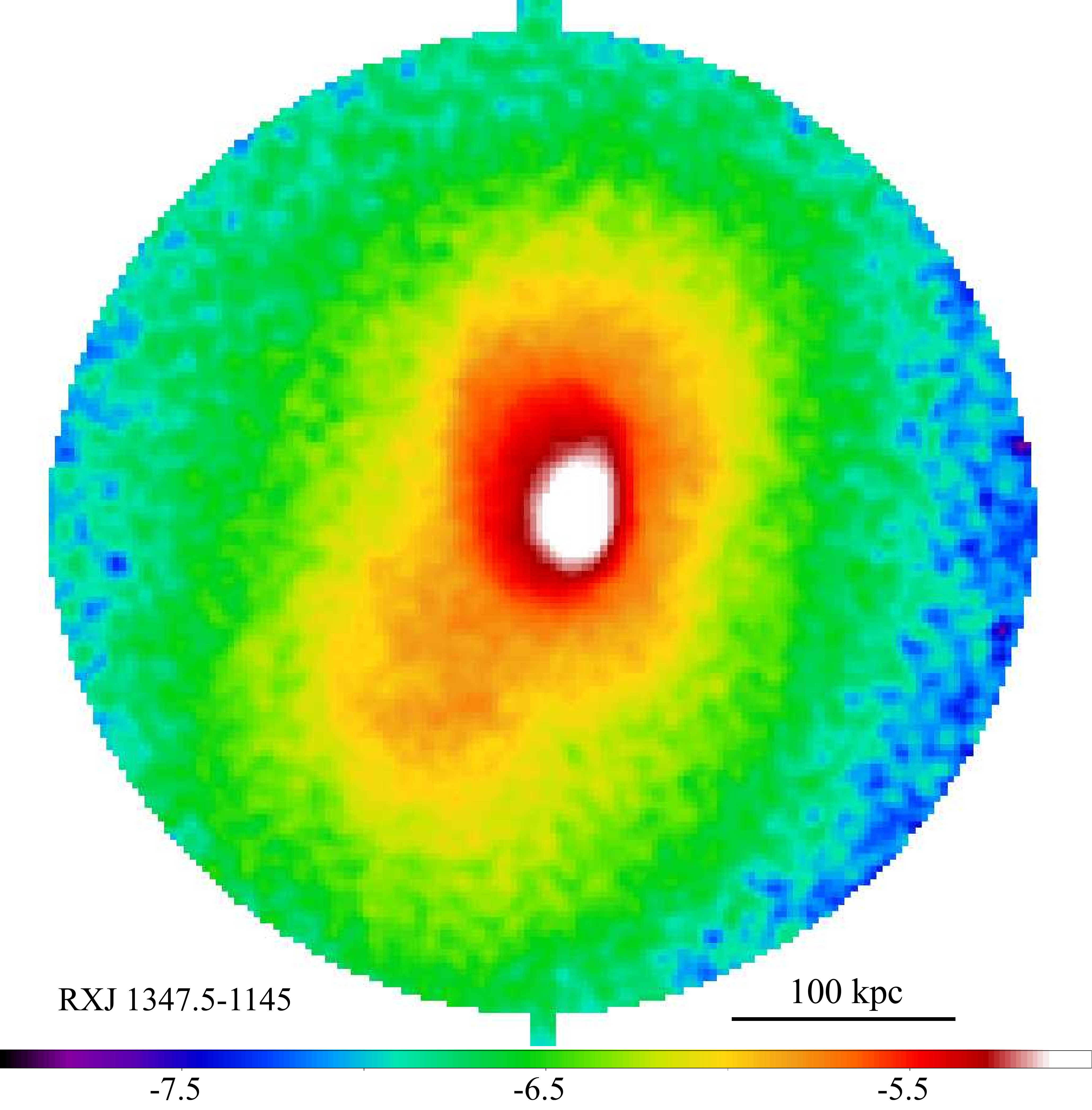}
  \includegraphics[width=7.5cm]{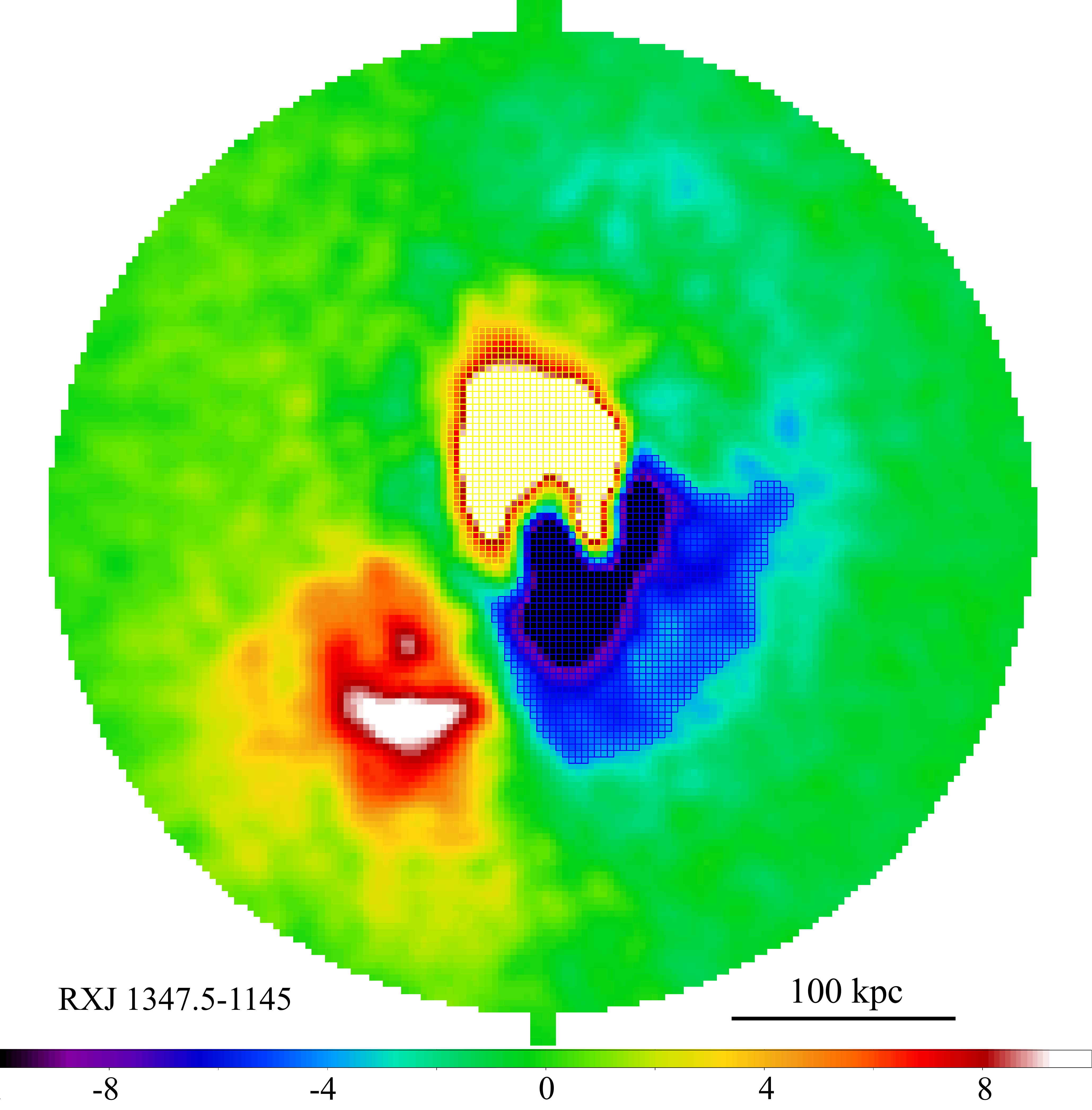} 
  \includegraphics[width=7.5cm]{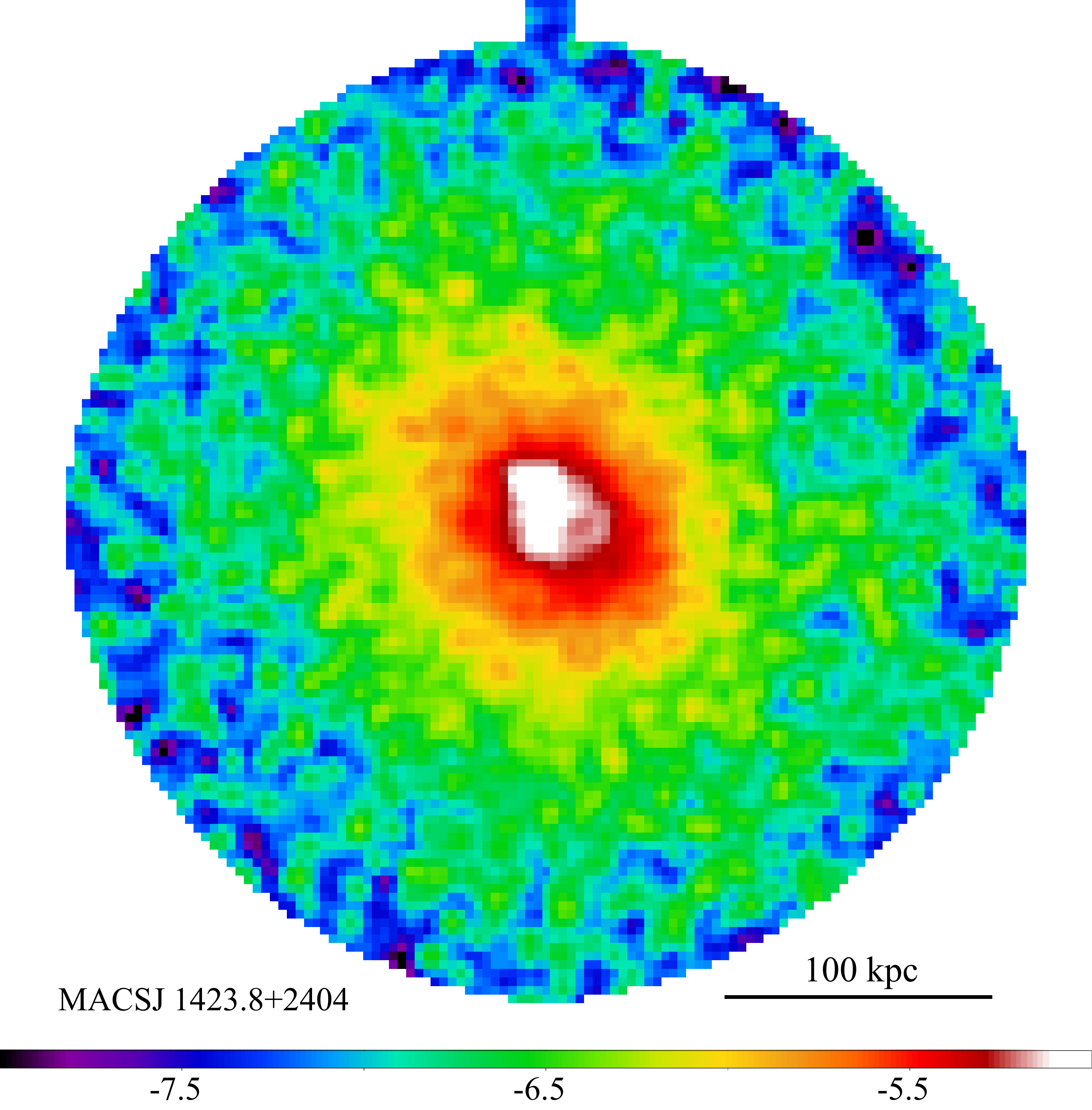}
  \includegraphics[width=7.5cm]{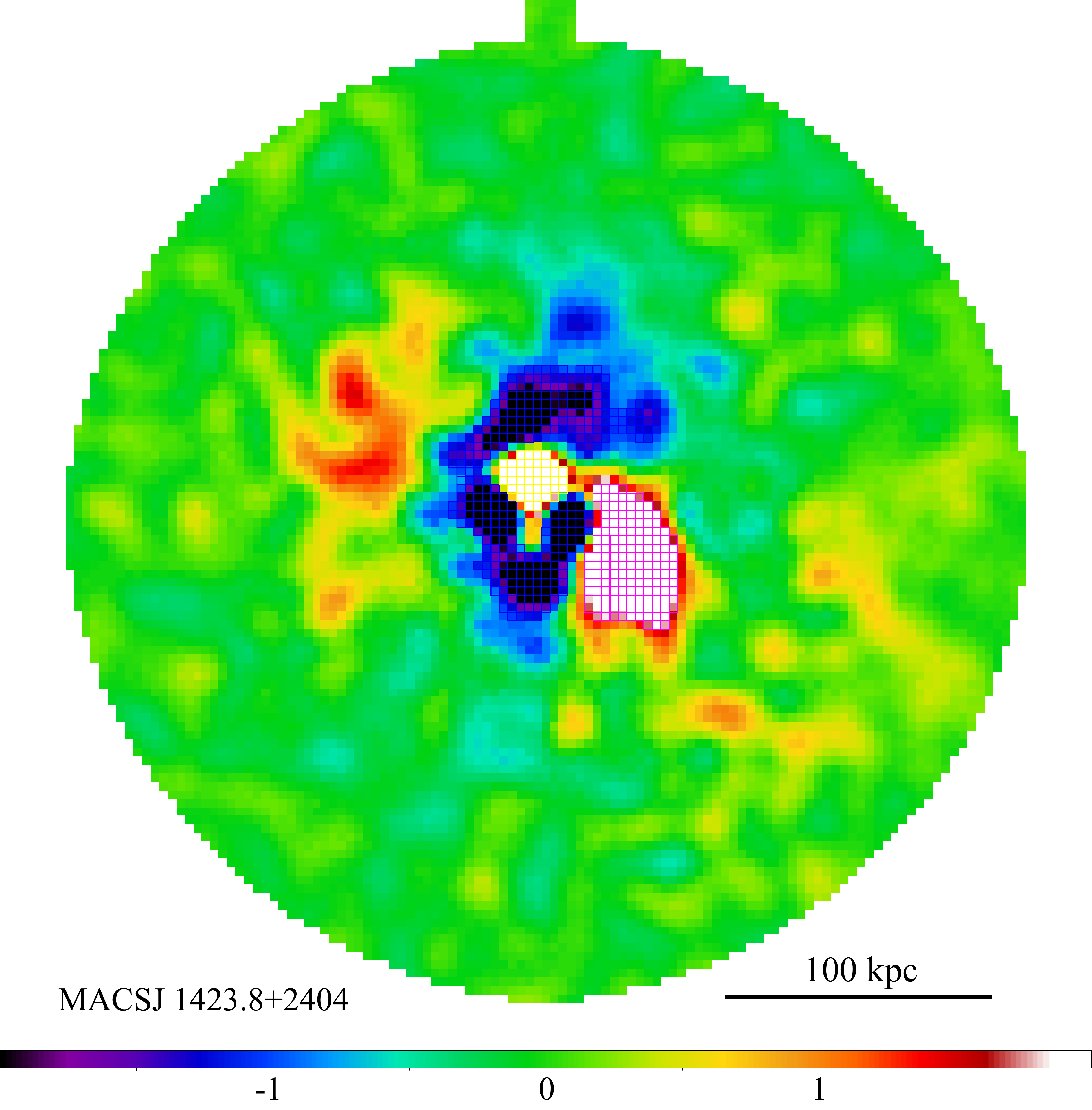} 
  \includegraphics[width=7.5cm]{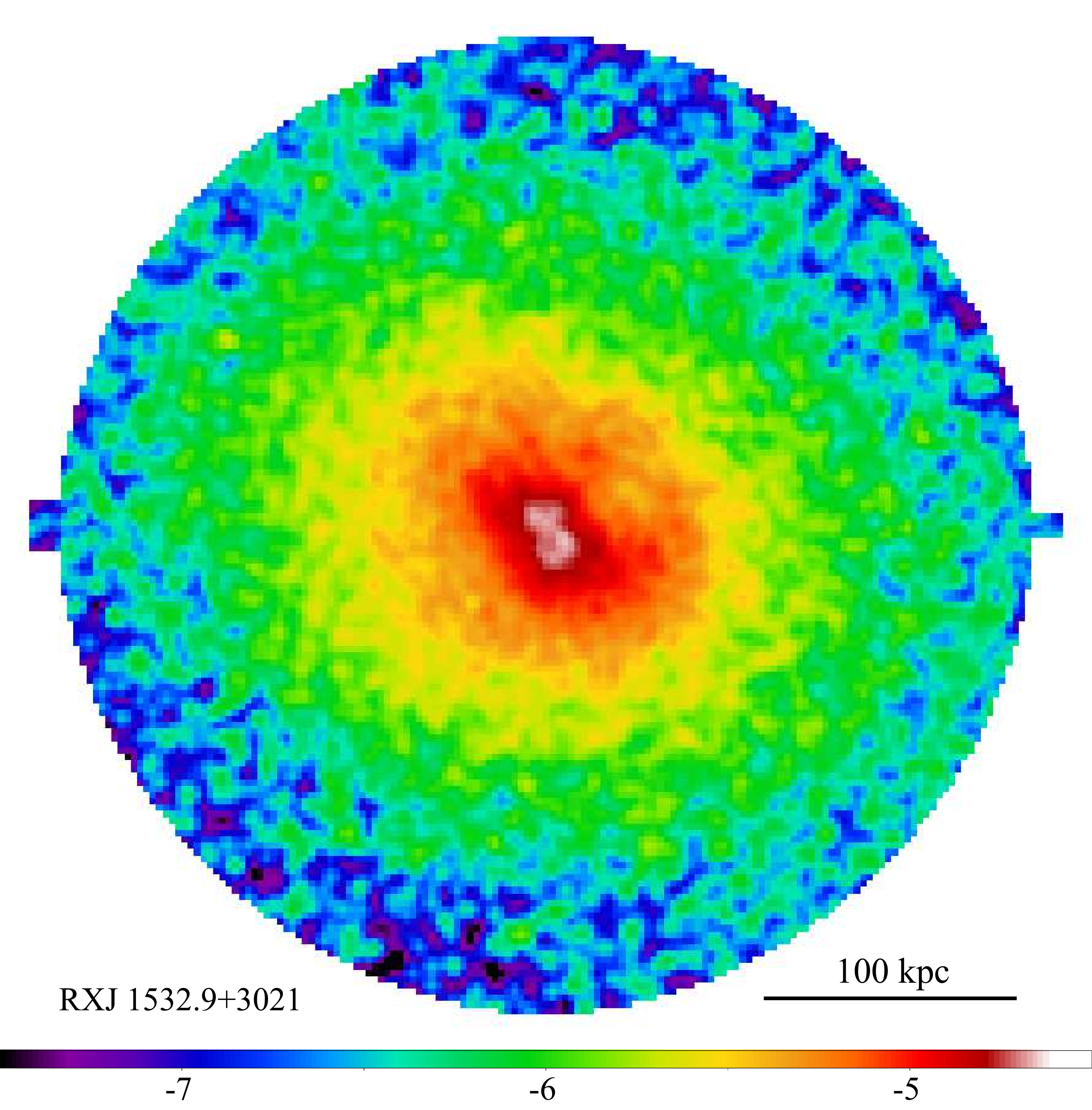}
  \includegraphics[width=7.5cm]{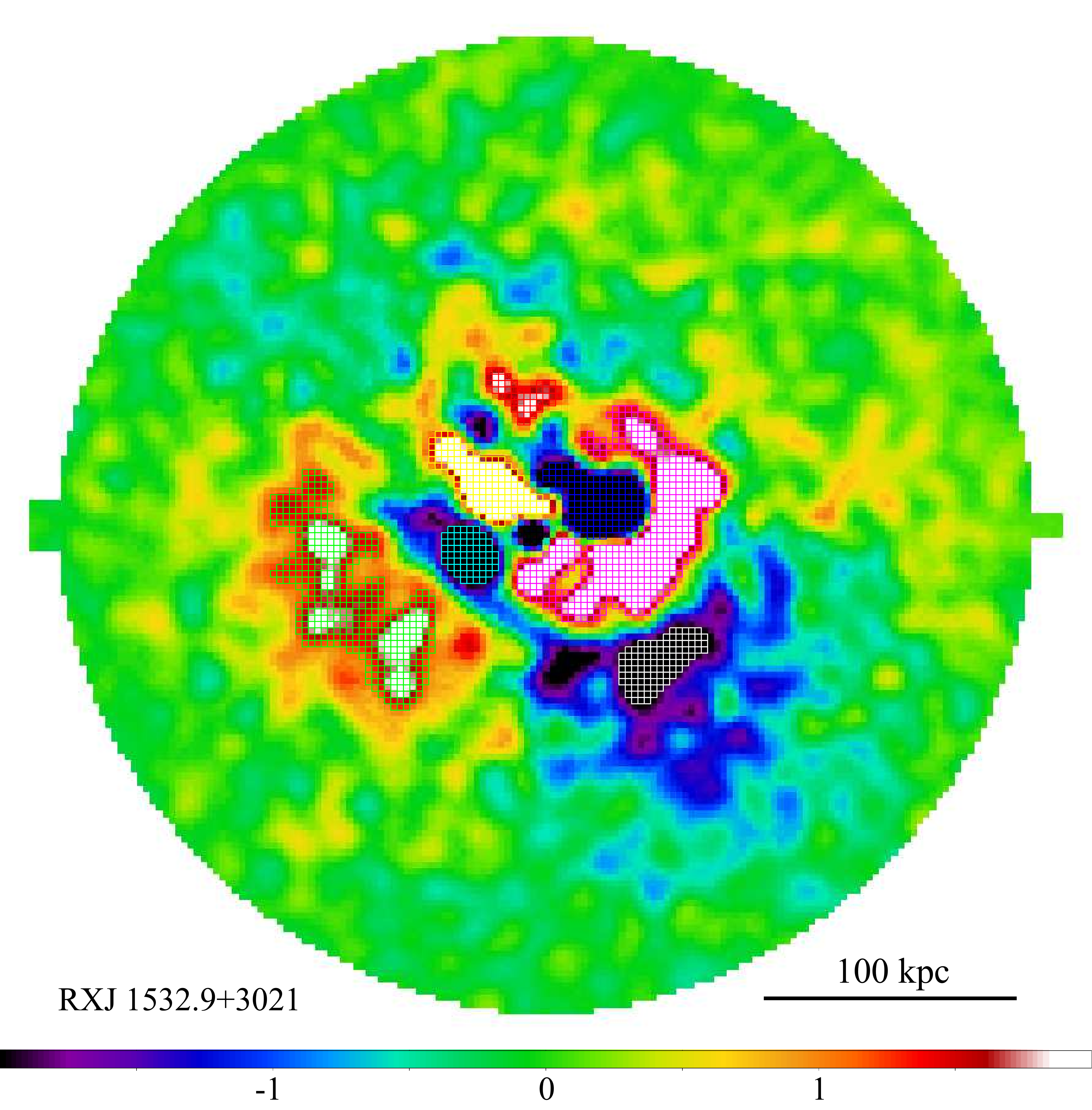}   
 \end{center}
\caption{Continued. RXJ\,1347.5-1145 (top), MACSJ\,1423.8+2404 (middle), and RXJ\,1532.9+3021 (bottom).
}
\end{figure*}

\begin{figure*}
\addtocounter{figure}{-1}
 \begin{center}
  \includegraphics[width=7.5cm]{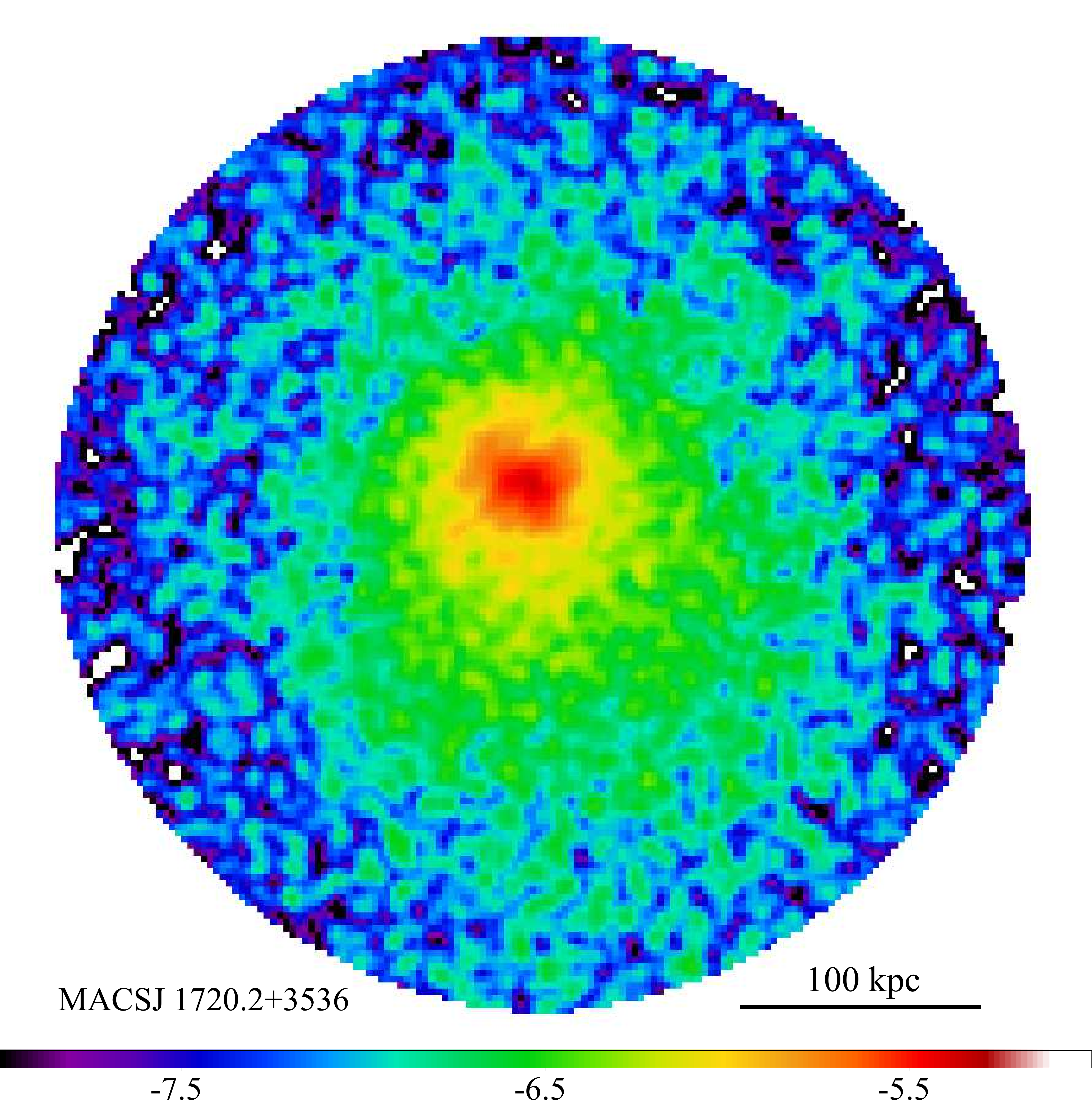}
  \includegraphics[width=7.5cm]{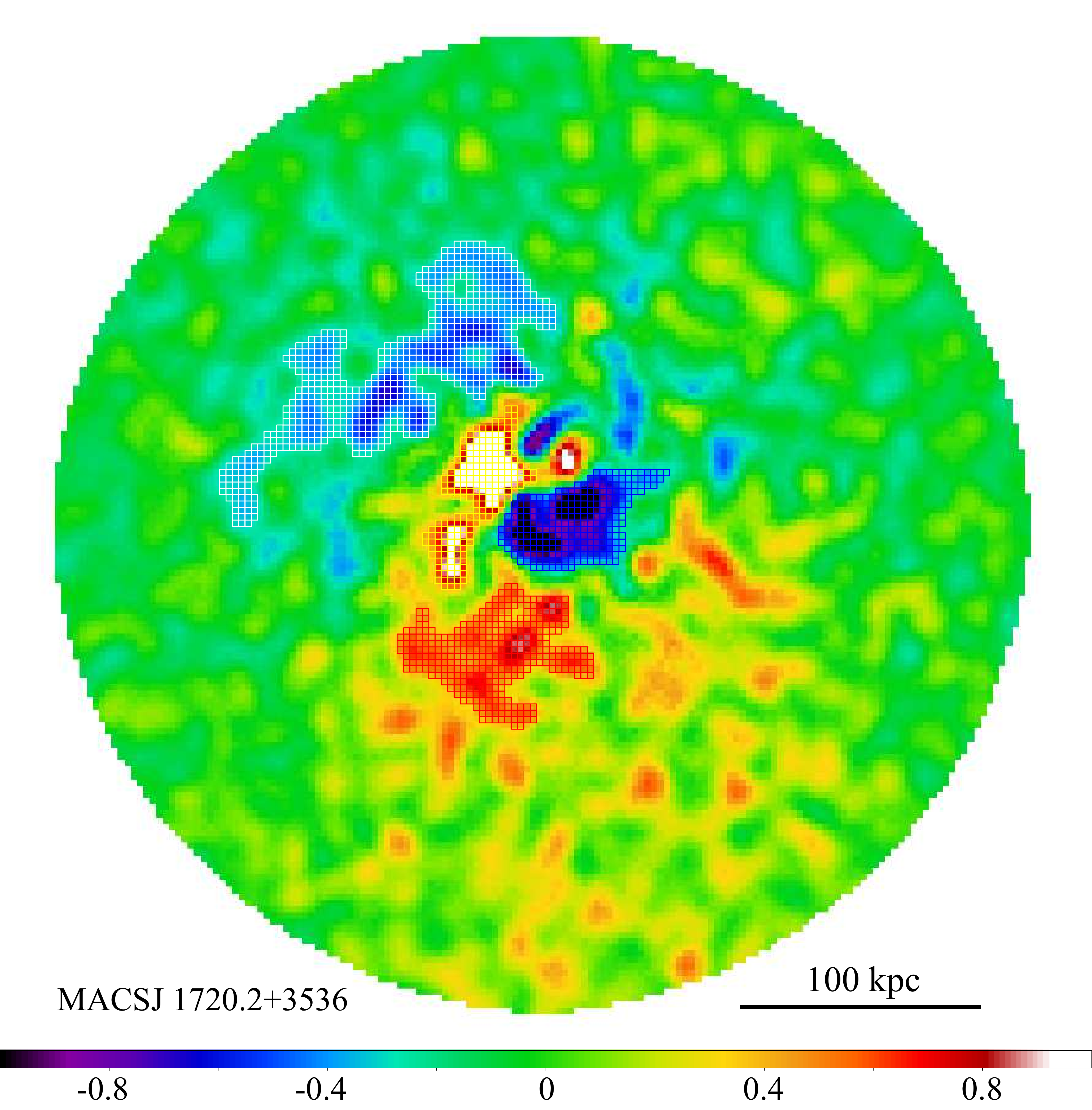}   
  \includegraphics[width=7.5cm]{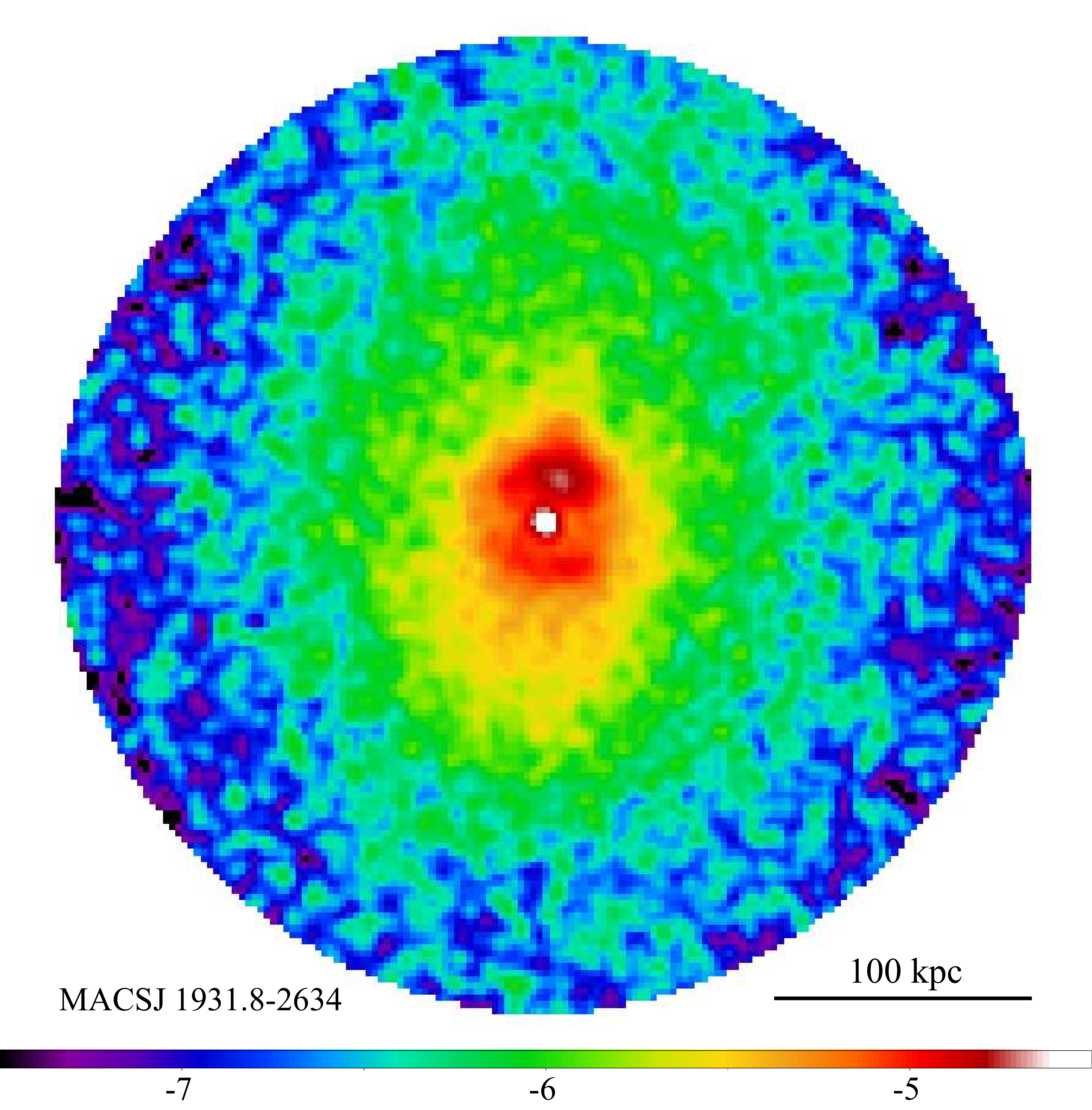}
  \includegraphics[width=7.5cm]{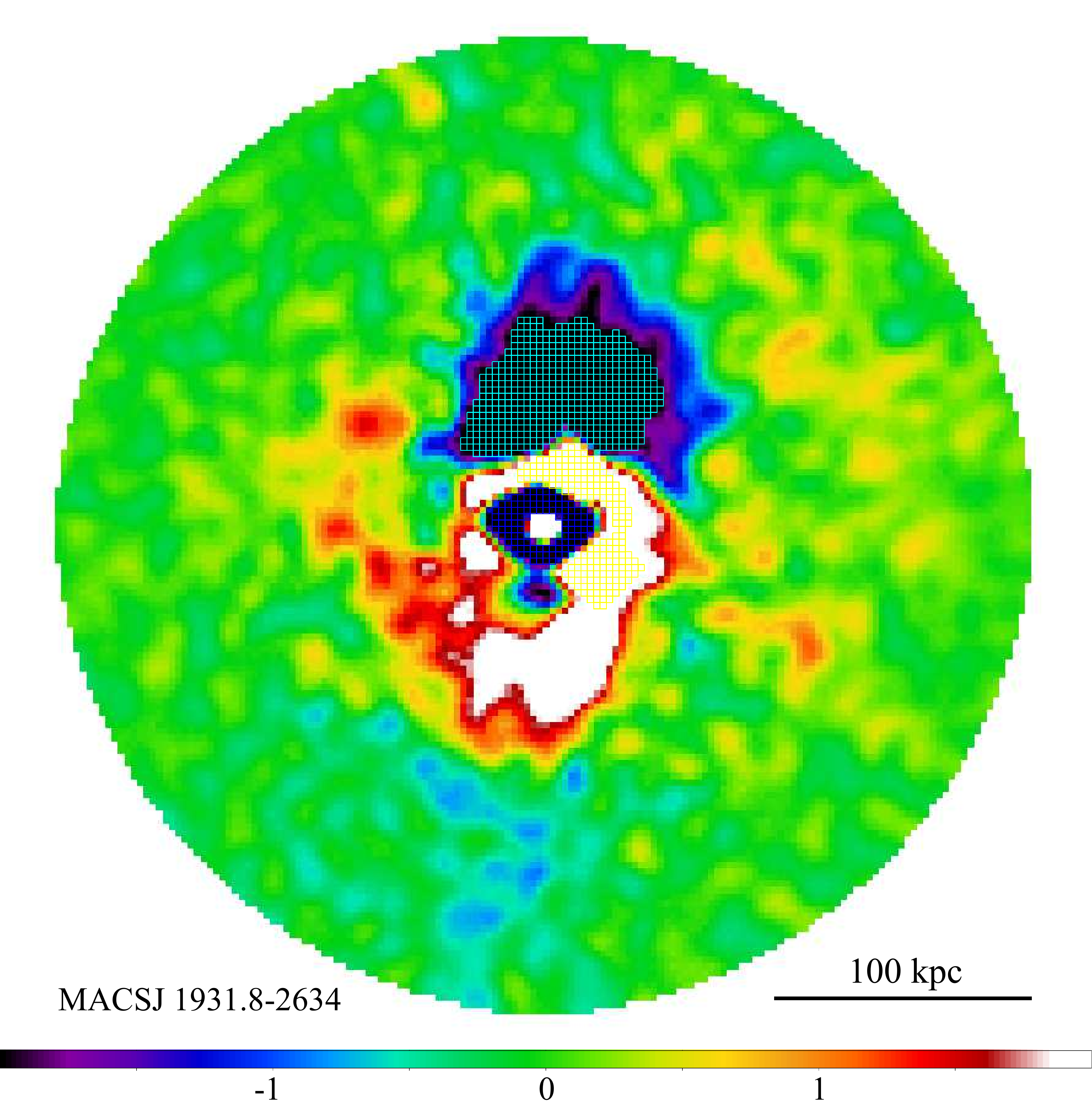} 
  \includegraphics[width=7.5cm]{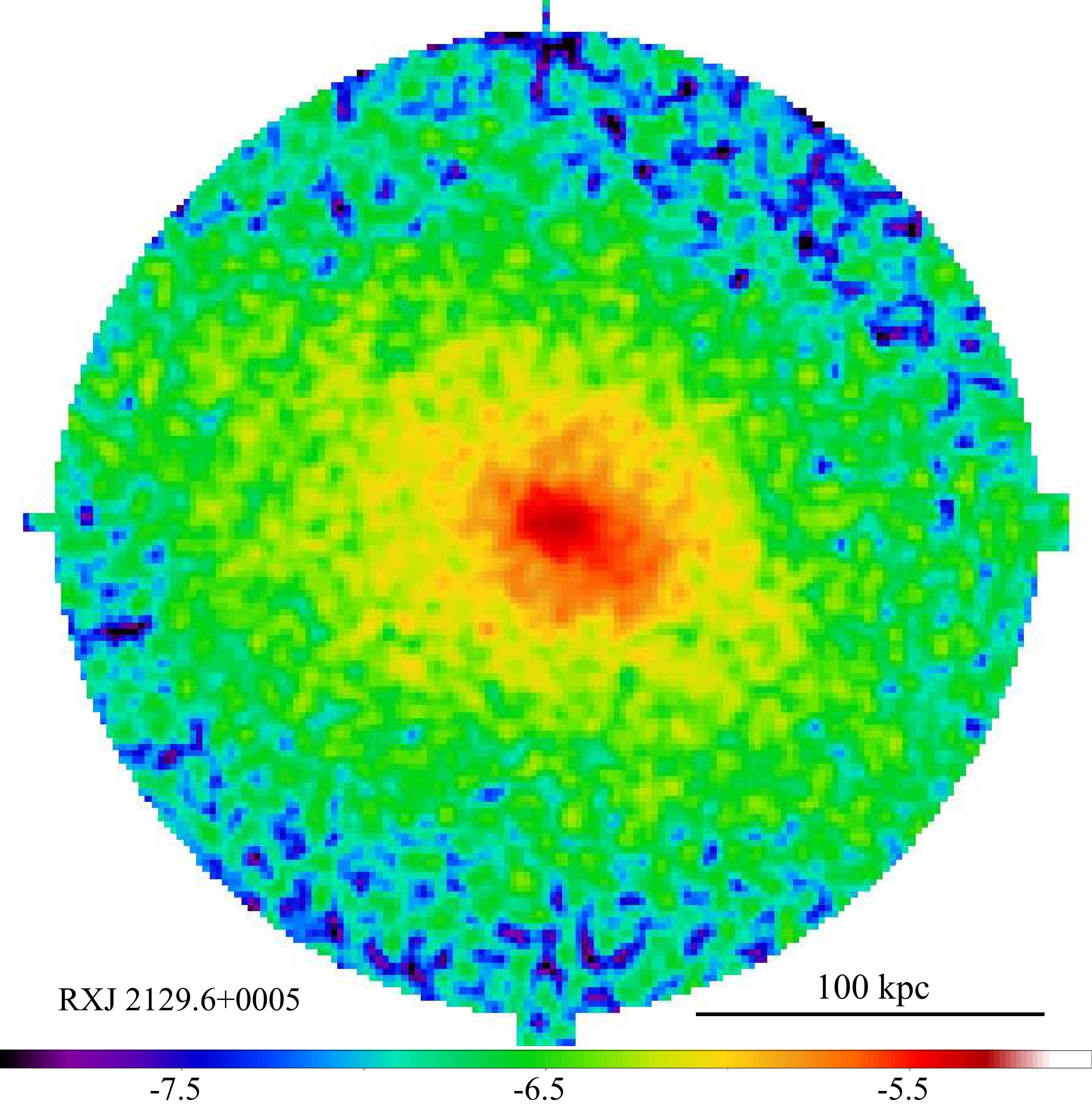}
  \includegraphics[width=7.5cm]{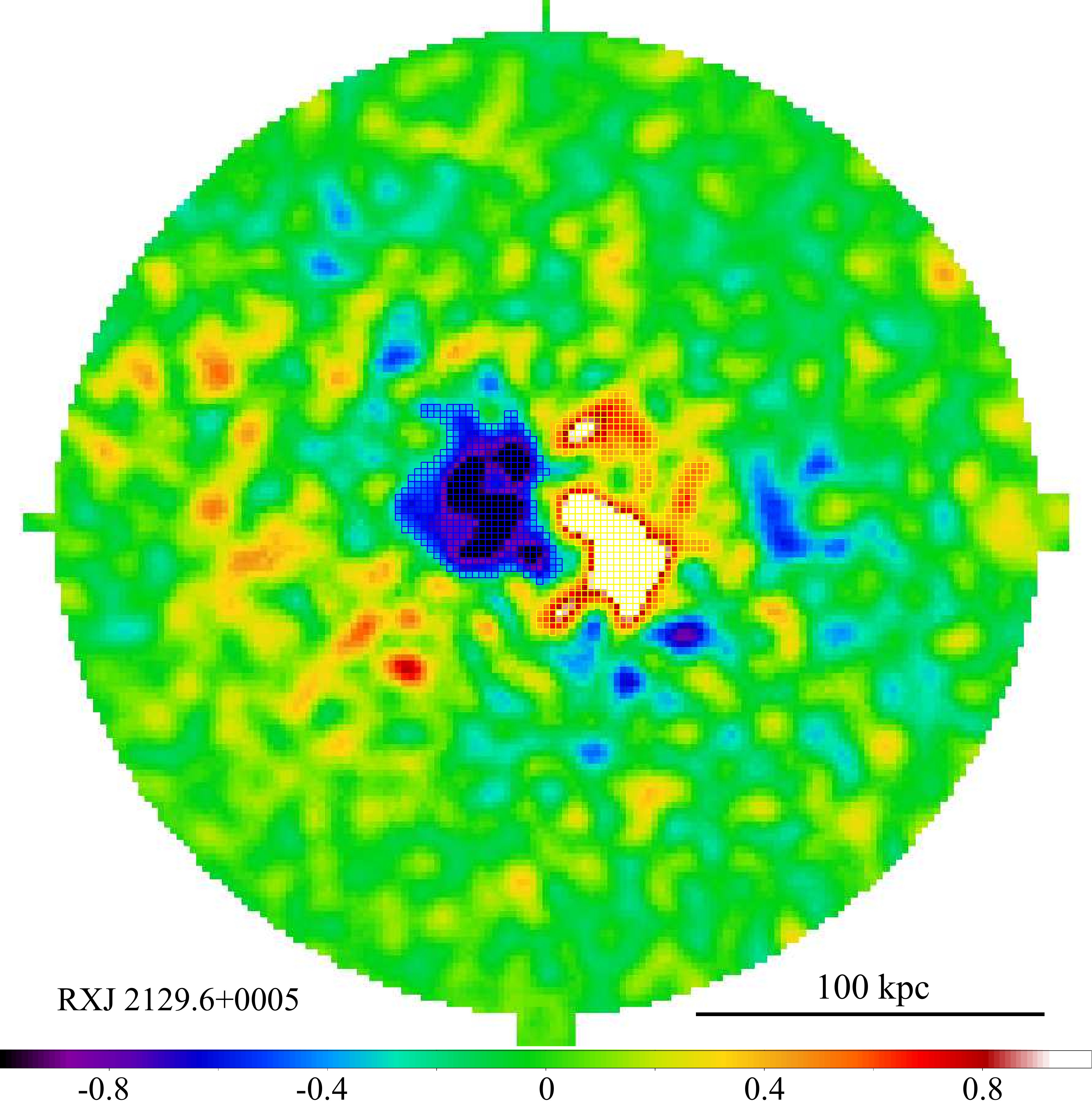}   
 \end{center}
\caption{Continued. MACS\,J1720.2+3536 (top), MACS\,J1931.8-2634 (middle), and RXJ\,2129.6+0005 (bottom).
}
\end{figure*}

\begin{figure*}
\addtocounter{figure}{-1}
 \begin{center}
  \includegraphics[width=7.5cm]{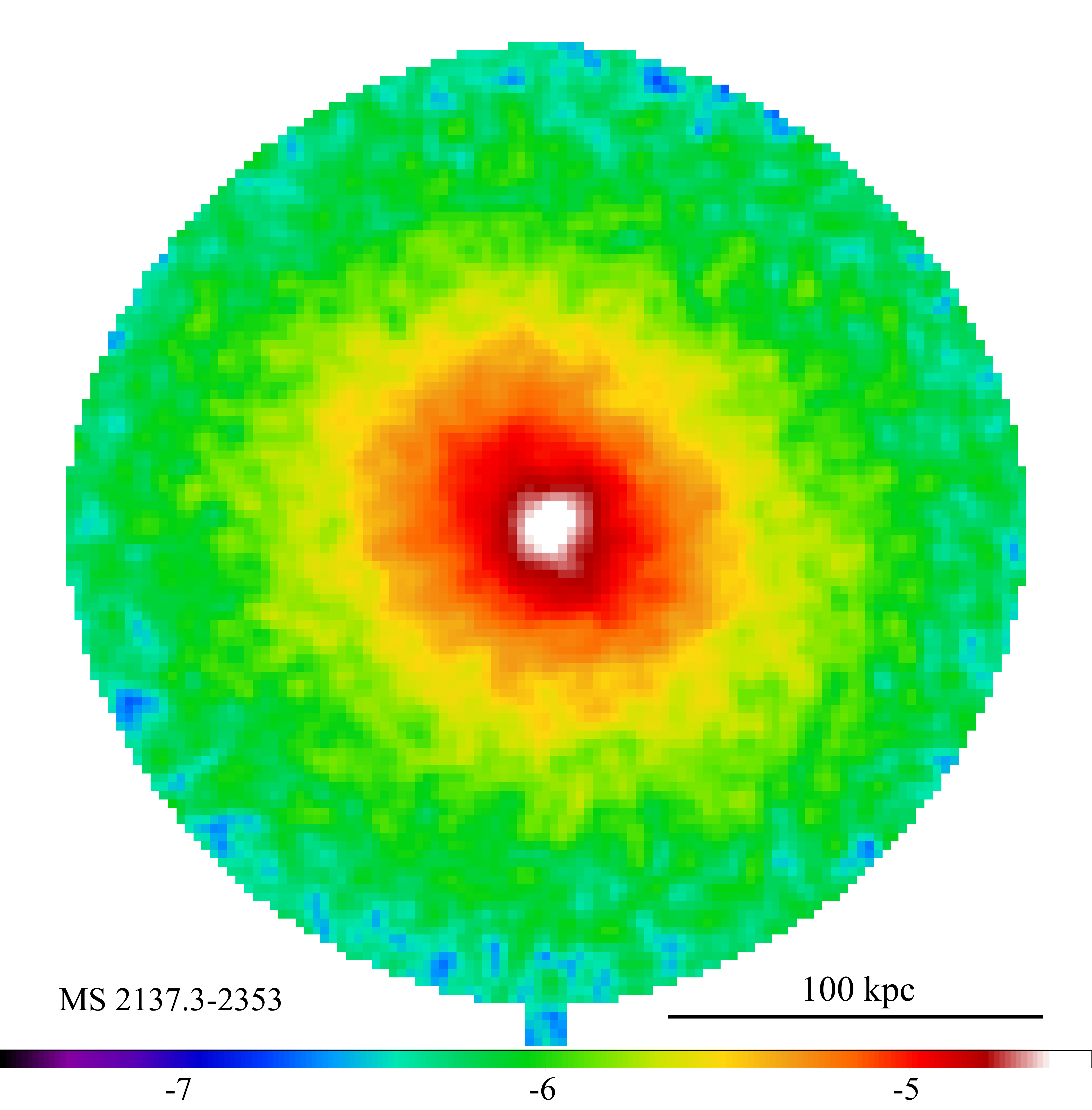}
  \includegraphics[width=7.5cm]{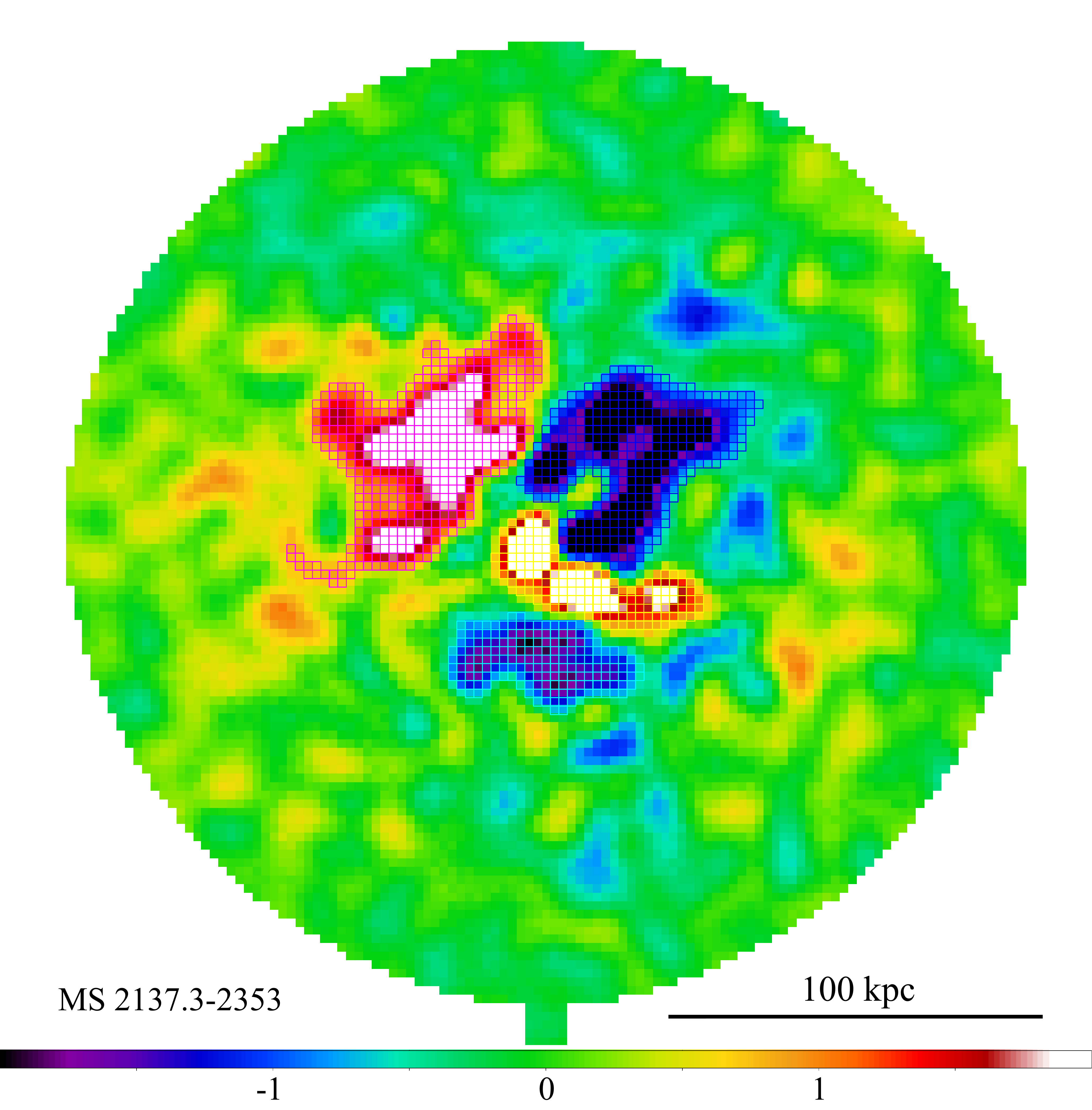}   
 \end{center}
\caption{Continued. MS\,2137.3-2353.
}
\end{figure*}

\section{Verification of our imaging analysis using simulation}
\label{sec:sim}

In Section~\ref{sec:img}, we have defined and applied our detection algorithm into the X-ray residual images to identify perturbed regions. To verify our methodology described in this paper, we carried out a hydrodynamic simulation. We tested our algorithm using datasets taken from the simulation.

\subsection{Configuration of simulation}

We performed a hydrodynamic simulation of a cluster merger to produce gas sloshing in a cool core by using the adaptive mesh refinement \citep[AMR;][]{Berger84, Berger89} code GAMER-2 \citep[][]{Schive18}. GAMER-2 implements GPU accelerations to speed up time consuming calculations. The box size of our simulation was 14.3\,Mpc on a side. We reached a very high spatial resolution of 0.85\,kpc using the GPU cluster at the National Center for High-Performance Computing (NCHC) in Taiwan (for technical details of this run, see Molnar \& Schive 2019, in prep). This simulation was semi-adiabatic, which means that only shock heating but no additional heating and cooling processes were considered.

The initial conditions of this simulation are as follows.
\begin{enumerate}
\item The masses of the primary cluster and the infalling subcluster are $6.9 \times 10^{14}$\,\MO ~and $2.3 \times 10^{14}$\,\MO, respectively.
\item The dark matter halo follows an NFW profile. We assumed an isotropic distribution for the direction of the dark matter particle velocities, and their magnitudes were drawn from a velocity distribution function based on the Eddington formula \citep{Eddington16}. The gas density and temperature profiles were calculated assuming a power law entropy distribution and hydrostatic equilibrium. We adopted a gas mass fraction of $f_{\rm gas} = 0.1056$. For more details of our simulation setup, we refer to \cite{Schive18} and \cite{ZuHone11b}.
\item The impact parameter of a merger is 1\,Mpc and the infall velocity is 1352\,km\,s$^{-1}$.
\item The initial distance between the primary and the infalling subcluster is 2.6\,Mpc.
\item Both clusters host a cool core. The gas temperature of the primary cluster increases from 2.1\,keV at the center to a maximum of 5.3\,keV at 210\,kpc, whereas that of the infalling subcluster increases from 1\,keV at the center to 2.6\,keV at 130\,kpc.
\end{enumerate}

\subsection{Tests of our imaging analysis using simulated X-ray data}
\label{sec:sim_ana}

\begin{figure*}
 \begin{center}
  \includegraphics[width=5.2cm]{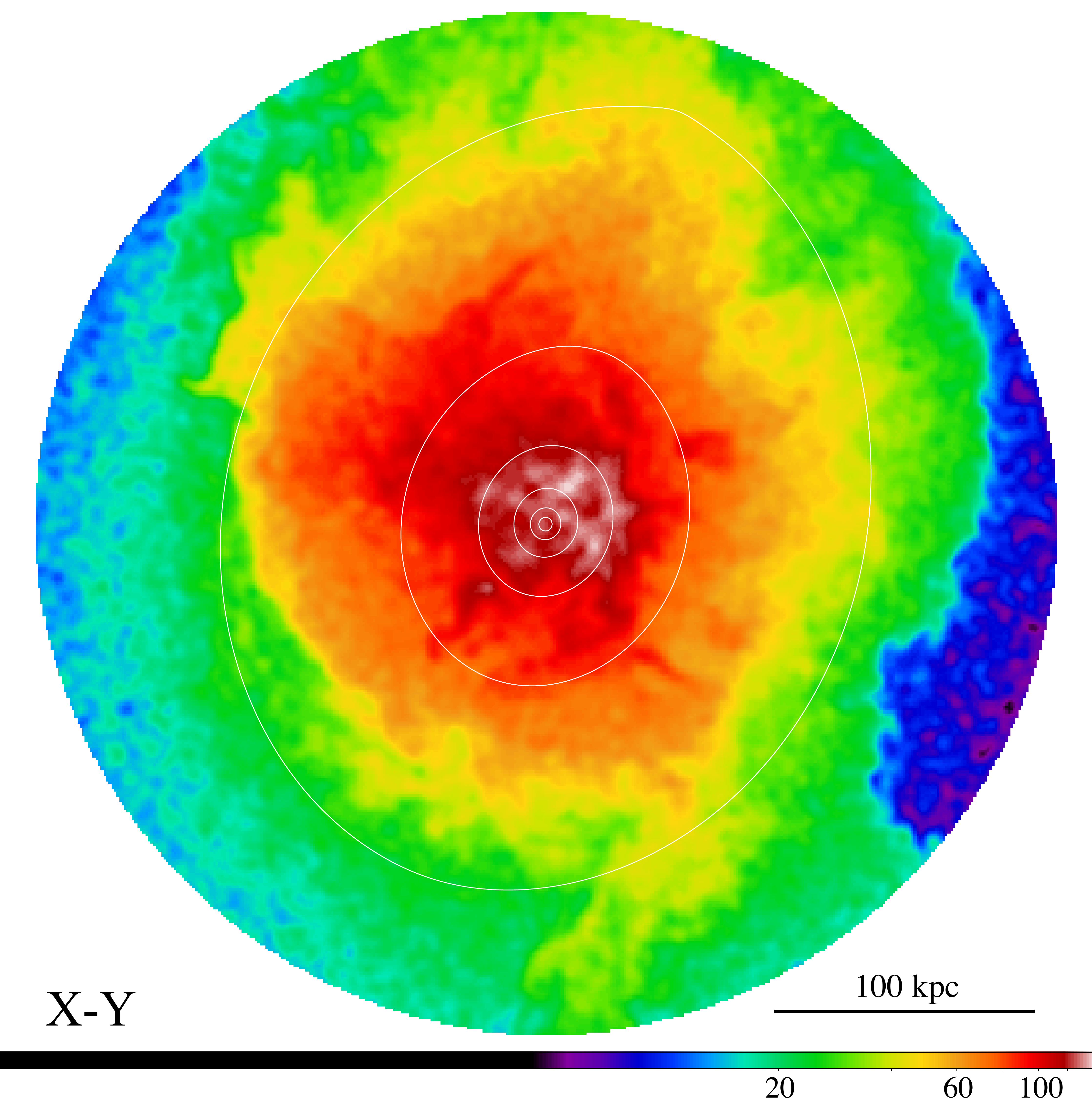}
  \includegraphics[width=5.2cm]{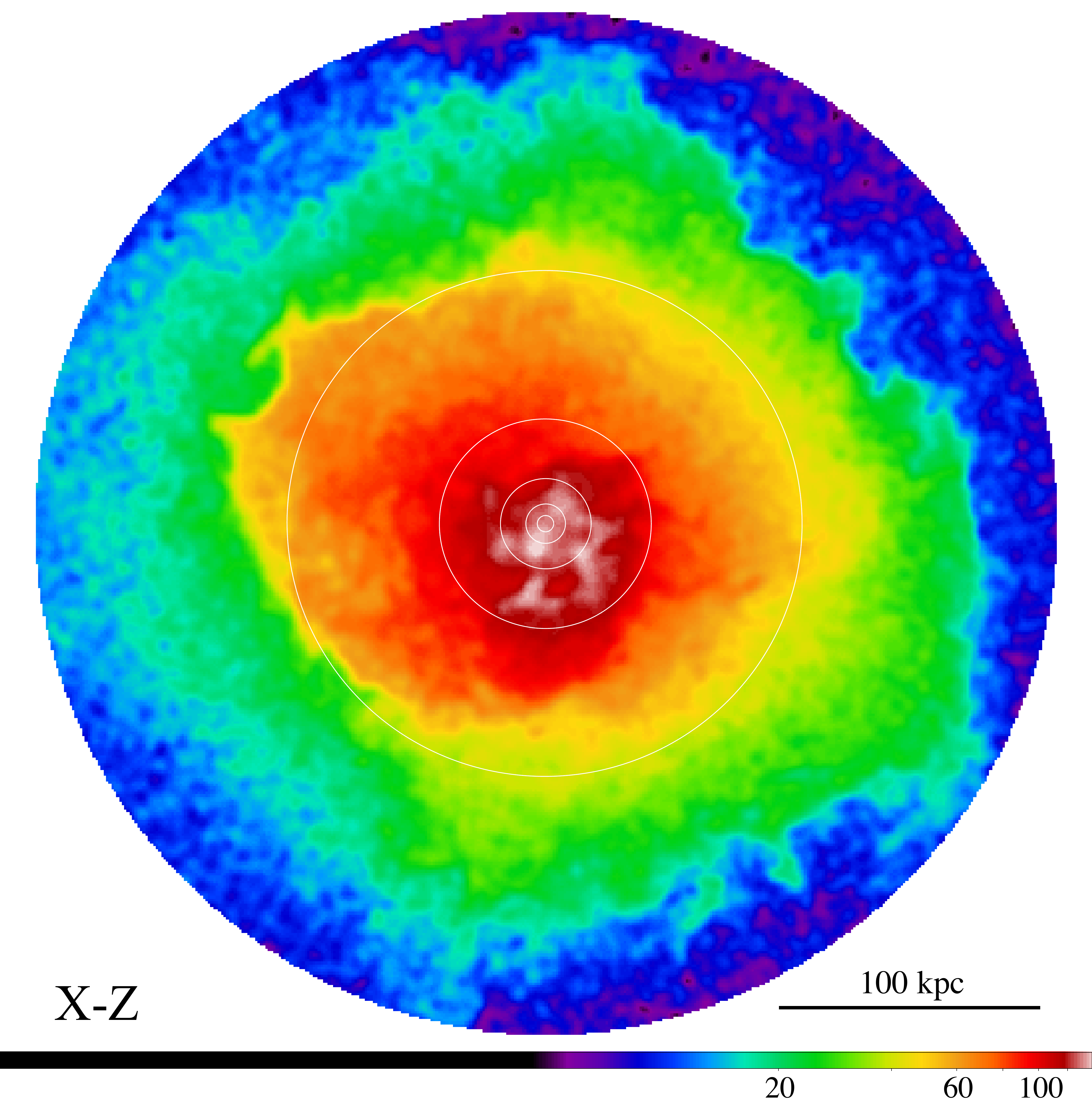}
  \includegraphics[width=5.2cm]{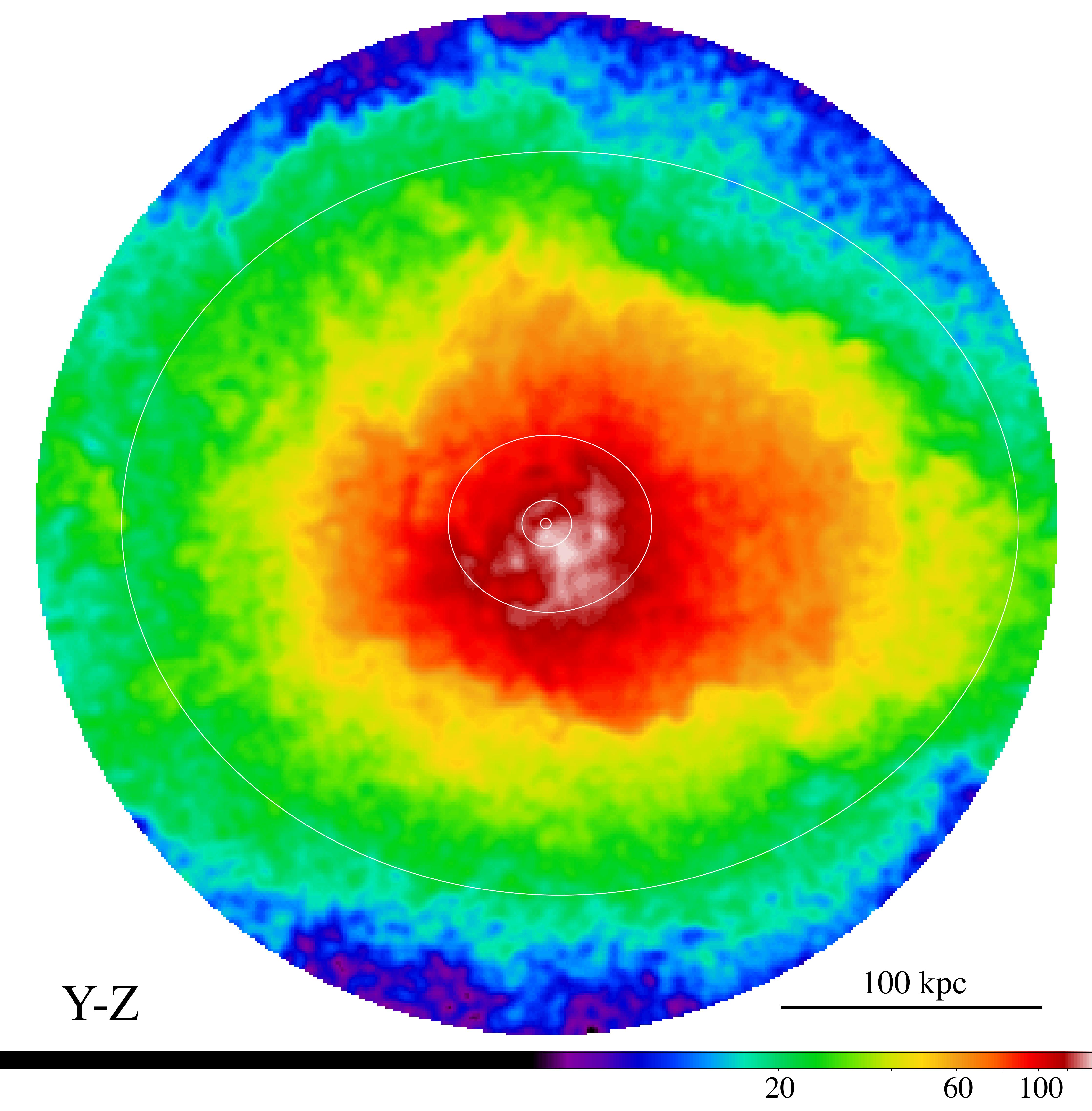}
  \includegraphics[width=5.2cm]{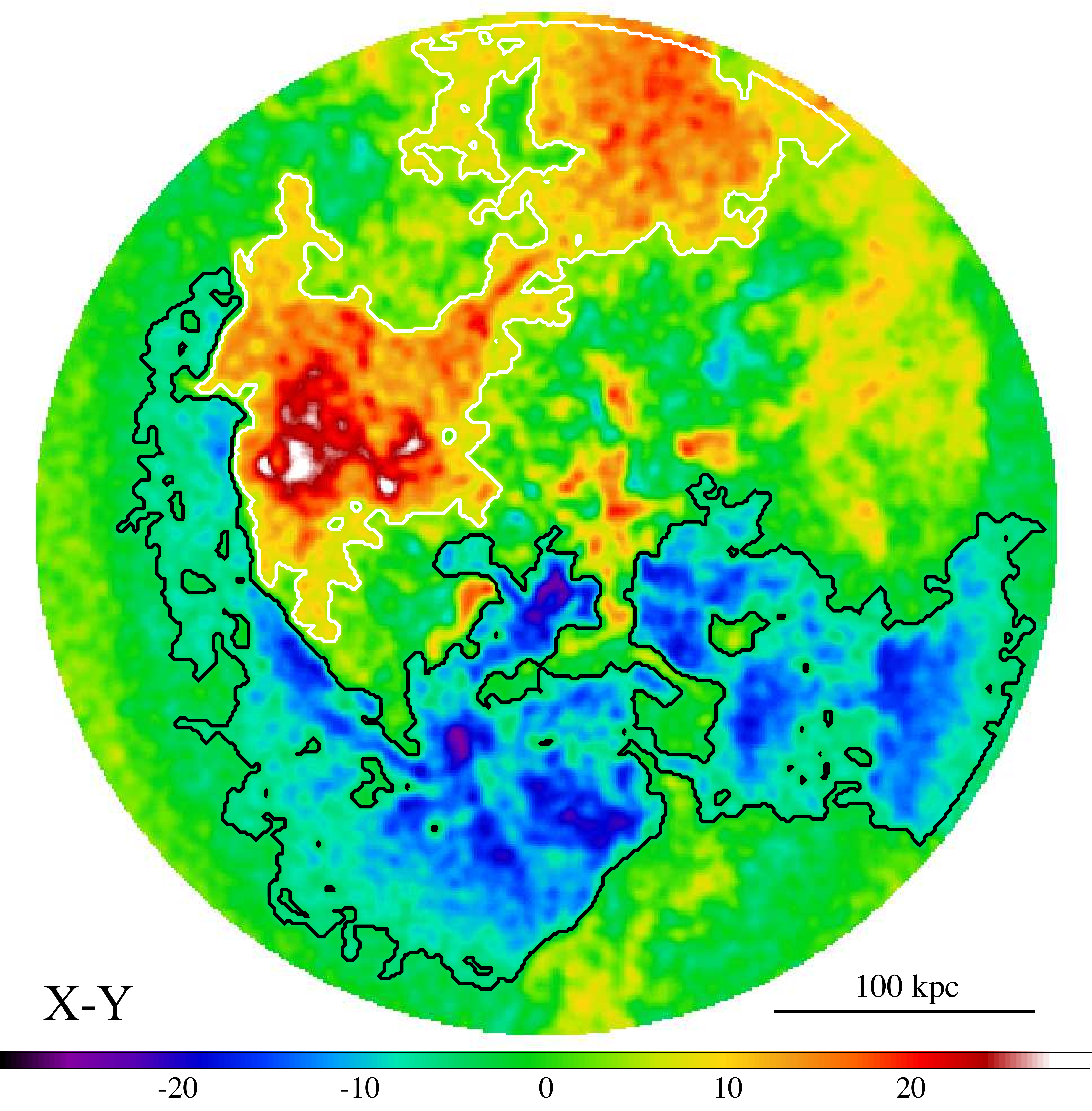}
  \includegraphics[width=5.2cm]{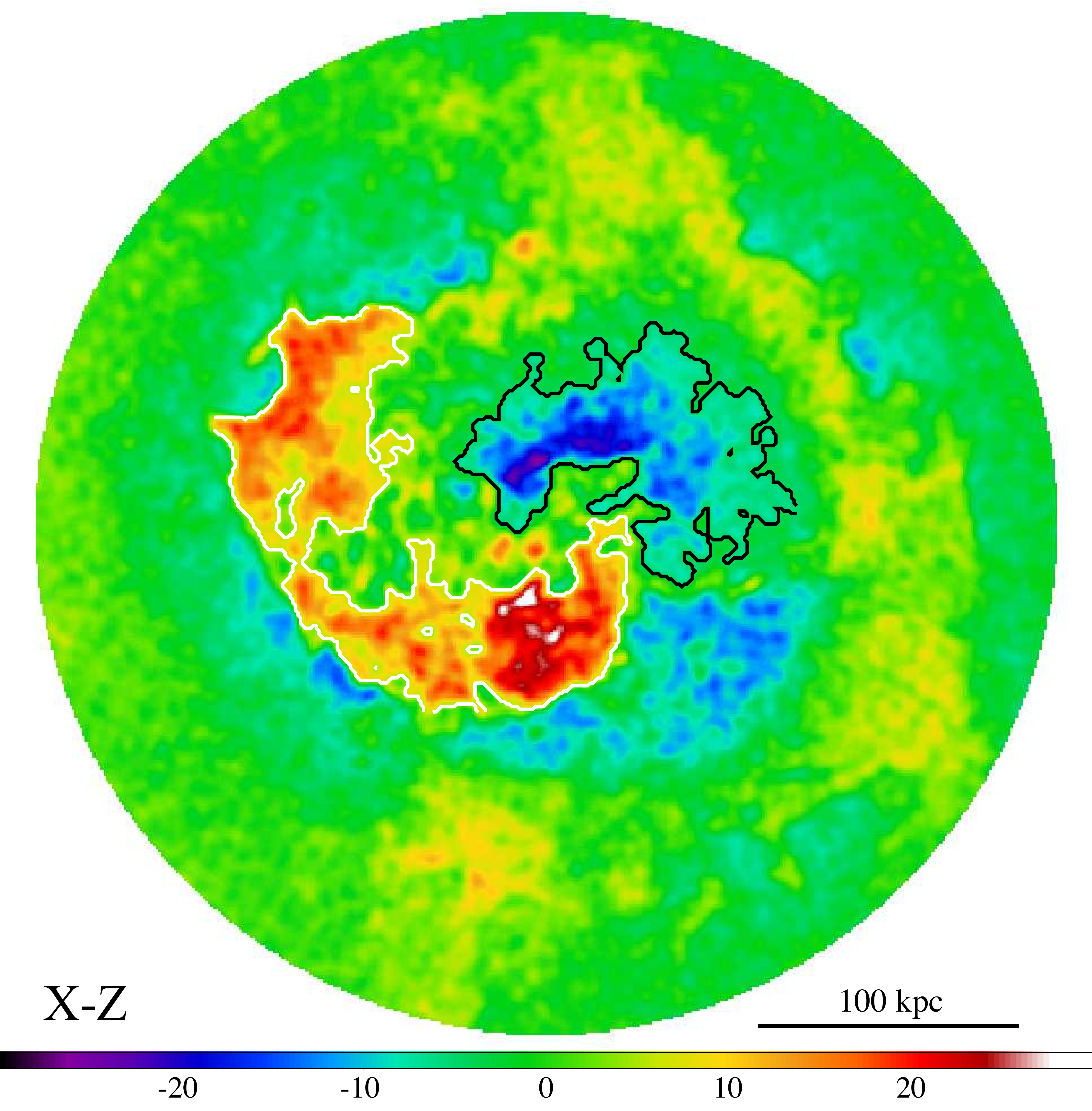} 
  \includegraphics[width=5.2cm]{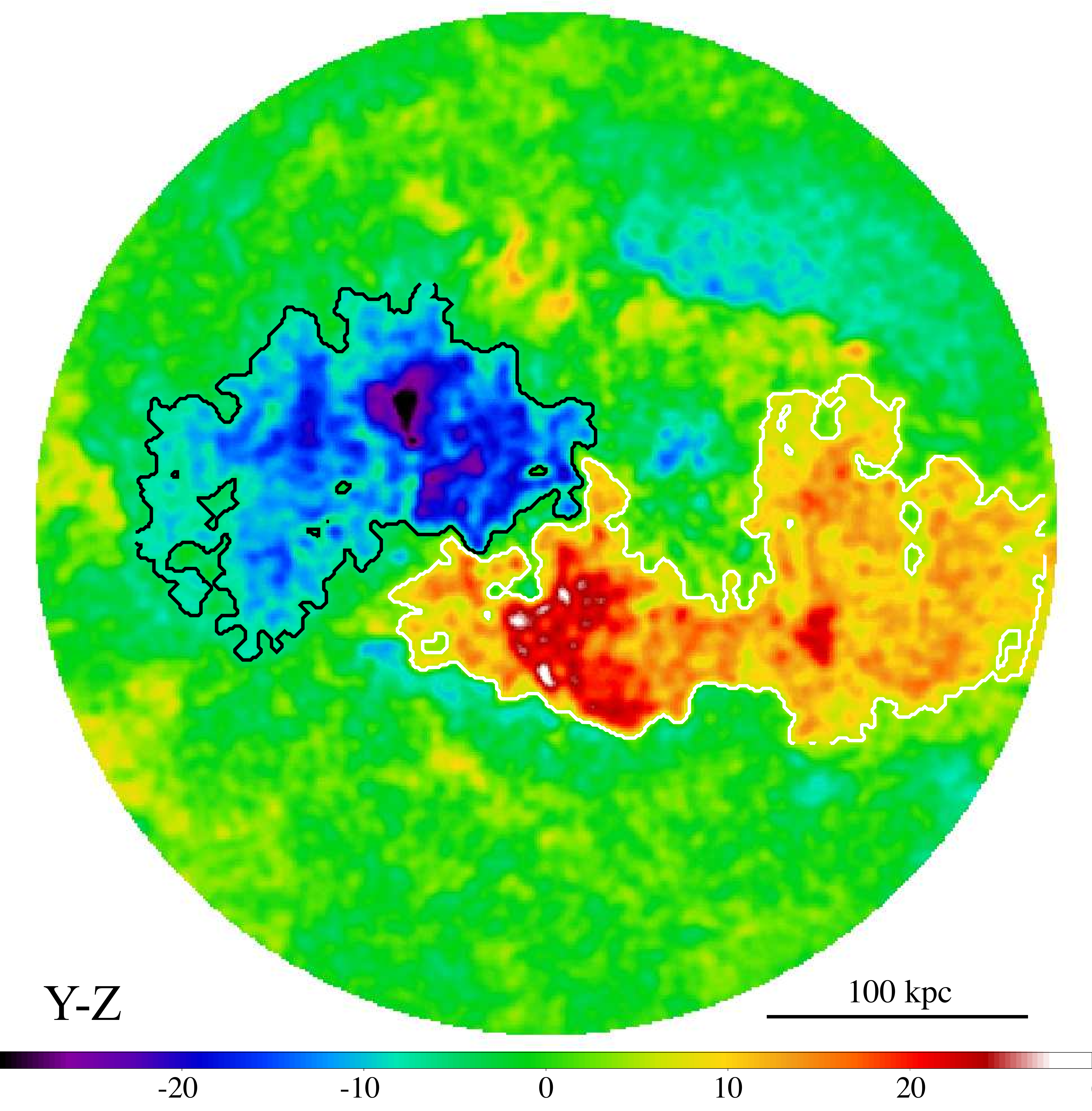} 
  \includegraphics[width=5.2cm]{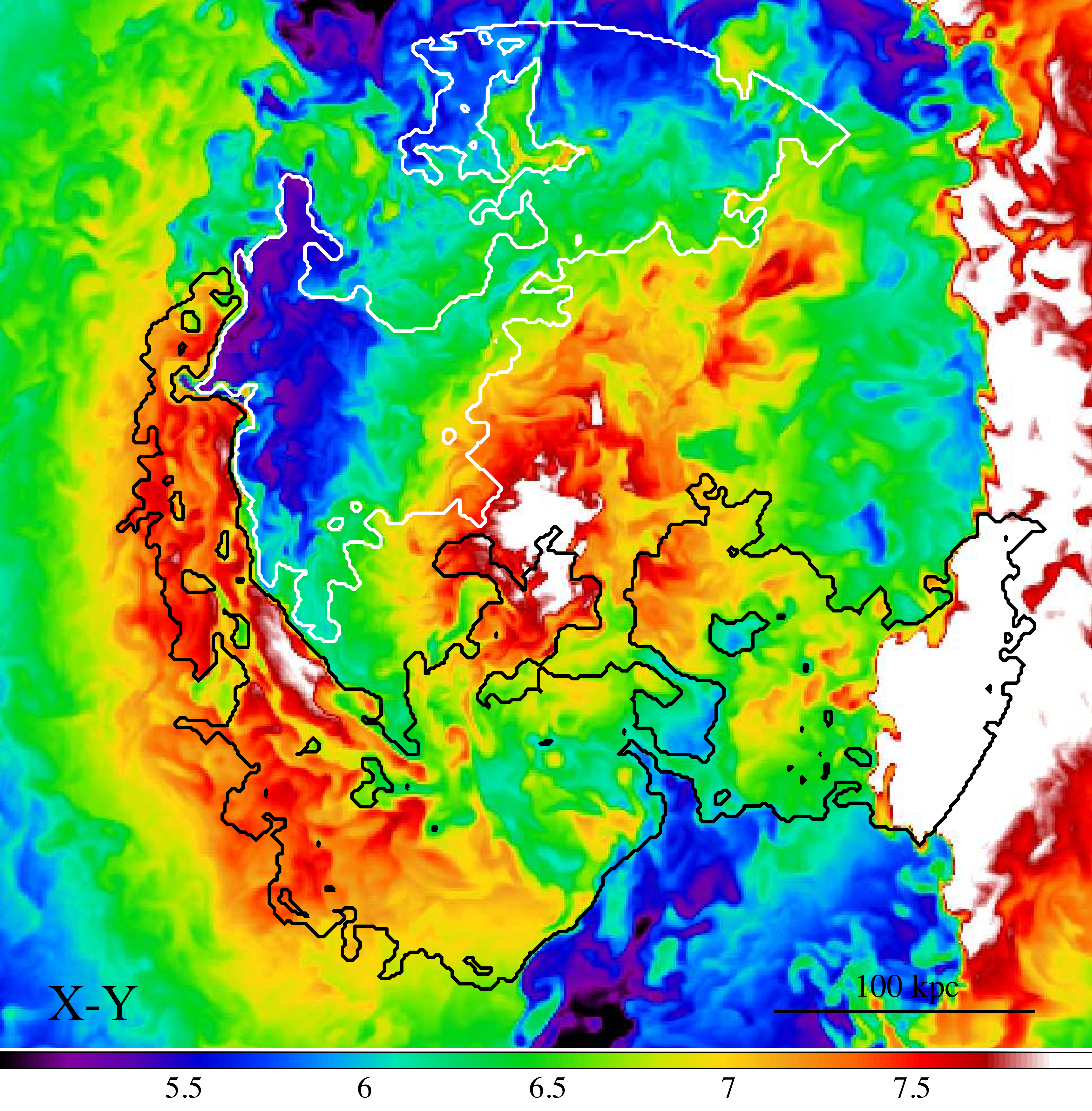}
  \includegraphics[width=5.2cm]{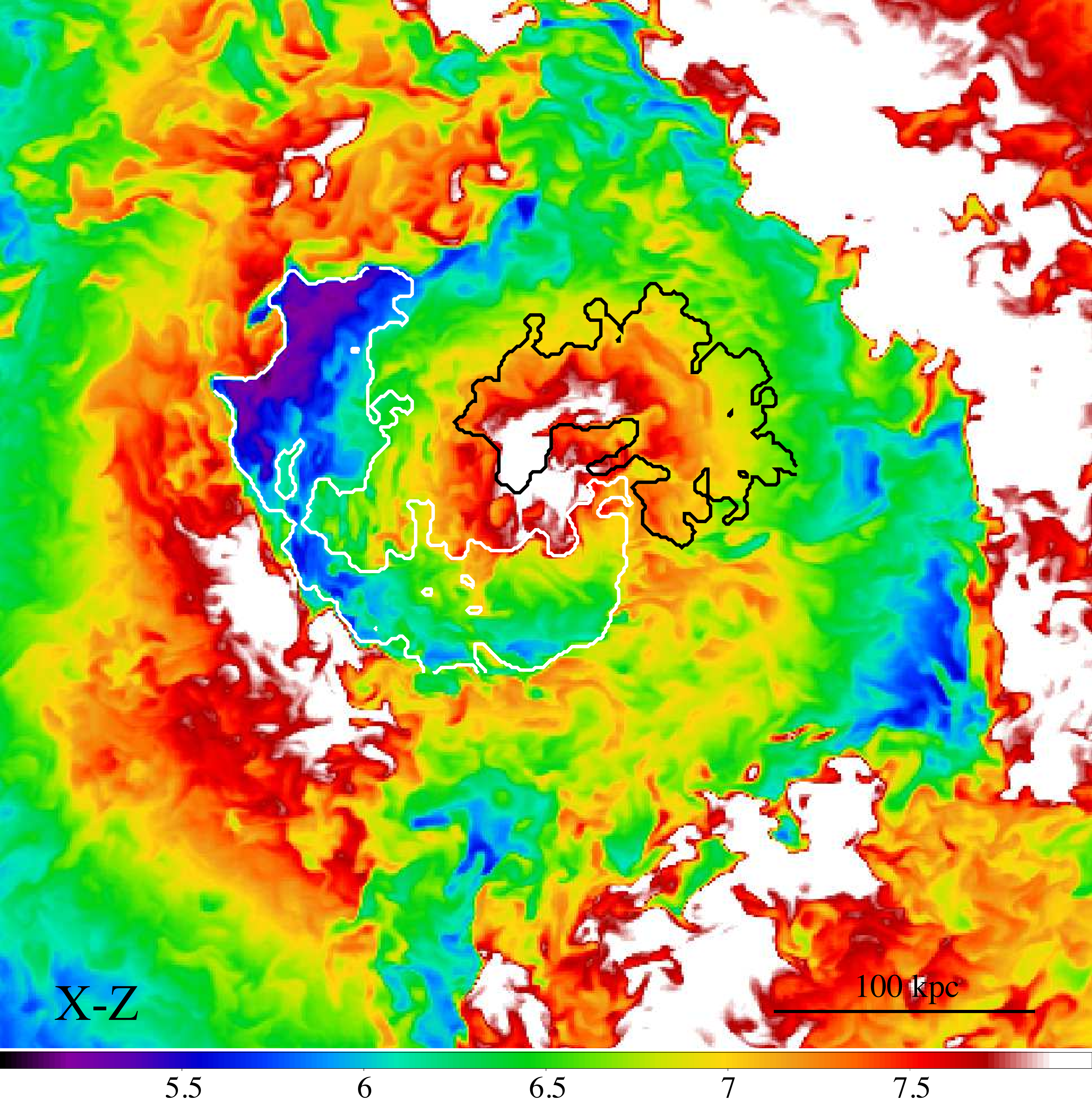} 
  \includegraphics[width=5.2cm]{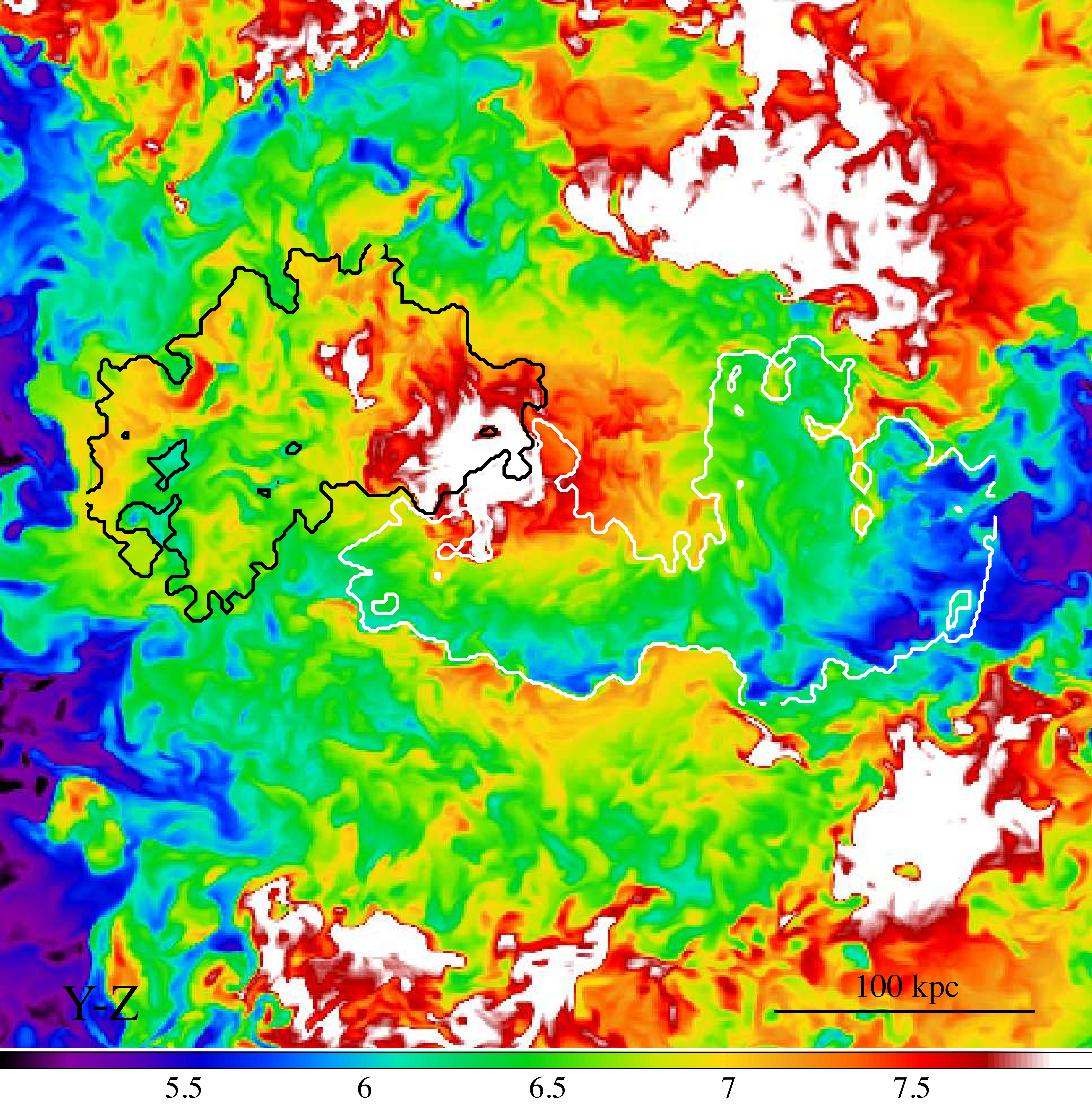} 
  \includegraphics[width=5.2cm]{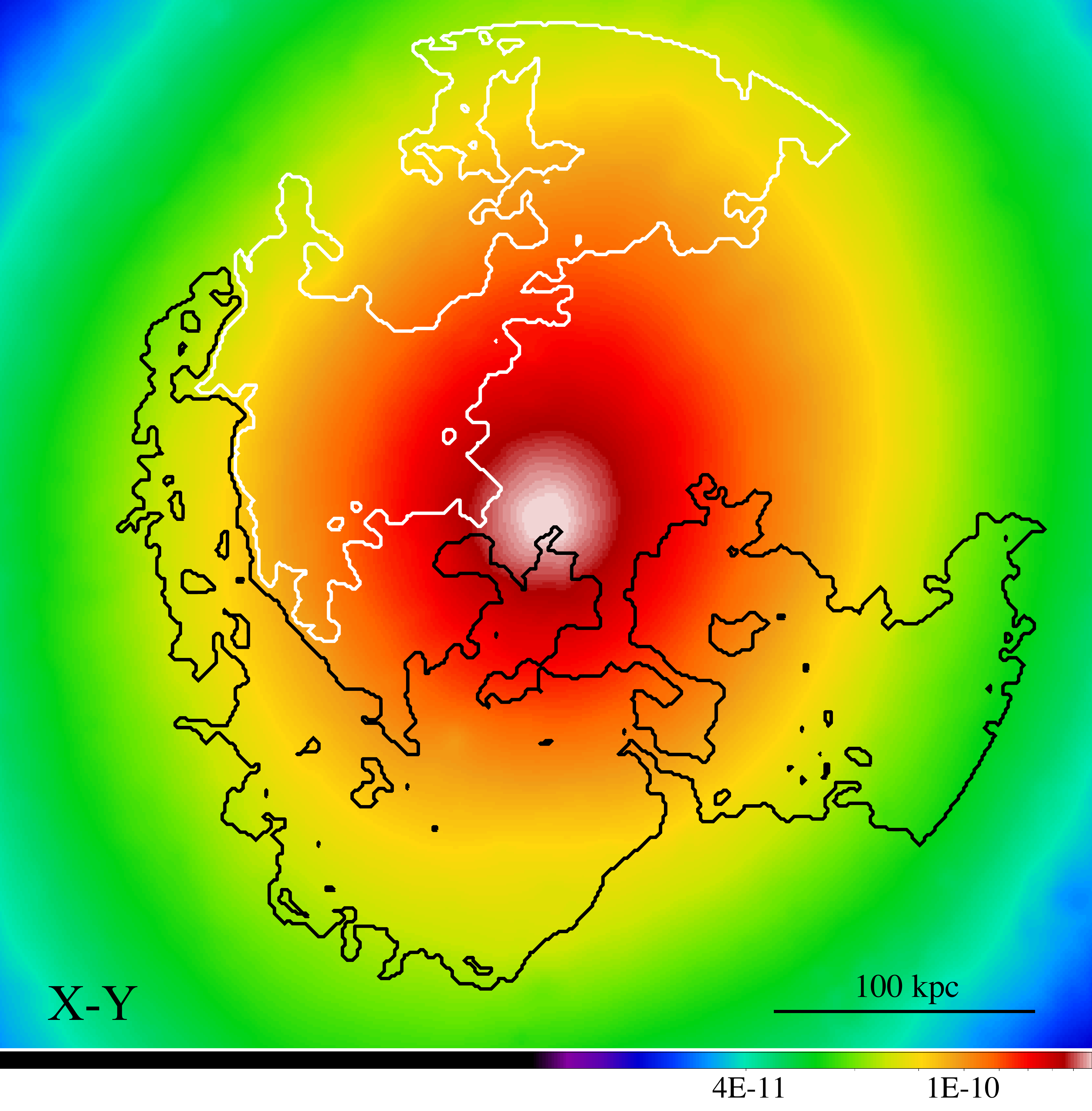}
  \includegraphics[width=5.2cm]{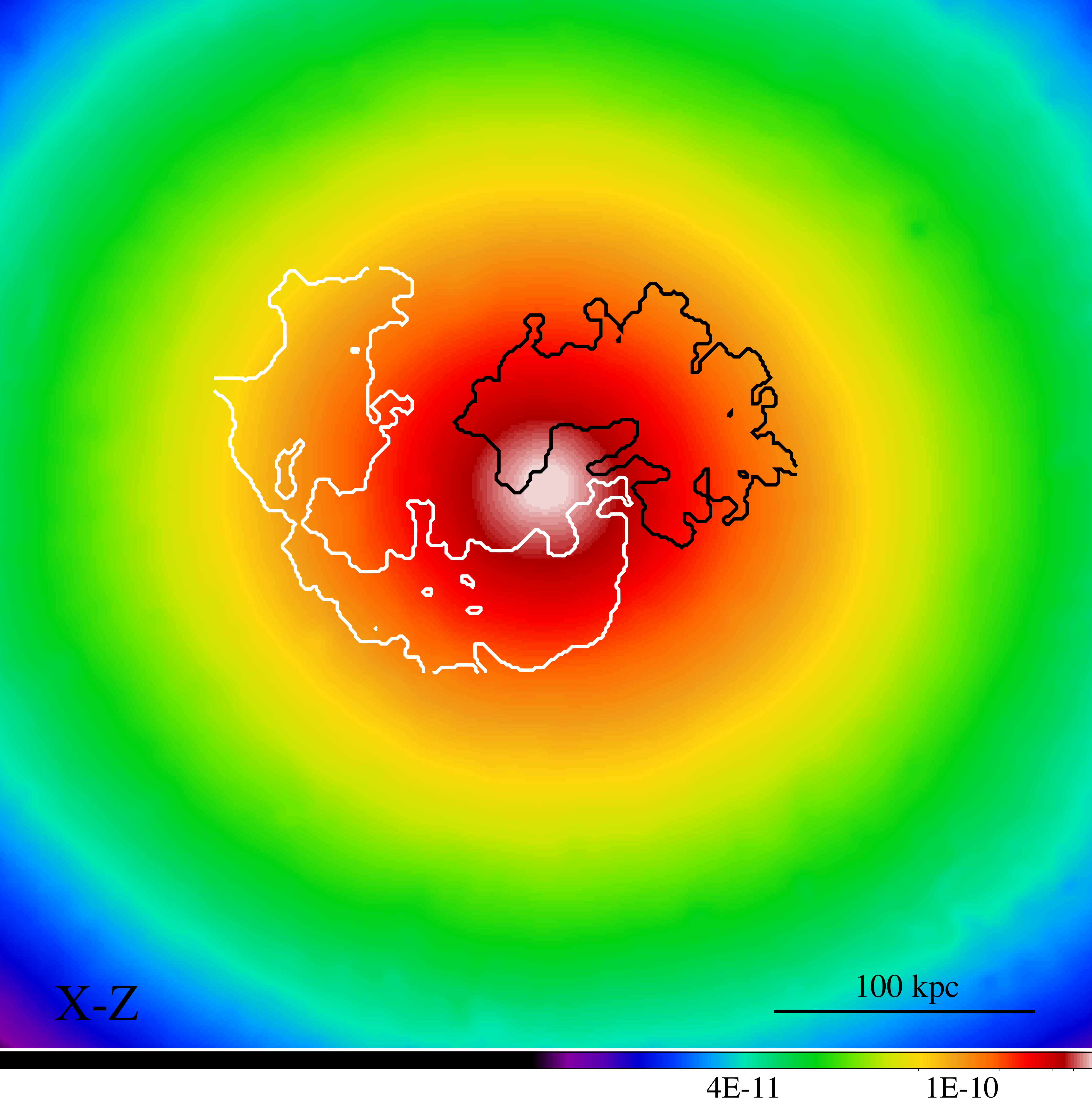} 
  \includegraphics[width=5.2cm]{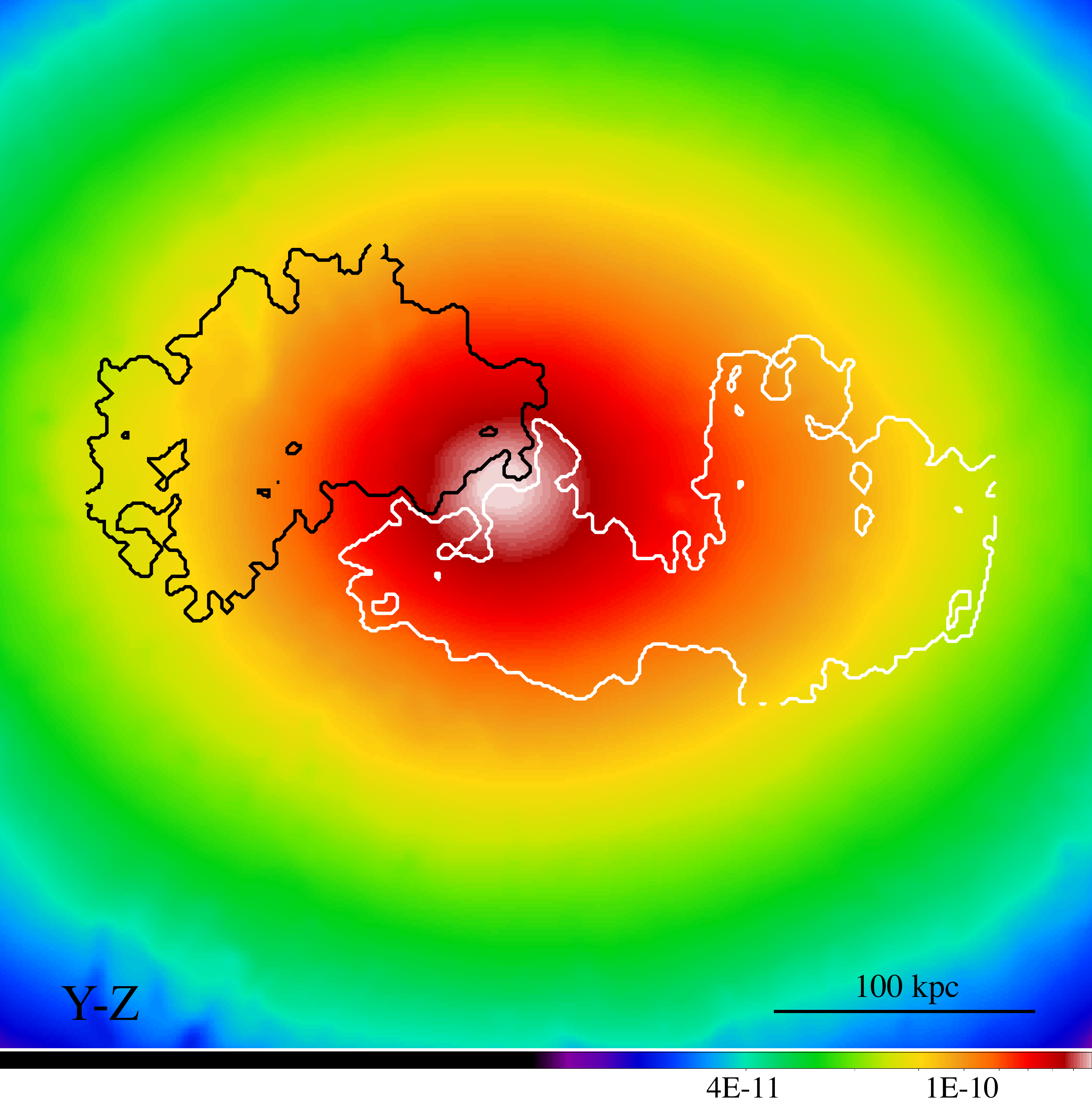} 
 \end{center}
\caption{
Results of a hydrodynamic simulation of a cluster merger in the three viewing angles. The merger has taken place in the plane of the sky (i.e., the X-Y plane). The left, middle, and right panels show the extracted images taken from the datasets of the X-Y, X-Z, and Y-Z planes, respectively.
Top: Simulated X-ray images within 200\,kpc. The white contours show the shape of the dark matter halo. These images are smoothed with a Gaussian kernel with 4\,pixel FWHM.
Second: X-ray residual images after removing the mean surface profile. The white and black areas correspond to the positive and negative excess regions, respectively. These images are also smoothed with a Gaussian kernel with 4\,pixel FWHM.
Third: Emission weighted temperature maps in units of keV. The white and black regions are the same as those in the second row.
Bottom: Same as the third row but for pressure maps in units of dyne.
}
\label{fig:sim}
\end{figure*}

Figure~\ref{fig:sim} shows the results of our cluster merger simulation in different viewing directions, i.e., X-Y, X-Z, and Y-Z planes, respectively. The merger has taken place in the plane of the sky (i.e., the X-Y plane) so that the majority of the disturbed ICM is moving in the plane of the sky. 
We extract a region 2.3\,Mpc $\times$ 2.3\,Mpc from the simulated data, and the pixel corresponds to 1\,kpc squared. We extracted a region of the central 200\,kpc to apply our detection algorithm. 

The top panels of Figure~\ref{fig:sim} show simulated X-ray images of the central 200\,kpc at three viewing angles. The centrally peaked feature in the X-ray surface brightness can be found in each simulated X-ray image so that this simulated cluster can be identified as a relaxed cluster regardless of the viewing angles. The white contours in the top panels of Figure~\ref{fig:sim} show the shape of the dark matter halo. Although the initial shape of the dark matter halo is circularly symmetric, its shape is slightly elongated due to this merger. The shape of the X-ray surface brightness is well-aligned with that of the dark matter halo. As described in Section~\ref{sec:img}, the observed shape of the X-ray surface brightness is in good agreement with that of the SL mass map. We confirmed this property by our simulations.

The second panels of Figure~\ref{fig:sim} show X-ray residual images of the three viewing angles after subtracting the mean surface profile using our method described in Section~\ref{sec:img}. Using the same definition to identify a perturbed region mentioned in Section~\ref{sec:img}, we found both positive and negative excess regions, indicating that they are a universal feature regardless of viewing angles. The white and black regions show the positive and negative excess regions, respectively. In this analysis, the center of an ellipse model is fixed to be that of the dark matter halo, which is comparable to the method that we used in Section~\ref{sec:img}. The measured PA and AR for each X-ray surface brightness image are $175 \pm 1^{\circ}$ and $0.81 \pm 0.01$ for the X-Y plane, $111 \pm 1^{\circ}$ and $0.86 \pm 0.01$ for the X-Z plane, and $94 \pm 1^{\circ}$ and $0.76 \pm 0.01$ for the Y-Z plane, respectively.

The third panels of Figure~\ref{fig:sim} show emission weighted temperature maps of the three viewing angles. The positive and negative excess regions are overlaid on each temperature map. To evaluate the ICM temperature, we calculated the mean and standard deviation of the temperature in each perturbed region. These results are summarized in Table~\ref{tab:sim_kT}. The positive and negative excess regions are associated with lower temperature gas and higher temperature gas, respectively. All the characteristic property is consistent with the observable  of gas sloshing. This result, therefore, demonstrates that our detection algorithm can detect and distinguish the lower gas temperature region as a positive excess and higher gas temperature region as a negative excess. 

\begin{table*}[ht]
\begin{center}
\caption{
Comparison of the temperature in the perturbed region in the three viewing angles extracted using the method with the mass peak or the X-ray peak, respectively.
}\label{tab:sim_kT}
\begin{tabular}{lcccc}
\hline\hline	
Viewing angle	& \multicolumn4c{Temperature (keV)	}									\\ 
			& \multicolumn2c{Mass peak}			& \multicolumn2c{X-ray peak}			\\
			& Positive			& Negative		& Positive			& Negative		\\ \hline
X-Y plane		& $6.04 \pm 0.44$	& $7.17 \pm 0.65$	& $6.09 \pm 0.49$	& $7.13 \pm 0.70$ 	\\
X-Z plane		& $6.14 \pm 0.55$	& $7.26 \pm 0.42$	& $6.26 \pm 0.42$	& $7.31 \pm 0.52$	\\
Y-Z plane		& $6.37 \pm 0.48$	& $7.05 \pm 0.49$	& $6.54 \pm 0.40$	& $6.78 \pm 0.40$	\\
\hline
\end{tabular}
\end{center}
\end{table*}

We checked pressure maps of the three viewing angles shown in the bottom panels of Figure~\ref{fig:sim}. We found no pressure discontinuity in the perturbed region and in between, which are consistent with the picture that gas motions induced by gas sloshing are subsonic and consistent with pressure equilibrium. 

We also investigated the ICM temperature in the perturbed region after ellipse modeling by choosing the peak position of the simulated X-ray image as the center of the ellipse model. This type of procedure is frequently used when only X-ray imaging data are available. We found that the X-ray peak in each viewing angle is located at 5.4\,kpc for the X-Y plane, 10.8\,kpc for the X-Z plane, and 16.6\,kpc for the Y-Z plane away from the center of dark matter halo. Hence, the measured PA and AR are $174 \pm 1^{\circ}$ and $0.81 \pm 0.01$,  $113 \pm 1^{\circ}$ and $0.87 \pm 0.01$, and  $91 \pm 1^{\circ}$ and $0.74 \pm 0.01$, respectively. We created an X-ray residual image of each viewing angle and measured the temperature in the positive and negative excess regions. The results are summarized in Table~\ref{tab:sim_kT}. The temperature difference is almost consistent with that shown in the above study, whereas the result of the Y-Z plane raises a caution that no temperature difference may be found in between the positive and negative excess regions. This result indicates that the ellipse model determined around the mass peak may be a better tracer for lower and higher temperature regions than that determined around the X-ray peak. However, a further study is needed to validate this hypothesis.

This sanity check using the hydrodynamic simulation indicates that (1) the positive and negative excess regions induced by a cluster merger can be identified as the perturbed regions using our imaging analysis, and (2) the ICM temperature in the positive excess region is systematically lower than that in the negative excess region. In addition, the ellipse model determined around the mass peak seems to be better than that determined around the X-ray peak.

On the other hand, we measured the contrast of gas density perturbations (i.e., $| \Delta I_{\rm X}| / \langle I_{\rm X} \rangle$) in each X-ray residual image using both ellipse models. The estimated values in each viewing angle are summarized in Table~\ref{tab:sim_Ix}. the contrast of gas density perturbations in the simulated X-ray image is comparable to the observed values shown in Table~\ref{tab:dIx}.

\begin{table}[ht]
\begin{center}
\caption{
Comparison of the contrast of gas density perturbations (i.e., $| \Delta I_{\rm X}| / \langle I_{\rm X} \rangle$) measured by the ellipse model with the mass peak or the X-ray peak.
}\label{tab:sim_Ix}
\begin{tabular}{lcc}
\hline\hline	
Viewing angle	& \multicolumn2c{$| \Delta I_{\rm X}| / \langle I_{\rm X} \rangle$}	\\
			& Mass peak			& X-ray peak			\\ \hline
X-Y plane		& $0.257 \pm 0.002$		& $0.261 \pm 0.002$ 	\\
X-Z plane		& $0.186 \pm 0.004$		& $0.267 \pm 0.002$		\\
Y-Z plane		& $0.222 \pm 0.003$		& $0.245 \pm 0.002$		\\
\hline
\end{tabular}
\end{center}
\end{table}

\section{Discussion}

\subsection{Thermodynamic properties of gas sloshing}
\label{sec:thermo}

\begin{figure*}
 \begin{center}
  \includegraphics[width=6.0cm,angle=90]{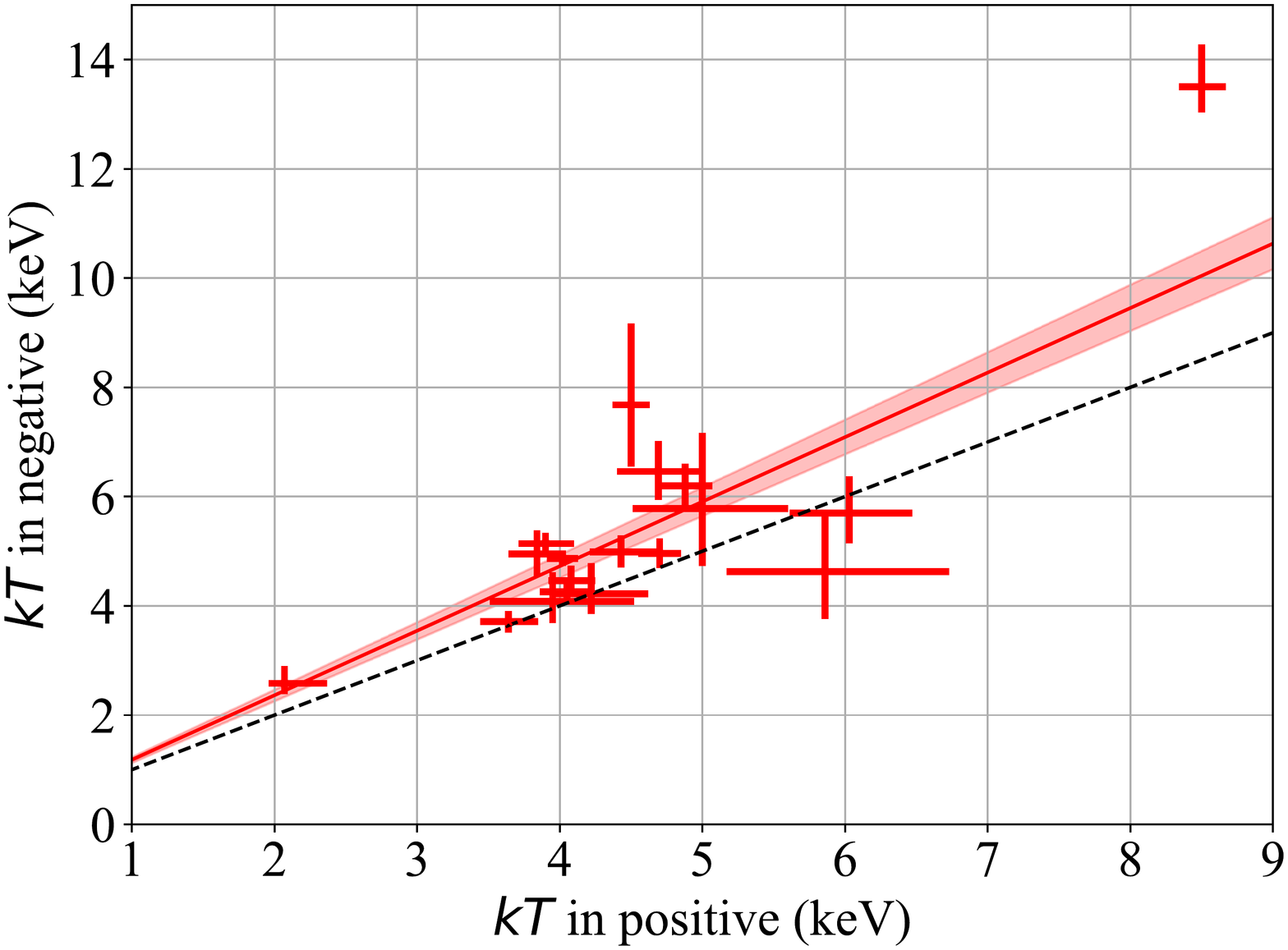}
  \includegraphics[width=6.0cm,angle=90]{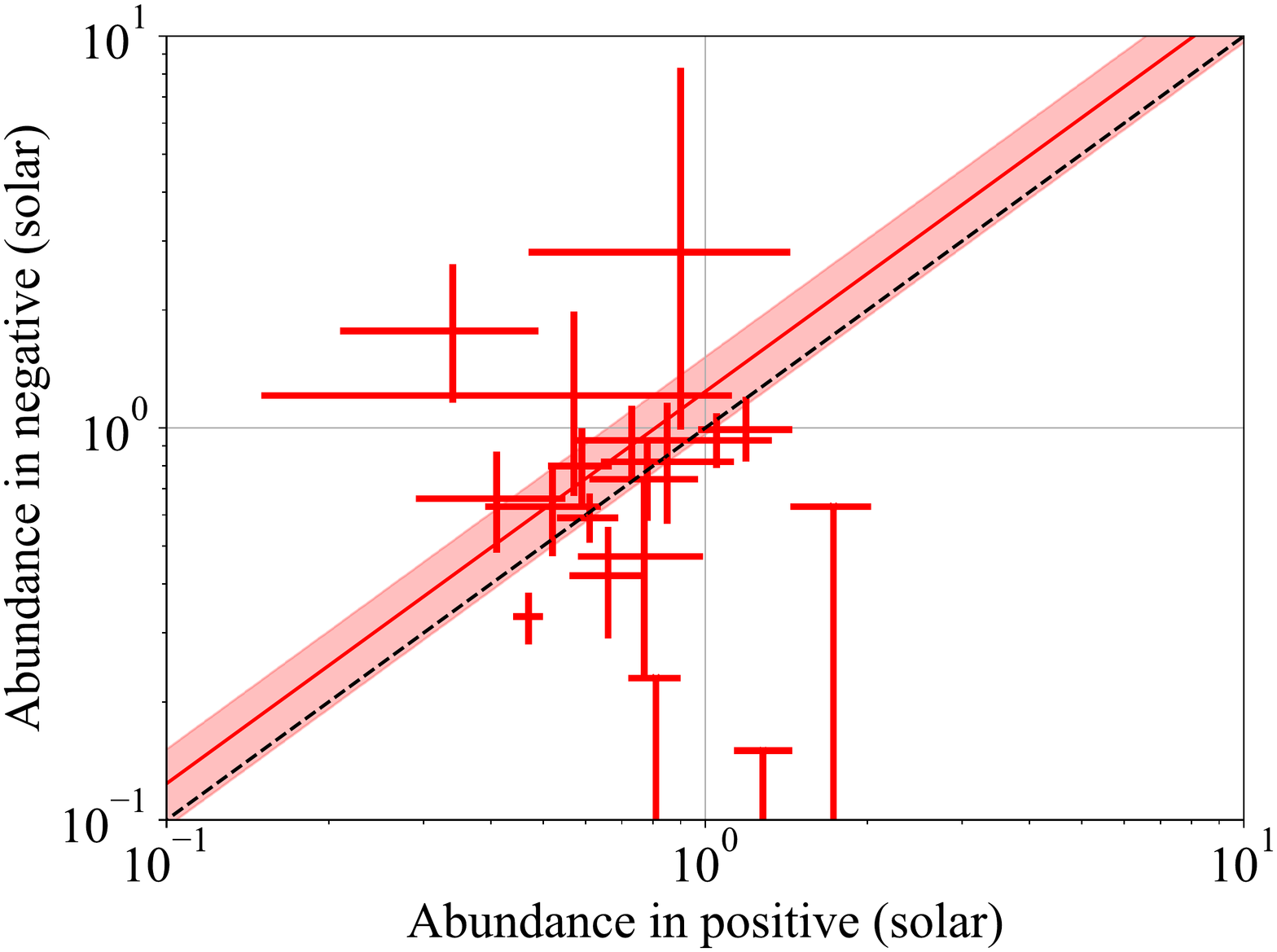}
  \includegraphics[width=6.0cm,angle=90]{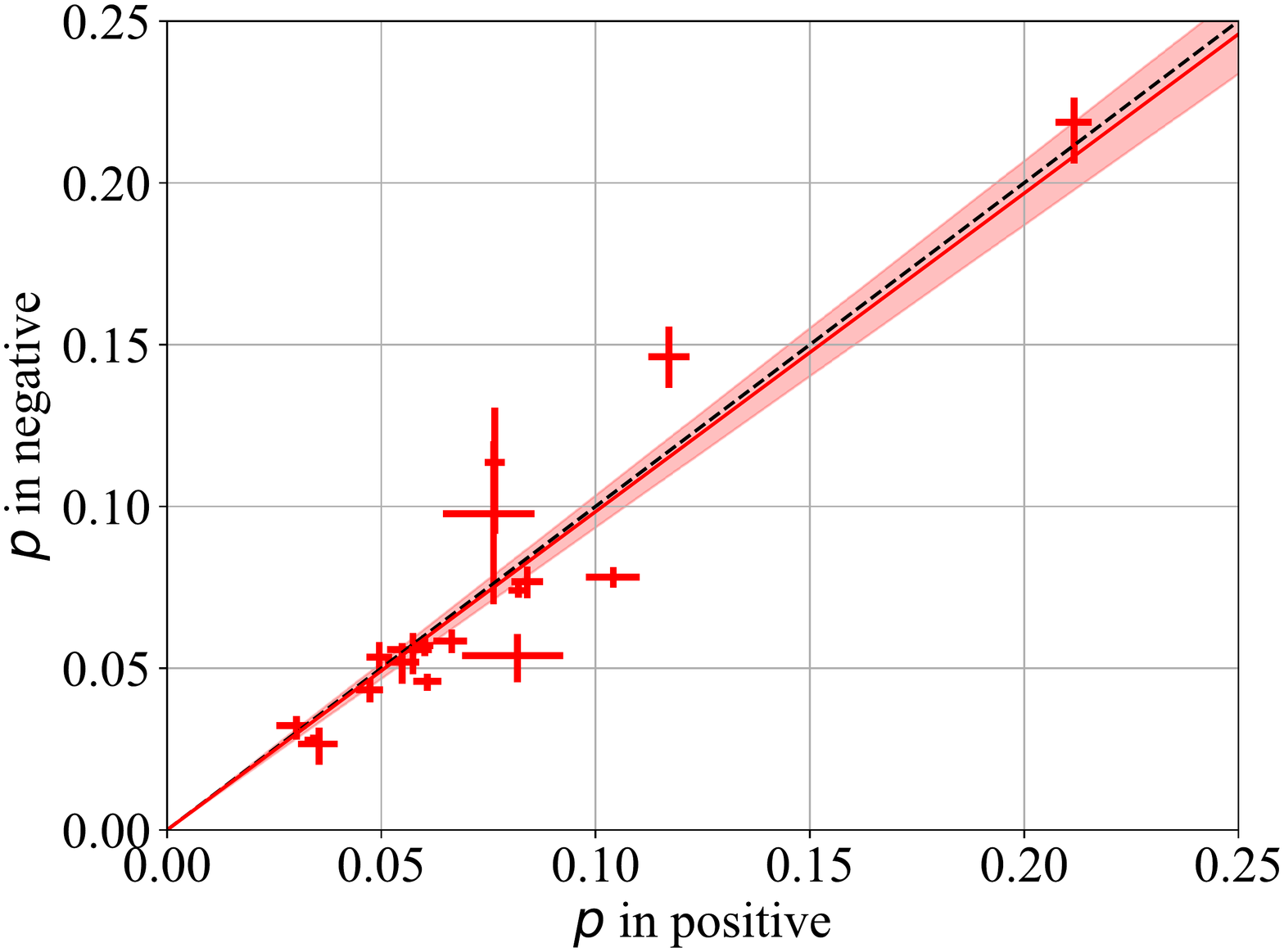}
  \includegraphics[width=6.0cm,angle=90]{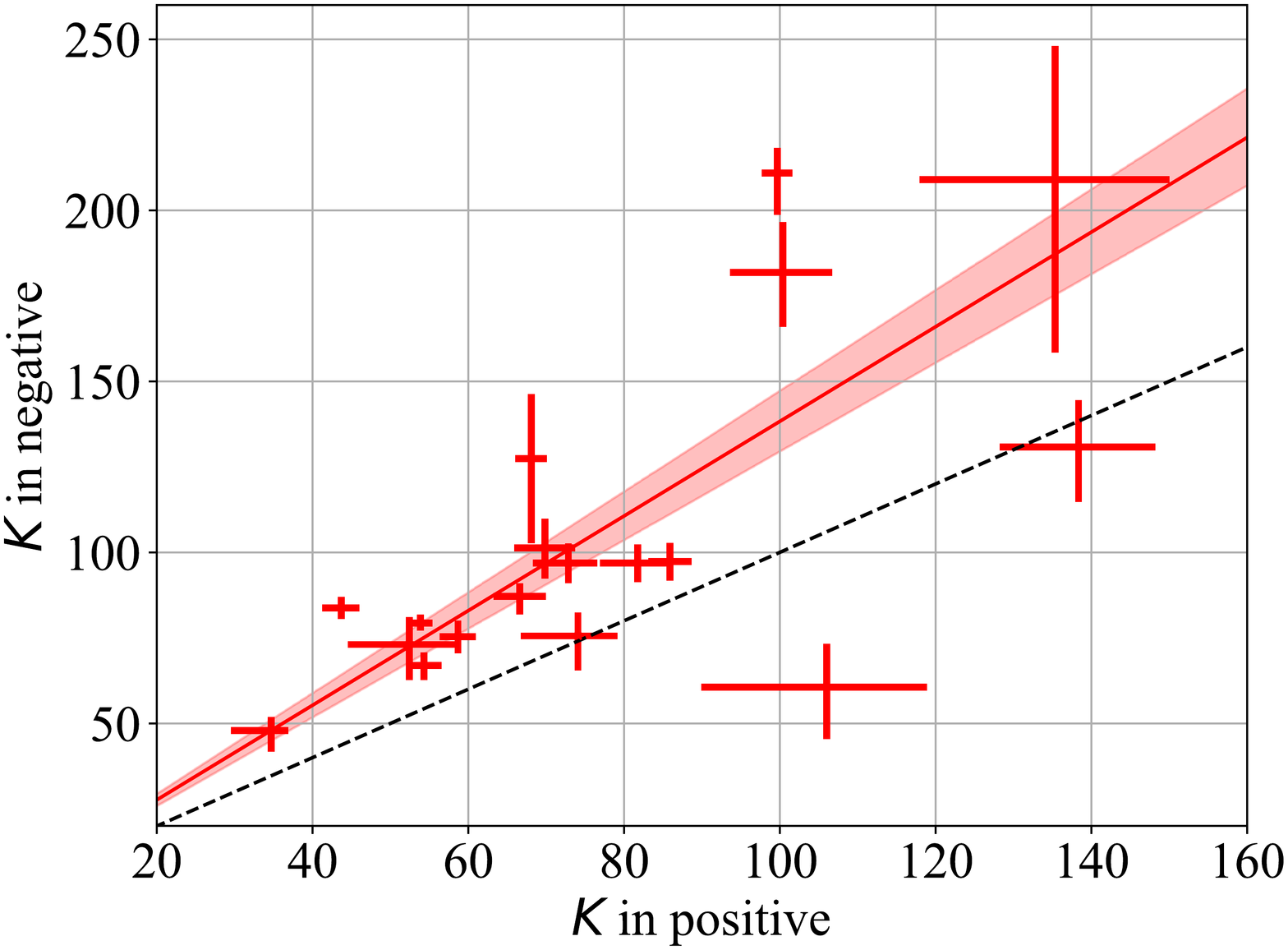}
 \end{center}
\caption{
Comparison of the ICM properties in the positive excess region with that in the negative excess region: temperature (top left), abundance (top right), pressure (bottom left), and entropy (bottom right), respectively. The red crosses in each figure show the observed values. The black dashed and red lines show unity and the best-fit model of $f(x) = ax$ (see Table~\ref{tab:best}), respectively. The red band corresponds to the 68\,\% confidence level of the model. The data with no counterpart region are excluded.
}
\label{fig:comp1}
\end{figure*}

We have characterized the thermodynamic properties of the ICM in the perturbed regions found in the cool cores. To reveal their origin and characteristics from the systematic manner, we first compared the ICM properties in the positive excess regions with that in the negative excess region. Figure~\ref{fig:comp1} shows their differences in terms of temperature, abundance, pressure, and entropy, respectively. Since the region \#2 of MACSJ\,1423.8+2404 and the region \#2 of MACS\,J1931.8-2634 have no counterpart region, we excluded these two regions from the comparison. 

We fitted each parameter using a linear model without an intercept, i.e., $f(x) = ax$. This is because there is expected to be no difference if no gas density perturbations are found in a cool core. As shown in Figure~\ref{fig:image}, the positive and negative excess regions are located at similar distances from the center. If there is no gas density perturbations, the temperature and abundance of the ICM in a cool core are expected to be azimuthally symmetric, and, within errors, there would be no differences in physical quantities between the positive and negative excess regions. If there is no density and temperature perturbation, no differences in pressure and entropy can be found either. 

The best-fit model of each parameter and its confidence level are shown in Figure~\ref{fig:comp1}. We also summarized the best-fit slope of the linear model in Table~\ref{tab:best}. We found that the ICM temperature in the negative excess regions is systematically higher than that in the positive excess regions, while the ICM abundance appears to be consistent between the positive and negative excess regions. In addition, we found that the best-fit model of the pressure is consistent with unity. In the case of the entropy, the best-fit model shows a deviation from unity clearly.

\begin{table}[ht]
\begin{center}
\caption{
Best-fit parameters of the linear model without an intercept, i.e., $f(x) = ax$.
}\label{tab:best}
\begin{tabular}{lc}
\hline\hline	
ICM property	& slope ($a$)		\\ \hline
Temperature	& $1.18 \pm 0.05$	\\
Abundance	& $1.24 \pm 0.28$	\\
Pressure		& $0.98 \pm 0.05$	\\
Entropy		& $1.38 \pm 0.09$	\\ \hline
\hline
\end{tabular}
\end{center}
\end{table}

The temperature difference between the positive and negative excess regions is one of the well-known features of gas sloshing. As mentioned in Section~\ref{sec:spec_ana}, if we focused on individual clusters, we found that a few clusters have the clear feature of gas sloshing in their ICM property. However, we confirmed that the temperature difference is found in all the galaxy clusters in our sample, suggesting that a systematic study is powerful to reveal the origin of gas density perturbations. We showed that the ratio of temperature difference is $1.18 \pm 0.05$. We need more data to confirm whether or not this ratio can be an indicator to identify gas sloshing. 

As mentioned in Section~\ref{sec:sim_ana}, we have also measured the temperature difference in the simulated data. The ratio found by the simulations in each viewing angle is $1.19 \pm 0.14$ for the X-Y plane, $1.18 \pm 0.14$ for the X-Z plane, $1.11 \pm 0.11$ for the Y-Z plane. Their ratios are consistent with those measured by the observed trend (see the top left panel of Figure~\ref{fig:comp1} or Table~\ref{tab:best}). This result supports that the observed temperature difference is generated by a merger, i.e., gas sloshing.

For the ICM abundance, we found no apparent offset between the positive and negative excess regions. The abundance gradient between them is reported in some cool cores \citep[e.g.,][]{Ghizzardi14}. Such gradient is predicted by gas sloshing because an inflowing hot gas has a lower abundance than the gas originally in a cool core. Such feature is however not reported in some other cool-core clusters \citep[e.g.,][]{Clarke04, Sanders14, Ueda17}. The observed trend in the ICM abundance is comparable to unity, i.e., no difference, implying that the inflow of a relatively hot gas comes from nearby a cool core so that the gradient could be relatively small. 

The important result is evidence of pressure equilibrium among the perturbed regions. This result is consistent with the picture of gas sloshing, as shown in the bottom panels of Figure~\ref{fig:sim}. Although \cite{Ueda18} showed the direct evidence of pressure equilibrium of sloshing gas in RXJ\,1347.5-1145 using the combined analysis of X-ray and SZE data, we extended this point to 12 cool cores. This result then indicates that the gas density perturbations generated by gas sloshing is isobaric, which is reported applying the effective equation of state \citep[e.g.,][]{Churazov16, Zhuravleva18} and solving the equation of state for the gas density perturbations based on the X-ray and SZE images \citep{Ueda18}. To confirm this property, we also applied a linear regression analysis into the data. We then found that the obtained confidence level is consistent with unity in the entire scale, i.e., no intercept is indicated. This result is in good agreement with that derived by the linear model fit without an intercept.

Based on the pressure equilibrium (see the bottom left panel of Figure~\ref{fig:comp1}) and the temperature difference (see the top left panel of Figure~\ref{fig:comp1}), a deviation from the mean density profile can be estimated. Following the slope for the temperature profile in Table~\ref{tab:best}, the number density in the positive excess region is a factor of 1.18 times larger than that in the negative excess region. This implies that the deviation from the mean density profile is considered to be $\sim 10$\,\%. According to Equation\,4 in \cite{Ueda18}, the value of $|\Delta I_{\rm X}| / \langle I_{\rm X} \rangle$ approximates $2 \Delta \rho / \sqrt{\langle \rho^2 \rangle}$, where $\langle \rho^2 \rangle$ is the mean square gas mass density. This means that if an amplitude of a density fluctuation is $\sim 20$\,\%, an amplitude of the surface brightness perturbation becomes $\sim 40$\,\%. This result is comparable to the observed value of the contrast of gas density perturbations (see Table~\ref{tab:dIx}). In addition to the observations, we found a similar amplitude in the simulated X-ray data. Using the same estimate as above, the density perturbation is inferred because of the pressure equilibrium. The average of temperature ratio in the simulations is $1.16 \pm 0.08$, which indicates that the density difference is $\sim 20$\,\%. The inferred value is consistent with that estimated by the observations. 

Owing to the observed temperature and density differences, the differences of the entropy are found to be significant, as shown in the bottom right panel of Figure~\ref{fig:comp1}. Through mixing of gas between the positive and negative excess regions induced by sloshing and/or turbulent gas motions, such an entropy difference can result in the increase of the gas temperature in the positive excess region.

We found that all the cool cores in our sample (i.e., 12 galaxy clusters) have feature suggesting gas sloshing. We then inferred a fraction of the sloshing cool cores to be $\sim 80$\,\% (95\,\% CL) at least in high mass, relaxed galaxy clusters like the CLASH sample.  

AGN feedback can also induce gas density perturbations in the cool core. \cite{Hlavacek-Larrondo13} reported an X-ray cavity created by an extreme AGN feedback in RX\,J1532.9+3021. For other clusters in our sample, \cite{Hlavacek-Larrondo12} searched for X-ray cavities in the X-ray surface brightness. However, \cite{Churazov16} found in the Perseus cluster that the ICM in X-ray cavities where are identified as bubbles is isothermal, solving the effective equation of state for small-scale perturbation. Since the ICM density in the X-ray cavities is smaller than the surroundings, there is identified as the negative excess region in our definition. Combining both properties, the thermal ICM pressure in the X-ray cavities is expected to be smaller than that in the positive excess region, which is inconsistent with the observed trend of the ICM pressure. The observed result favors that an isobaric process is reasonable to explain no pressure difference between the positive and negative excess regions. Even though some gas density perturbations are associated with the AGN feedback, our results indicate that most of gas density perturbation is induced by gas sloshing from the systematic manner.

\subsection{Contrast of gas density perturbations and its implications}
\label{sec:dIx}

In Section~\ref{sec:mdIx}, we have introduced $| \Delta I_{\rm X}| / \langle I_{\rm X} \rangle$ as a proxy of the contrast of gas density perturbations, following \cite{Ueda18}. Here we compare this X-ray brightness contrast with cluster mass and structural parameters, namely the total cluster mass ($M_{\rm 200c}$), the concentration parameter ($c_{\rm 200c}$), and the Einstein radius obtained by previous CLASH lensing studies \citep{Zitrin15, Merten15, Umetsu16, Umetsu18}. We adopt $M_{\rm 200c}$ and $c_{\rm 200c}$ of each cluster from \cite{Umetsu16}. Since MACSJ\,1311.0-0310 and MACSJ\,1423.8+2404 \citep[for which weak-lensing magnification data were not available; see][]{Umetsu14} were not included in the analysis of \cite{Umetsu16}, we adopt $M_{\rm 200c}$ and $c_{\rm 200c}$ of these two clusters from \cite{Merten15}.  Since \cite{Merten15} obtained lensing mass estimates that are, on average, 7\,\% smaller than those of \cite{Umetsu16}, we apply a correction factor of 7\,\% (i.e., $M_{\rm 200c}$ to 1.07$M_{\rm 200c}$) to the lensing mass estimates from \cite{Merten15}.

\begin{figure*}
 \begin{center}
  \includegraphics[width=6.0cm,angle=0]{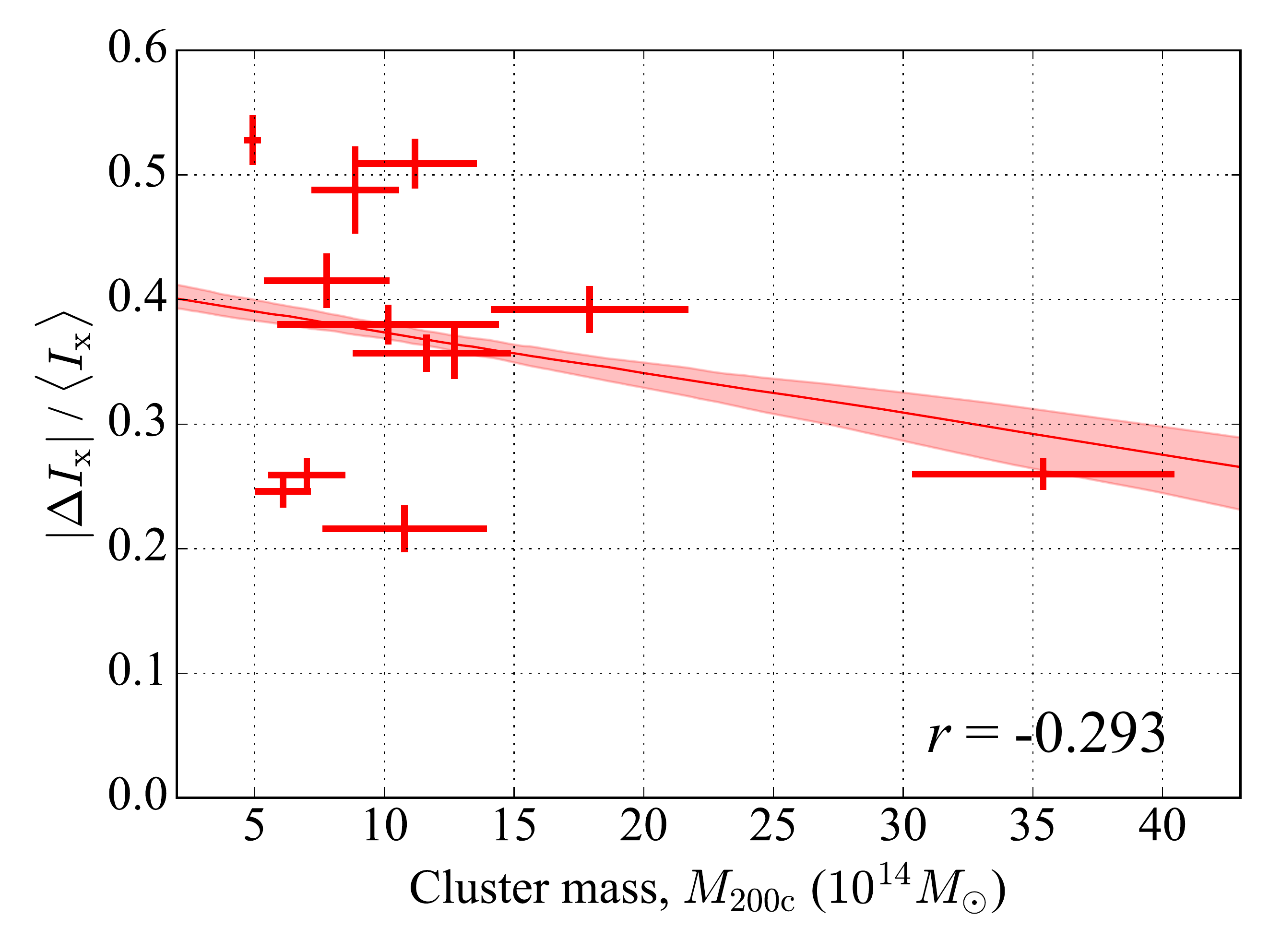}
  \includegraphics[width=6.0cm,angle=0]{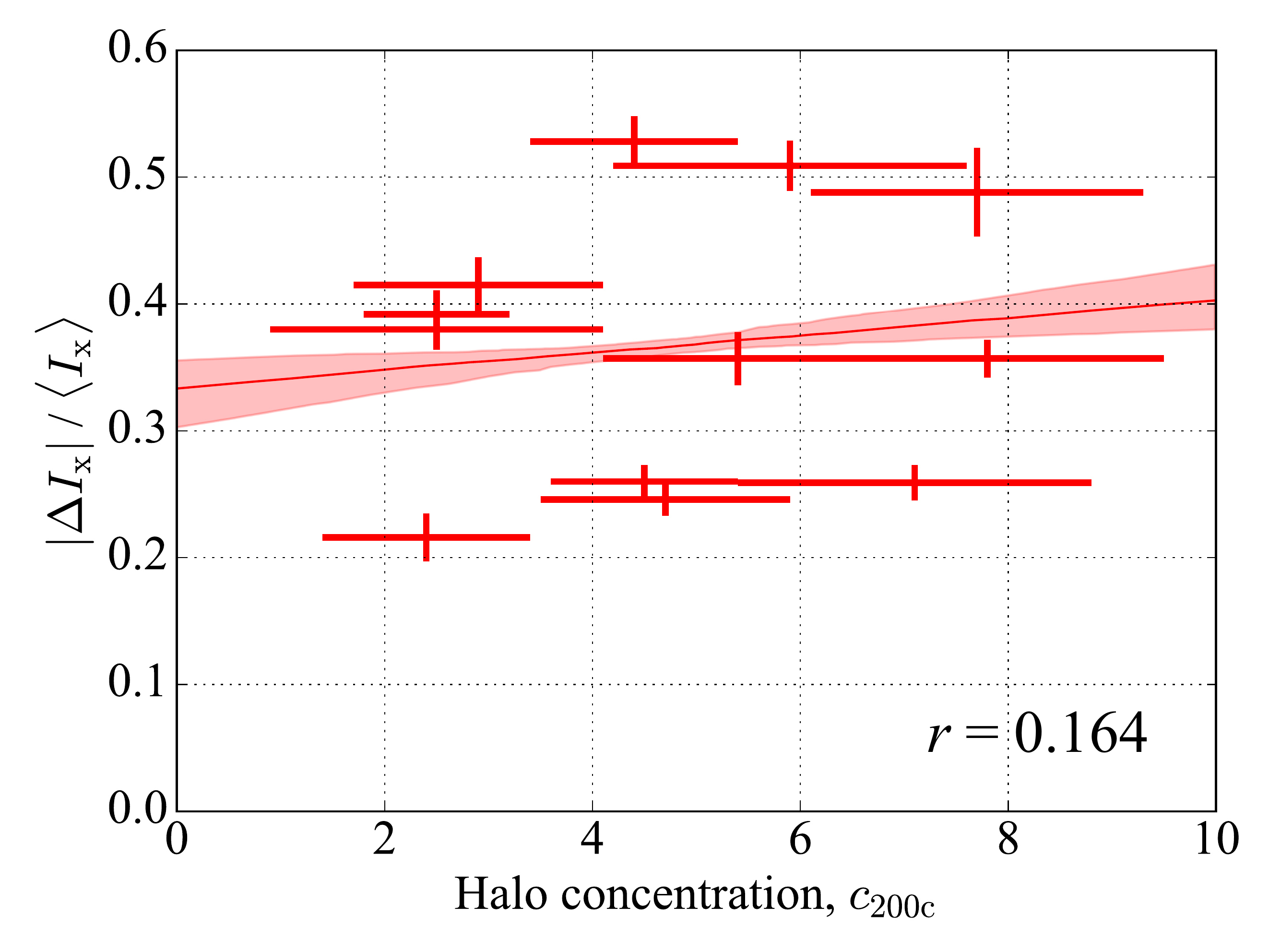}
  \includegraphics[width=6.0cm,angle=0]{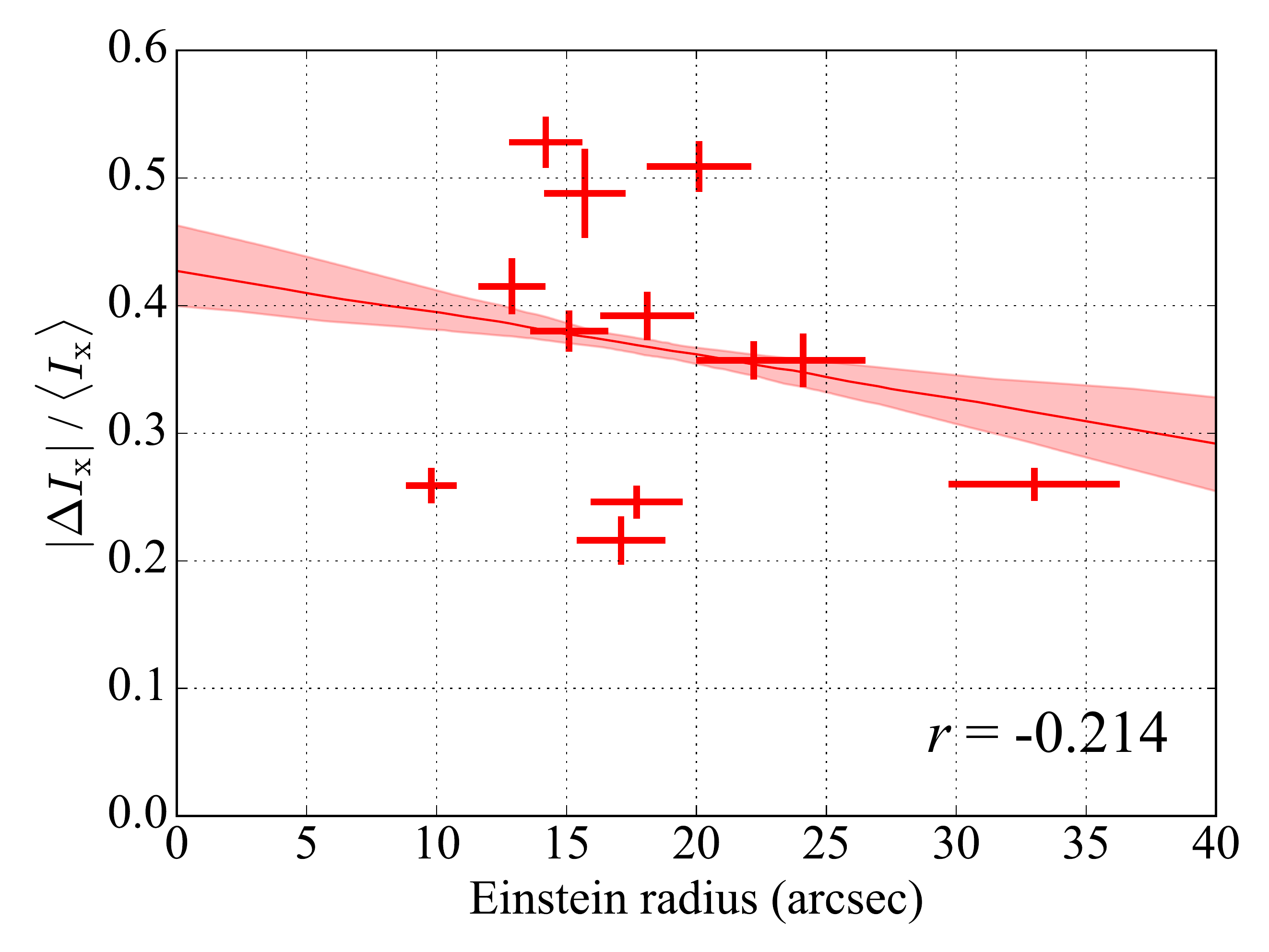}
  \includegraphics[width=6.0cm,angle=0]{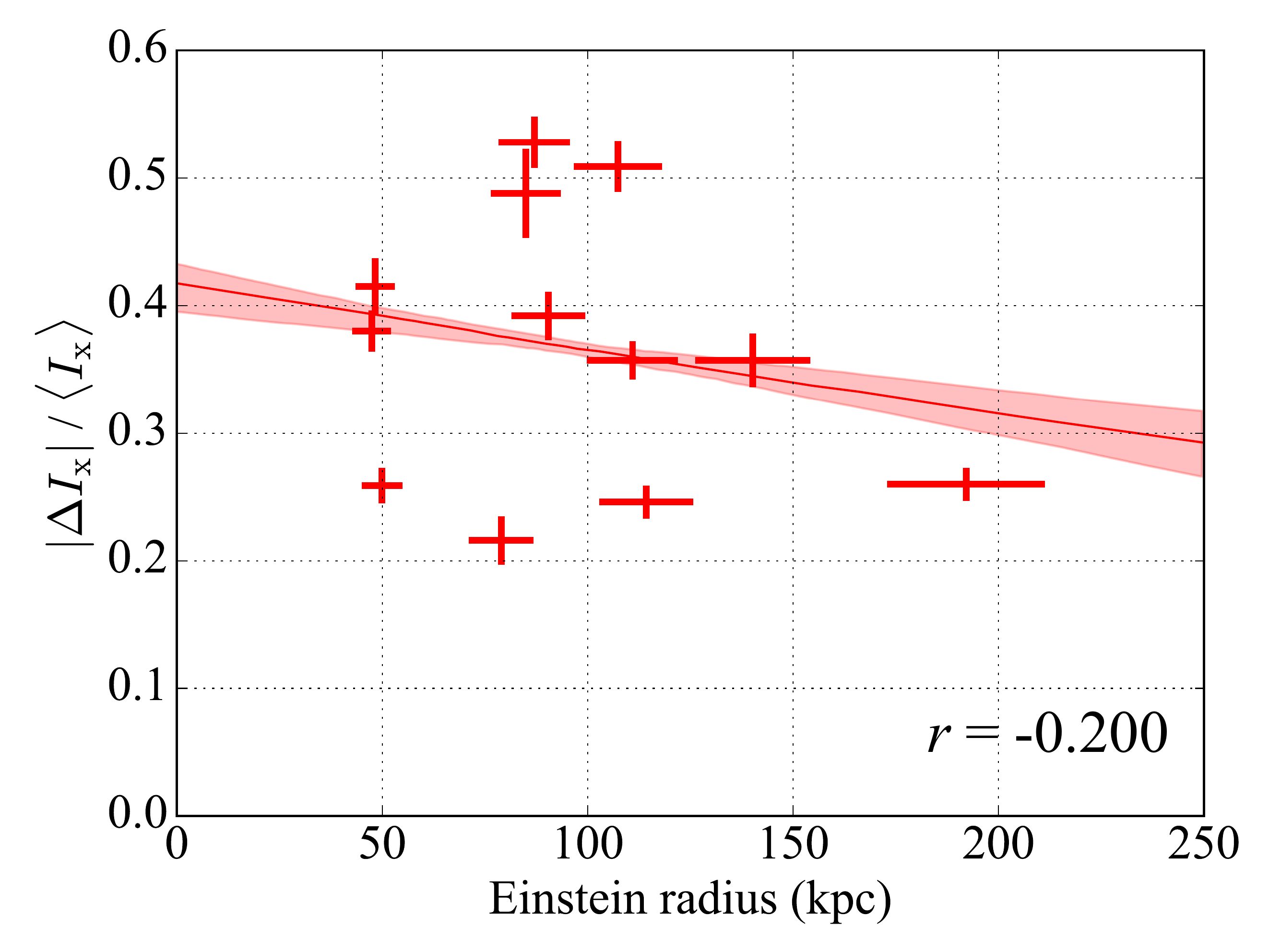}
  \includegraphics[width=6.0cm,angle=0]{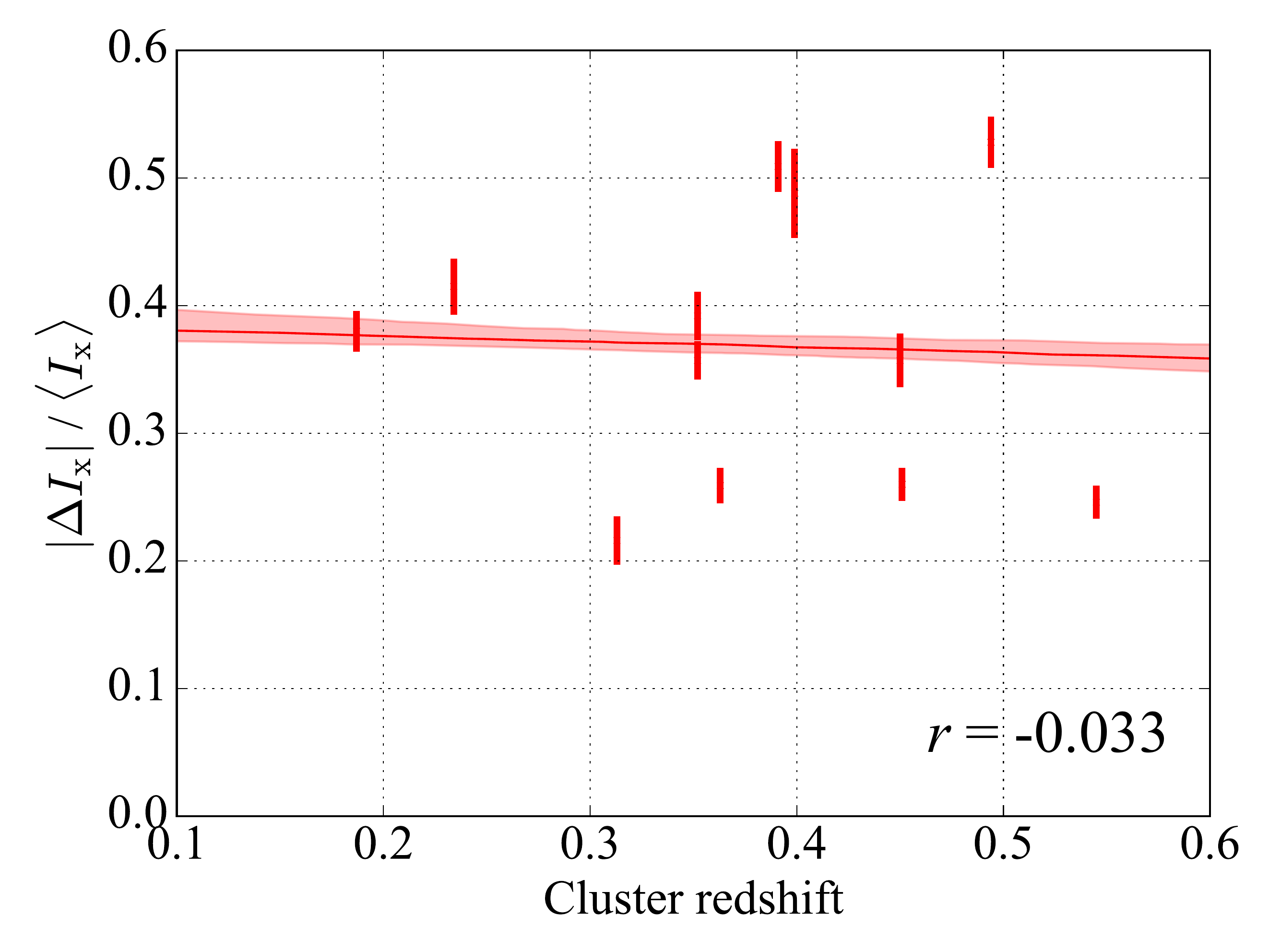}
 \end{center}
\caption{
Contrast of gas density perturbations, $| \Delta I_{\rm X}| / \langle I_{\rm X} \rangle$, as a function of the cluster mass ($M_{200c}$) (top left), the concentration parameter (top right), the Einstein radius in units of arcsec (middle left), or in units of kpc (middle right), and the cluster redshift (bottom), respectively. The data given by the lensing analyses are taken from \cite{Zitrin15}, \cite{Merten15}, \cite{Umetsu16}, and \cite{Umetsu18}. The red crosses represent the data of each cluster. The red line is the best-fit model of each data and the red band shows its 68\,\% confidence level. The sample correlation coefficients, $r$, of each pair are also shown in each panel. Their coefficients are calculated using the best-fit parameters.
}
\label{fig:comp3}
\end{figure*}

Figure~\ref{fig:comp3} shows the contrast of gas density perturbations in terms of each lensing parameter. The best-fit model and its confidence level are also displayed in Figure~\ref{fig:comp3}. The sample correlation coefficients of each pair were calculated using the best-fit parameters. Their coefficients indicate that the contrast of gas density perturbations is not correlated strongly with the lensing parameters. For example, the correlation coefficient between $| \Delta I_{\rm X}| / \langle I_{\rm X} \rangle$ and $M_{\rm 200c}$ (see the top left panel of Figure~\ref{fig:comp3}) is $r = -0.293$, suggesting an anti-correlation. Although the current sample size is small, we discuss possible implications of these results below.  

Cluster mergers with similar masses are expected to to produce larger gas density fluctuations in cool cores \citep{Ricker01}. High-mass primary clusters tend to have frequent minor mergers with small mass ratios. In this context, if gas sloshing is the mechanism responsible for the observed gas density fluctuations, an anti-correlation is expected in the mass profile (the top left panel of Figure~\ref{fig:comp3}). In fact, their relation shows a slight anti-correlation even though the number of data points is limited. Other parameters such as the impact parameter can also affect the amplitude of gas density perturbations. A large scatter in the mass range of $(5 - 20) \times 10^{14}$\,\MO ~might reflect a variety of a mass of an infalling cluster and its impact parameter. In future work, we will investigate which merger parameters dominate the generation of gas density perturbations in cool cores of galaxy clusters.

On the other hand, the relation between the contrast of gas density perturbations and the concentration parameter  implies that it has a slight positive correlation (see the top right panel of Figure~\ref{fig:comp3}). If a galaxy cluster has a large concentration parameter, then it has a large amount of matter in the cluster core compared to that with a small concentration parameter. The total amount of movable matter by gas sloshing is important to create gas density perturbations. In this sense, the observed trend is consistent with that expected by gas sloshing. This idea is also supported by an anti-correlation between the contrast of gas density perturbations and the Einstein radius. The clusters with centrally concentrated mass host larger gas density perturbations. However, it is hard to arrive at a firm conclusion regarding the origin of their correlations from the current dataset with a limited sample size. More observations and numerical simulations are needed to clearly understand the origin of the correlations. 

In the plot of the cluster redshift (the bottom panel of Figure~\ref{fig:comp3}), we found no strong correlation.

\begin{figure*}
 \begin{center}
  \includegraphics[width=6.0cm,angle=0]{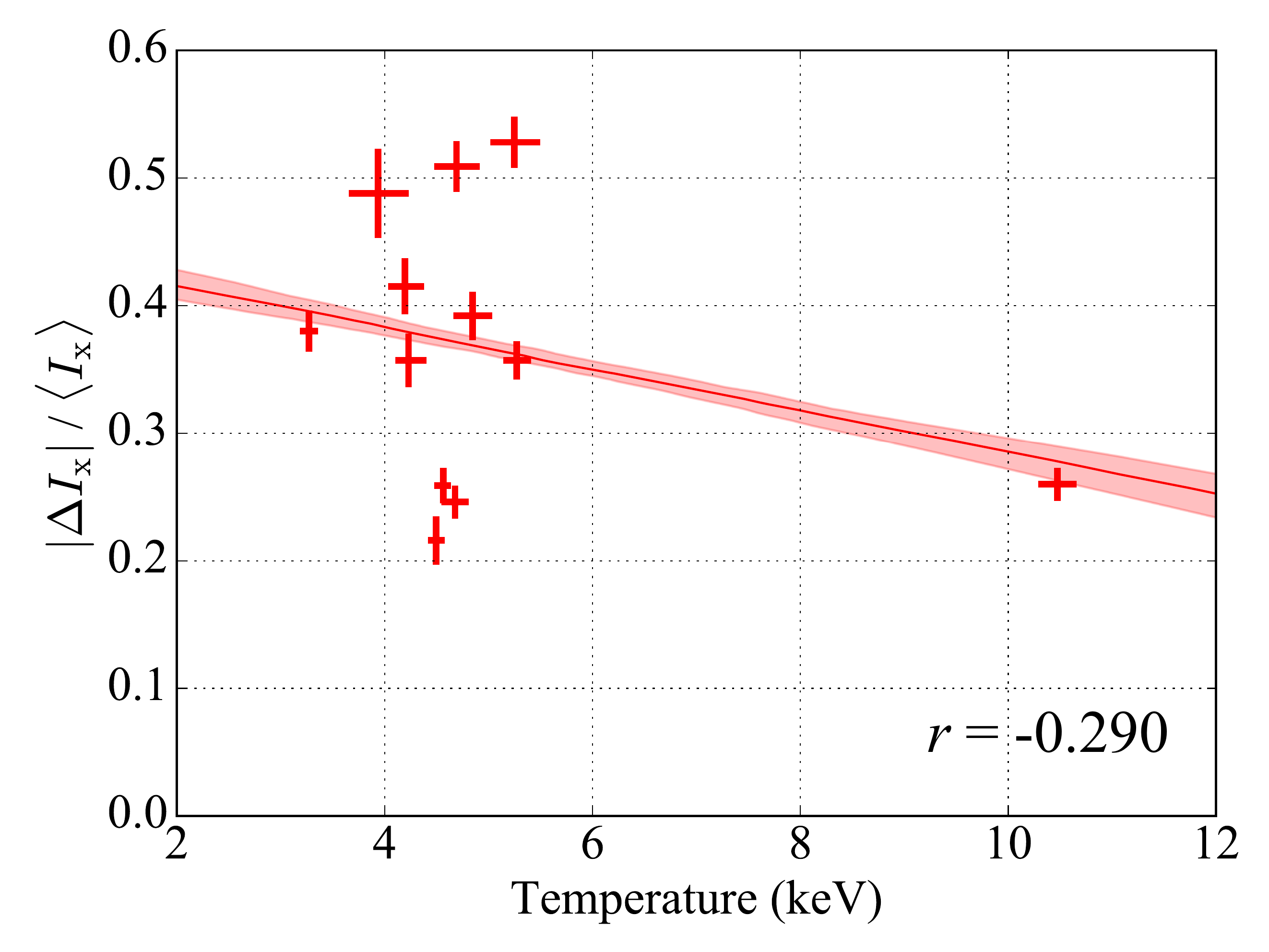}
  \includegraphics[width=6.0cm,angle=0]{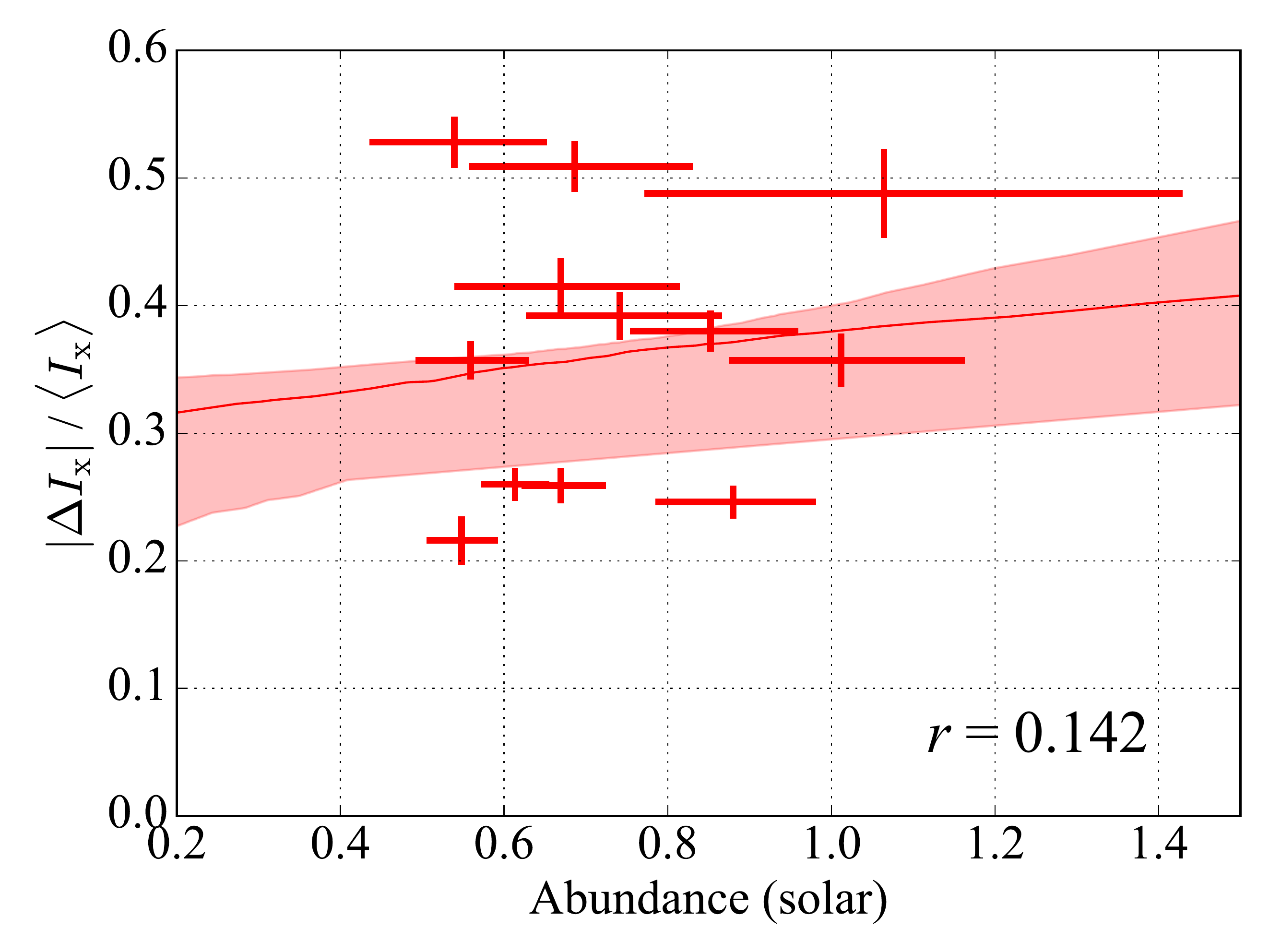}
  \includegraphics[width=6.0cm,angle=0]{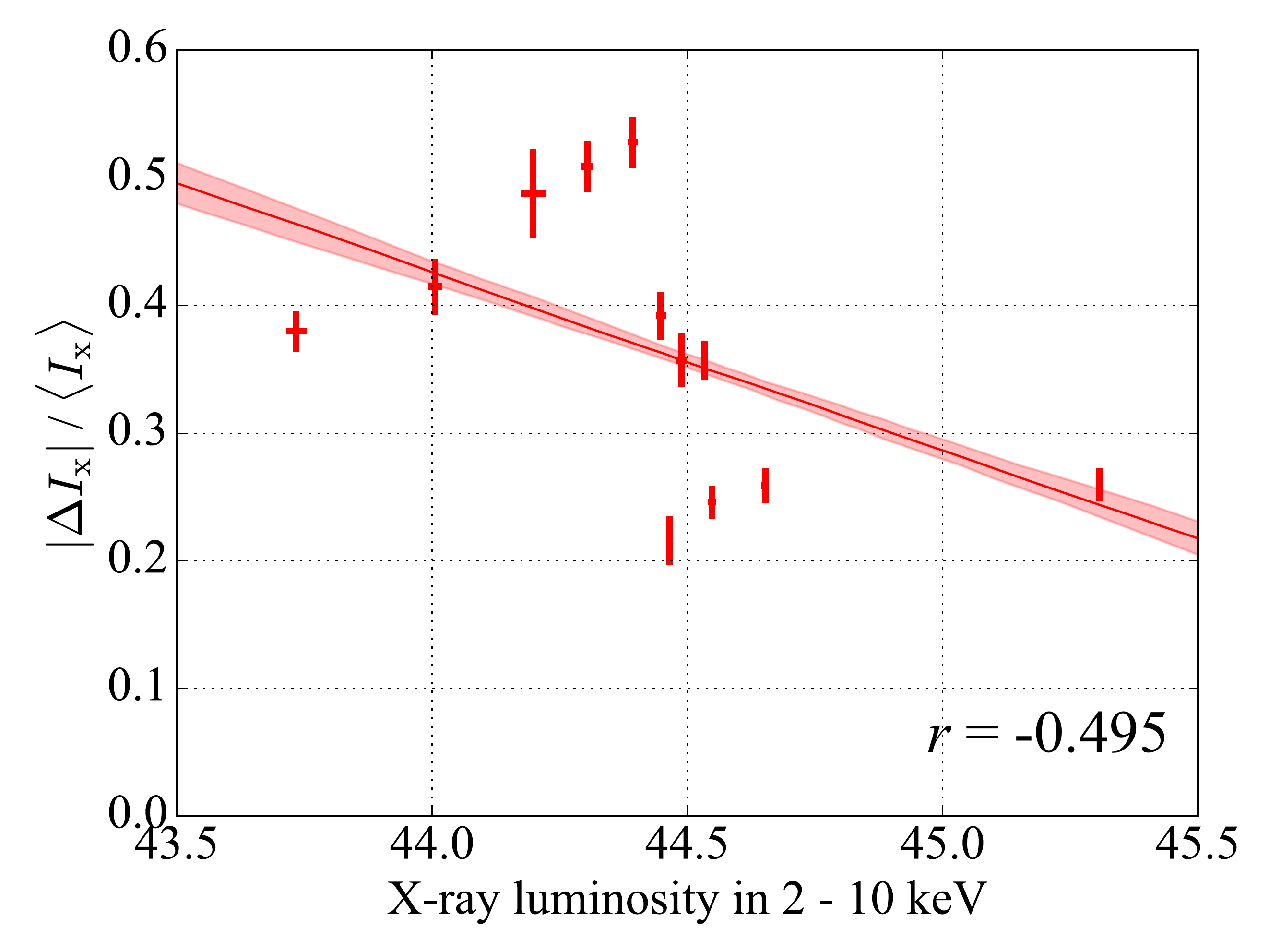}
  \includegraphics[width=6.0cm,angle=0]{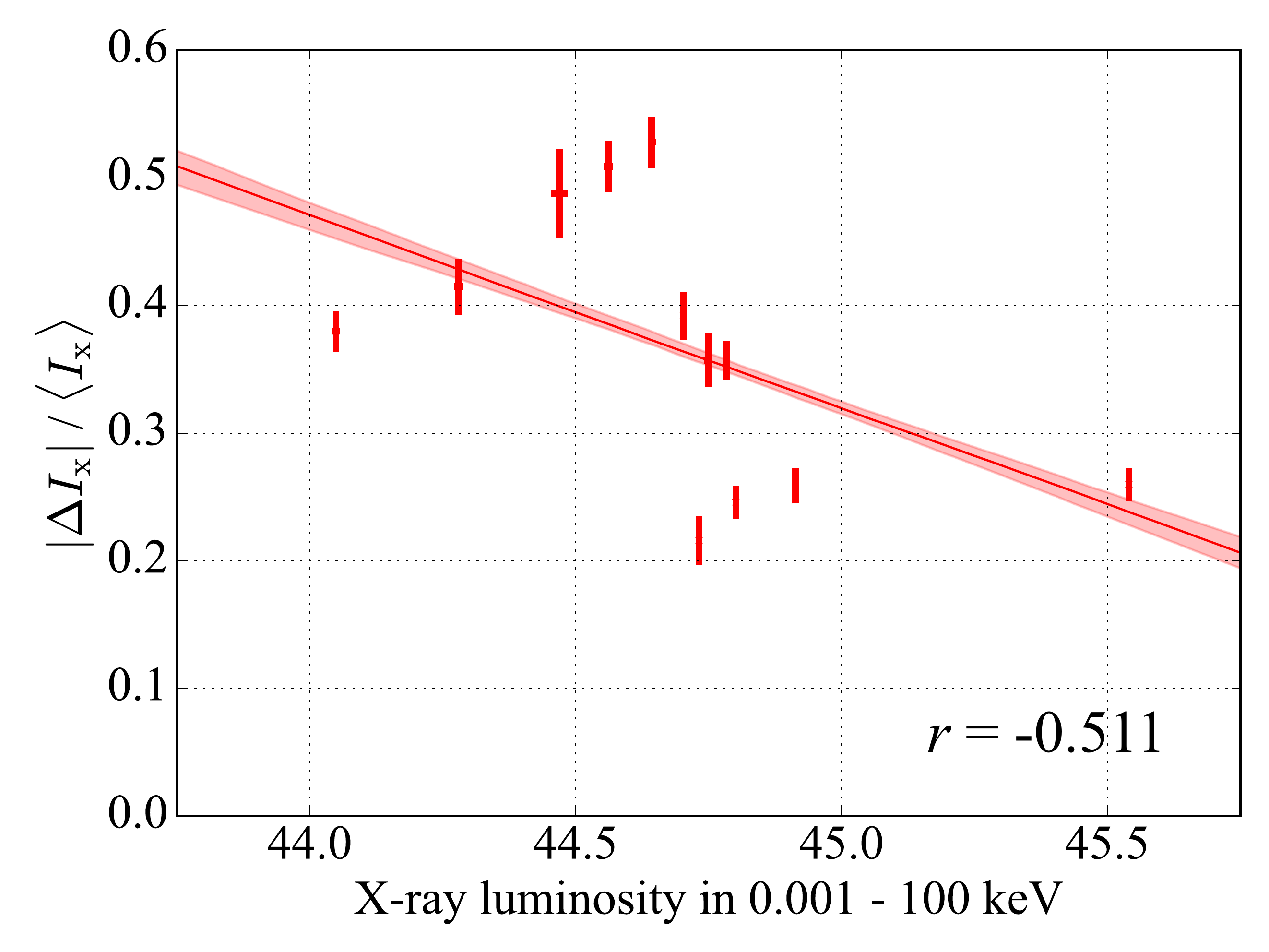}
 \end{center}
\caption{
Same as Figure~\ref{fig:comp3} but for the ICM temperature in units of keV in the overall perturbed region (top left), the ICM abundance in units of solar (top right), the logarithm of X-ray luminosity in 2 - 10\,keV in units of erg\,s$^{-1}$ (bottom left), and the logarithm values of X-ray luminosity in 0.001 - 100 keV in units of erg\,s$^{-1}$ corresponding to the bolometric X-ray luminosity (bottom right).
}
\label{fig:comp4}
\end{figure*}

In addition to the lensing associated parameters, we showed the contrast of gas density perturbations against the ICM parameters such as the temperature, abundance and X-ray luminosity in the whole perturbed region in Figure~\ref{fig:comp4}. The sample correlation coefficients of each pair are also shown in Figure~\ref{fig:comp4}. The trends in the ICM temperature and X-ray luminosity are similar to those in the mass, i.e., an anti-correlation. Since a massive cluster is luminous and its ICM temperature is high, these results prefer the scenario that a massive cluster has a smaller gas density perturbations in the cool core, which is consistent with the result from the comparison with the lensing parameters. On the other hand, the ICM abundance seems to have a slightly positive correlation. No correlation is however acceptable. Since how the metal is concentrated toward the cluster center depends on the evolution of the BCG, it seems to be hard to characterize gas sloshing.

\subsection{Sample of strong gas sloshing candidates}
\label{sec:vs}

As mentioned in Section~\ref{sec:thermo}, the observed trend of each ICM property is consistent with the scenario that a majority of the gas density perturbations is created by gas sloshing. However, RXJ\,1532.9+3021 and MACS\,J1931.8-2634, in fact, host an apparent X-ray substructure in their cool cores most likely due to the AGN feedback \citep[e.g.,][]{Ehlert11, Hlavacek-Larrondo13}. To investigate the characteristics of gas sloshing, we selected strong gas sloshing candidates out of our sample that they have only one positive and one negative excess regions in their X-ray residual images and their morphology seems to be a spiral, which is a typical pattern of gas sloshing taking place in the plane of the sky shown by numerical simulations \citep[e.g.,][]{Ascasibar06, ZuHone10, Roediger11}. That pattern is also shown in our simulations (see Figure~\ref{fig:sim}).

\begin{figure*}
 \begin{center}
  \includegraphics[width=6.0cm,angle=0]{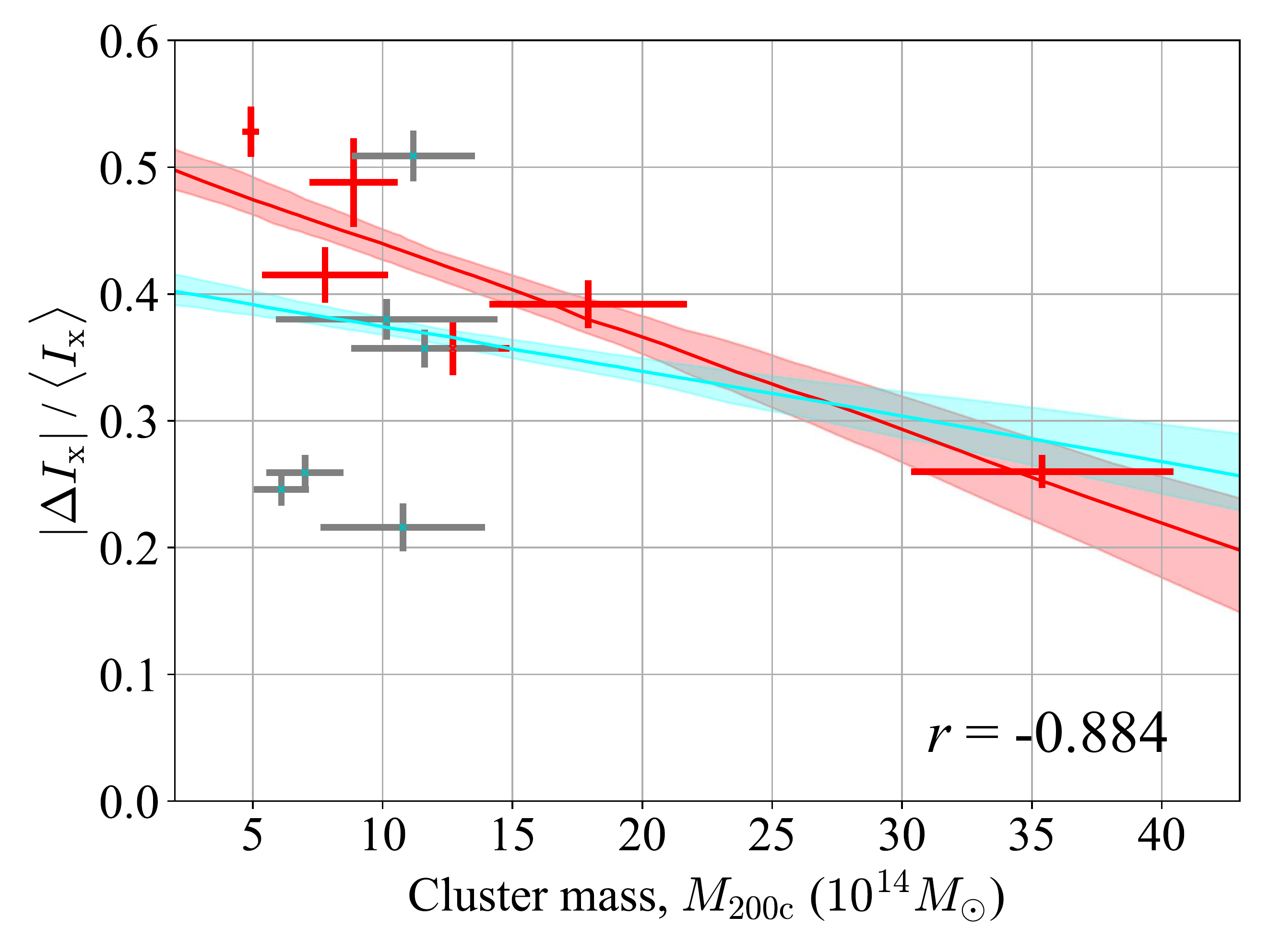}
  \includegraphics[width=6.0cm,angle=0]{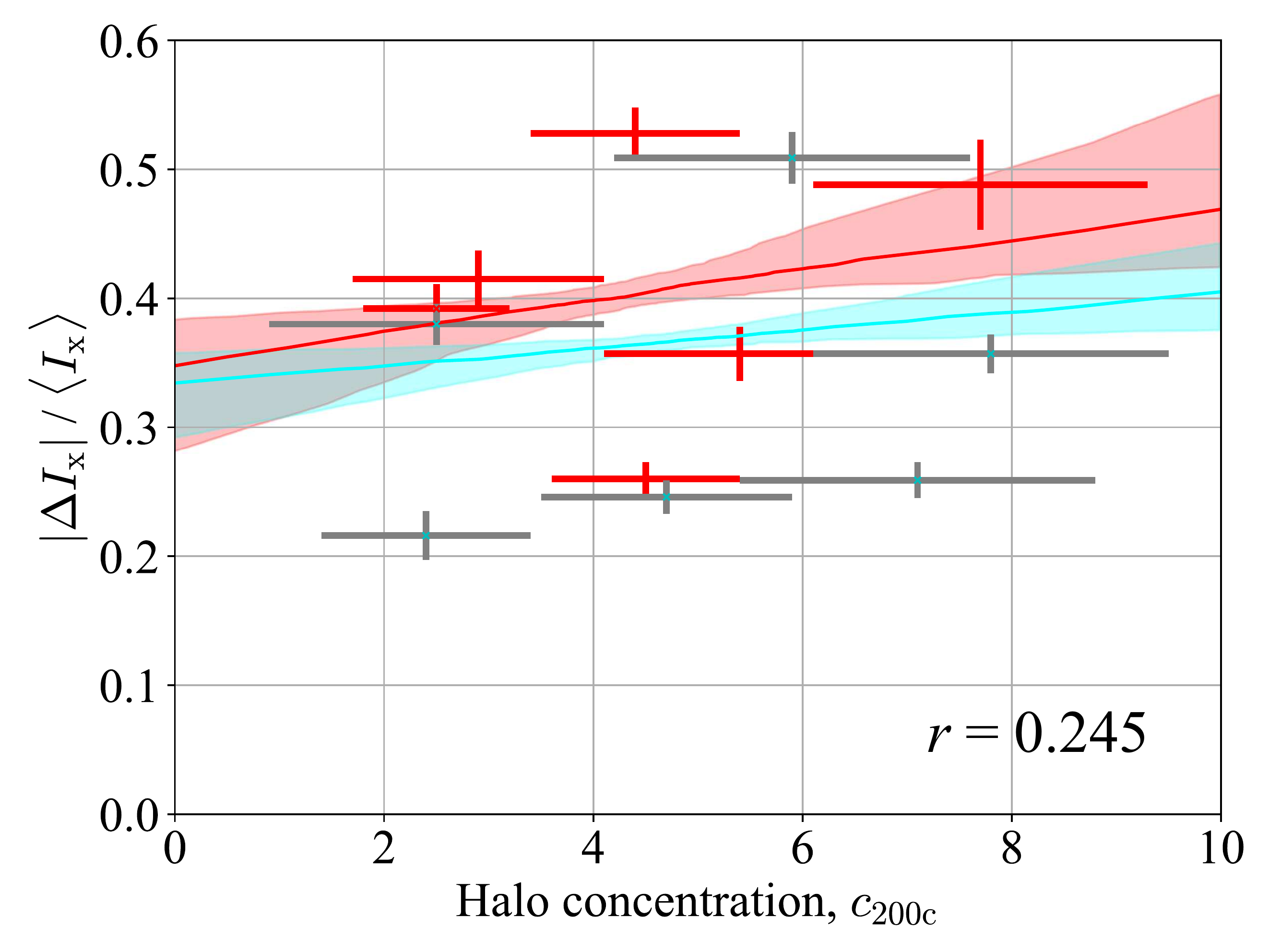}
  \includegraphics[width=6.0cm,angle=0]{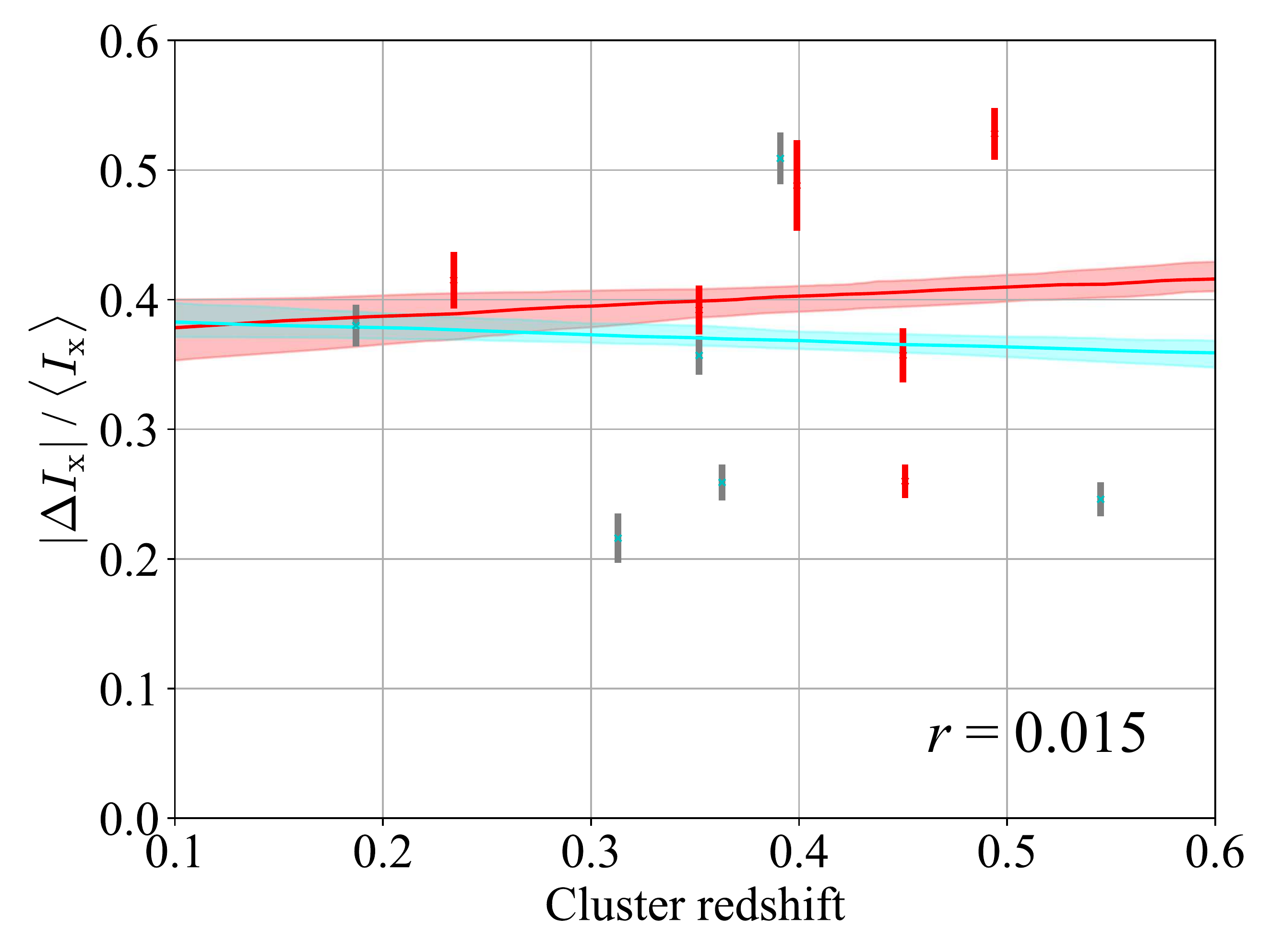}
 \end{center}
\caption{
Same as the top left, top right, and bottom panels of Figure~\ref{fig:comp3} but for the limited sample of the strong gas sloshing candidates. The red and gray crosses are the data included in and excluded from the limited sample, respectively. The red and cyan lines show the best-fit model for the limited sample and for all the sample that corresponds to the red band in Figure~\ref{fig:comp3}, respectively. The red and cyan bands are the 68\,\% confidence level of each model. The sample correlation coefficients, $r$, in each panel are calculated from the red data points.
}
\label{fig:comp5}
\end{figure*}

We chose six out of 12 galaxy clusters, i.e., MACSJ\,0329.6-0211, MACSJ\,0429.6-0253, MACSJ\,1115.8+0129, MACSJ\,1311.0-0310, RXJ\,1347.5-1145, and RXJ\,2129.6+0005, based on the X-ray residual image and the thermal properties of the ICM listed in Table~\ref{tab:fit}. Using this limited sample, we refitted the contrast of gas density perturbations with respect to the total cluster mass, concentration parameter, and cluster redshift, respectively. Figure~\ref{fig:comp5} shows the new best-fit model. The trend in the total cluster mass has more explicit anti-correlation than that shown in the previous one. In addition, the trend in the concentration parameter is slightly steeper than that in the original one. These trends allow us to emphasize that the majority of gas density perturbations are associated with gas sloshing, implying that the contrast of gas density perturbations is a good tracer for gas sloshing. On the other hand, the trend in the cluster redshift is offset from the previous one, indicating that the distribution of this limited sample is the same as that of the original sample.

\subsection{Comparison with the Strong-Lensing Mass Map}
\label{sec:mass}

To search for substructures in the central mass distribution and to identify their origin, we compared the X-ray surface brightness with the central mass distribution from the CLASH SL modeling by \cite{Zitrin15}. We employed two mass maps, i.e., the NFW model and the light-traces-mass (LTM) model. Figure~\ref{fig:wSL} shows the contours of both mass maps overlaid on the X-ray surface brightness. As mentioned in Section~\ref{sec:img}, the morphology of X-ray surface brightness is in good agreement with the dark matter distribution. 

We found that MACSJ\,0329.6-0211 has a substructure in the central mass distribution in the northwest direction. Figure~\ref{fig:sub0329} shows the zooming-out mass map of both the NFW and LTM models overlaid on its X-ray surface brightness. As shown in Figure~\ref{fig:image}, the spiral-like pattern is found in the X-ray residual image. The thermodynamic properties of the ICM in the positive and negative excess regions are compatible with the nature of gas sloshing. If this substructure in the mass map is associated with an infalling subcluster, gas sloshing in this cluster may be induced by this secondary component. On the other hand, we found no apparent feature of gas stripping around the secondary component, indicating that a merger is in the second passage or more. We will discuss this feature in detail in Section~\ref{sec:sub0329}. 

For RXJ\,1347.5-1145, \cite{Zitrin15} found the secondary component in the central mass distribution in the southeast region. As already studied by e.g., \cite{Ueda18}, this substructure is considered to be in the first passage because of the presence of the stripped and shock-heated gas behind this secondary component. No association of the secondary component with the X-ray surface brightness in MACSJ\,0329.6-0211 implies that the merger time scale in this cluster is longer than that RXJ\,1347.5-1145.

No substructure in the central mass distribution is found in the other clusters. This indicates that the substructure in the cluster center is rare even though gas sloshing is taking place in their cool cores. An infalling subcluster is then considered to be located far away from the cluster center. The weak-lensing (WL) analysis thus enables us to search for such substructures, while it is hard to identify which substructure is responsible for gas sloshing \citep[e.g., Figure~1 of][for the CLASH sample]{Umetsu14}. On the other hand, the presence of several substructures in the WL mass map indicates that gas sloshing can take place continuously in different time scale. Namely, each substructure allows to induce gas sloshing until its kinetic energy is lost. In addition, numerical simulations indicate that sloshing gas motions are long-lived structures and they can persist for several to many Gyrs after the subclusters merged with the primary cluster. This can be another reason for the absence of substructures in the core.

\subsection{Gas sloshing as a possible mechanism to suppress runaway cooling}
\label{sec:heat}

As indicated by previous studies of numerical simulations \citep[e.g.,][]{Fujita04b, ZuHone10} and recent observations \citep[e.g.,][]{Su17}, gas sloshing can be a possible candidate to suppress runaway cooling. It is still under debate whether or not all of cool-core clusters experiences gas sloshing, while our results indicate that gas sloshing occurs in most of cool cores. Our results, therefore, support this hypothesis. 

The redshift range of our sample is $0.187 < z < 0.545$. The time scale in this redshift range then corresponds to over $3$\,Gyr, which indicates that gas sloshing can be effective in more than 3\,Gyr. On the other hand, we inferred the fraction of the sloshing cool core to be 80\,\% (95\,\% CL). Since the cool core fraction in the CLASH clusters is 48\,\% (i.e., 12/25), which is consistent with that measured by X-ray-selected samples \citep[e.g.,][]{Andrade-Santos17}. The time scale inferred from the fraction is then 3.7\,Gyr assuming that a cluster is formed since $z = 1$, i.e., 7.8\,Gyrs ago. Both estimates are in agreement with each other. Since it can be considered that gas sloshing injects heat into the ICM in $\sim 3$\,Gyr, gas sloshing is possible to suppress runaway cooling of the ICM, as suggested by numerical simulations. However, it is still unclear how an amount of energy can be deposited by a merger through gas sloshing. An energy injection rate is important to understand the capability of gas sloshing. We will investigate this subject using hydrodynamic simulation in future.

Gas sloshing is capable of inducing the KHI. The kinetic energy of gas sloshing can be dissipated through the KHI-induced turbulence. For the Perseus cluster, \cite{Ichinohe19} estimated the KHI-induced turbulent heating rate per unit volume to be $Q_{\rm turb} \sim 10^{-26}$\,erg\,cm$^{-3}$\,s$^{-1}$. If this heating rate is the case, the heating rate of gas sloshing, therefore, can be inferred to be $\sim 1.2 \times 10^{45} (r/100\,$\rm kpc$)^3$\,erg\,s$^{-1}$, assuming that the volume of the sloshing region is spherical and its radius is 100\,kpc. The inferred heating rate is larger than the observed cooling rate of the sample, i.e., the bolometric X-ray luminosity ($< 10^{45}$\,erg\,s$^{-1}$; see the bottom right panel of Figure~\ref{fig:comp4}), except for RXJ\,1347.5-1145. In spite of a rough estimate, this result is supportive of the capability of gas sloshing.

In addition, we estimated the total energy injected by gas sloshing using the time scale of sloshing (i.e., $\sim 3$\,Gyr). The total energy is thus estimated as $\sim 1.2 \times 10^{62} (r/100\,$\rm kpc$)^3 (t / 3\,{\rm Gyr})$\,erg. If the initial velocity of an infalling subcluster is 1000\,km\,s$^{-1}$, the required total mass is then $\sim 1.2 \times 10^{13}$\,\MO, which is comparable to a massive galaxy if the efficiency is unity. In a realistic case, the efficiency would be smaller than unity. If the efficiency would be 10\,\%, the required mass becomes the group scale. Our estimate indicates that one minor merger induced by an infalling galaxy group per 3\,Gyr enables to suppress runaway cooling of the ICM through gas sloshing. However, this estimate is too rough to propose that gas sloshing is responsible for the heating source for the entire cluster age. 

On the other hand, numerical simulations suggest that gas sloshing can weaken a strong centrally-peaked X-ray surface brightness \citep[e.g.,][]{ZuHone10}. This effect reduces the number density of the ICM in the center so that the effective cooling time is expected to be longer than the previous value. Since larger concentration parameters lead to larger contrast of gas density perturbations (see the top right panel of Figure~\ref{fig:comp3}), the centrally peaked X-ray surface brightness is expected to be reduced after gas sloshing. In addition, turbulent heat conduction \citep{Cho03, Kim03, Voigt04} induced by sloshing gas motions may work effectively to enhance mixing between hotter and cooler gas. If the viscosity of the ICM is low, this effect works well in cool cores except at the very center. Both dissipation of turbulent kinetic energy and heat transport by turbulent mixing and electron conduction play an important role in gas sloshing as a heating source.

We note here that some limitations of the capability of gas sloshing to stop runaway cooling have been reported. \cite{ZuHone10} carried out one of the first numerical studies of gas sloshing and have pointed out that gas sloshing is unable to prevent a cooling catastrophe for more than 1-2\,Gyrs under some merger conditions, suggesting that the capability of gas sloshing strongly depends on the cluster mass ratio and impact parameter of mergers. In addition, magnetic fields in the ICM may suppress not only mixing of gas but also perturbations induced by the KHI, as shown in e.g., \cite{ZuHone11}. Even though magnetic fields of the ICM and thermal conduction are taken into account, the capability of gas sloshing to stop runaway cooling is still not clear due to such complex uncertainties \citep[e.g.,][]{ZuHone13b}. Further investigations using cosmological hydrodynamic simulations are needed to understand the effect of gas sloshing on gas cooling in cool-core clusters.

\subsection{Gas sloshing and radio mini-halos}
\label{sec:radio}

Some cool-core clusters host a diffuse radio emission feature, a so-called radio mini-halo, in their cores \citep[][for reviews]{Feretti12, van_Weeren19}. Radio mini-haloes are extended on a scale of $\le 500$\,kpc. The morphology of radio mini-haloes appears to be associated with that of cool cores in X-rays. \cite{Mazzotta08} found a spatial correlation between the substructure induced by gas sloshing and a radio mini-halo. Such correlation is also found in the Perseus cluster \citep{Walker17}. \cite{ZuHone13} proposed using numerical simulations that sloshing motions generate turbulence and such turbulence is potentially enough to produce diffuse radio emission. 

Our results indicate that most of the cool cores have experienced gas sloshing. If gas sloshing is responsible for creating a radio mini-halo, our results support this assumption. In fact, a diffuse radio emission has been detected in seven clusters out of our sample: MACSJ\,0329.6-0211 \citep{Giacintucci14}, MACSJ\,1115.8+0129 \citep{Pandey-Pommier16}, RXJ\,1347.5-1145 \citep{Gitti07}, RXJ\,1532.9+3021 \citep{Kale13}, MACS\,J1720.2+3536 \citep{Giacintucci17}, MACS\,J1931.8-2634 \citep{Giacintucci14}, and RXJ\,2129.6+0005 \citep{Kale15}. We note that MACS\,J1931.8-2634 is classified into an unclassified source \citep{Giacintucci14, van_Weeren19}, and MACSJ\,1115.8+0129 and MACS\,J1720.2+3536 are recognized as a candidate for radio mini-halo \citep{Pandey-Pommier16, Giacintucci17, van_Weeren19}. The four clusters out of the limited sample introduced in Section~\ref{sec:vs} are included in the above list. A deep radio observation of the remaining clusters will provide us with a good opportunity to test this assumption.

\subsection{Importance of high-resolution, cosmological hydrodynamic simulations}

High-resolution, cosmological hydrodynamic simulations provide a useful tool to understand gas dynamics and thermodynamics in galaxy clusters,
and guide us to study the physical properties of X-ray substructures \citep[e.g.,][]{Lyskova19}. Even though there have been numerous numerical studies to interpret multi-wavelength observations of galaxy clusters, there is a significant lack of understanding of how the thermodynamic properties of the ICM co-evolve with the dark-matter potential through merging processes. For example, if a secondary merger occurs while the cool core is still experiencing gas sloshing, what kind of gas density perturbations can we observe? In addition, how can gas sloshing impact on the AGN feedback and its remnant in the X-ray surface brightness? Gas sloshing probably coexists with the AGN feedback in the cool core, as shown in this paper. As demonstrated by the {\it Hitomi} observations of the Perseus cluster, a gas motion induced by gas sloshing and the AGN feedback coexists in the cool core \citep{Hitomi16, Hitomi18d}. In fact, \cite{Fujita19} have proposed that ubiquitous turbulence induced by mass accretions and AGN feedback are both important to maintain the balance between cooling and heating in cool cores on a long time scale. High-resolution spectroscopy with high-resolution imaging allows us to measure the thermodynamics and dynamics of the ICM in the cool core in detail. The next generation X-ray observatory such as {\it Athena} \citep{Nandra13} and {\it Lynx} \citep{Lynx18} will be capable of resolving such gas motion more accurately. The bridge between gas sloshing and the AGN feedback will help us to understand the thermal evolution of the ICM in terms of a cosmological point of view in future. 

As discussed in Section~\ref{sec:mass}, galaxy clusters host several substructures in their WL mass distributions, which are expected to induce gas sloshing features in their cool cores. Numerical simulations are needed to interpret multi-wavelength observations of such a complex system. The frequency of gas sloshing and the total kinetic energy of infalling subclusters are key factors to understand the role and importance of gas sloshing in suppressing runaway cooling of the ICM in the cool core. Further observational and theoretical studies are required to understand the nature of gas sloshing in context of the evolution of galaxy clusters. 

We will investigate this subject in our future work using high-resolution cosmological hydrodynamic simulations. As demonstrated in Section~\ref{sec:sim}, numerical simulations can be used to test and validate our methodology and analysis tools, and guide us to understand the ICM properties in sloshing cores. We will study the dependence of ICM parameters, such as the temperature difference and the contrast of gas density perturbations, as a function of the merger geometry and initial conditions of the primary cluster. In a future work, we will study key parameters related to gas sloshing, i.e., the heating rate, time scale, and efficiency converted into heat.

\section{Conclusions}

In this paper, we carried out a systematic study of gas density perturbations in cool cores of 12 high-mass relaxed clusters selected from the CLASH sample. Our analysis is based on archival X-ray data from the {\it Chandra X-ray Observatory} and the publicly available strong-lens mass models of \cite{Zitrin15}. The main conclusions of this paper are summarized as follows.   

\begin{enumerate}

\item All 12 relaxed clusters in our sample exhibit gas density perturbations in their cool cores. They have regions with, at least, one positive excess and one negative excess regions in their X-ray residual image after removing their mean X-ray surface brightness distribution. The observed position angle and axis ratio of the X-ray surface brightness are in good agreement with those of the central mass distribution obtained from the {\it HST} SL modeling.

\item We identified locally perturbed regions on the basis of gas density fluctuations in the X-ray residual image. We extracted and analyzed X-ray spectra of the ICM in the perturbed regions, finding that the thermodynamic properties of the ICM are consistent with those perturbed by gas sloshing. In particular, the ICM in the perturbed regions is found to be in pressure equilibrium. Our results indicate that gas sloshing in cool cores takes place in more than 80\,\% of relaxed galaxy clusters (95\,\% CL). 

\item We carried out a hydrodynamic simulation of a cluster merger to produce the synthetic X-ray observations of cool-core clusters in gas sloshing. By analyzing the simulated X-ray images, we found that our detection algorithm can accurately extract both the positive and negative excess regions and can reproduce the temperature difference between them, which is consistent with that found in the observations.  

\item We quantified the contrast of gas density perturbations (i.e., $|\Delta I_{\rm X}| / \langle I_{\rm X} \rangle$) and compared it with X-ray observables and lensing properties, such as the cluster mass ($M_{200c}$) and the concentration parameter ($c_{\rm 200c}$). the contrast of gas density perturbations is found to anti-correlate with the total cluster mass, whereas it has a positive correlation with the concentration parameter. The observed trends indicate that the gas density perturbations are most likely induced by gas sloshing.  

\item We found no apparent secondary dark-matter component in the central mass distribution, except for the cases of MACSJ\,0329.6-0211 and RXJ\,1347.5-1145. This indicates that, for the majority of the sample, the original perturber of gas sloshing, i.e., infalling subcluster(s), is likely located in the outer regions. Another possible reason is that, as indicated by numerical simulations, sloshing gas motions are long-lived structures. They persist for several to many Gyrs, perhaps even after the subclusters merged with the primary cluster.

\item Our results indicate that the majority of density perturbations in cool cores of our sample are induced by gas sloshing. On the other hand, some perturbations appear to arise from AGN feedback (e.g., X-ray cavities), as found in the X-ray residual image of RXJ\,1532.9+3021. High-resolution hydrodynamic simulations including AGN feedback physics will allow us to quantify the relative impact of gas sloshing and feedback mechanisms on the thermodynamic properties of the ICM in cool cores.

\item The heating rate inferred from KHI-induced turbulence due to gas sloshing is found to be larger than the bolometric X-ray luminosity in the perturbed region, except for RXJ\,1347.5-1145, suggesting that gas sloshing is a possible heating source. The total energy injected by gas sloshing over 3\,Gyr is estimated as $\sim 10^{62}$\,erg. This indicates that a minor merger induced by a galaxy group could be enough to suppress runaway cooling of the ICM, if the efficiency is as high as 10\,\%. 

\item In agreement with previous studies, our results indicate that turbulence induced by gas sloshing provides a possible origin of radio mini-haloes. In fact, seven galaxy clusters in our sample show diffuse radio emission in their center. Deep radio observations are needed to test this hypothesis. 

\end{enumerate}

\acknowledgments
We are grateful to the anonymous referee for helpful suggestions and comments.
We thank John ZuHone for providing us the initial models of the clusters and Justin Schive for helping with the runs on the NCHC GPU cluster.
SU thanks Ignacio Ferreras for his helpful comments and constructive suggestions.
The scientific results of this paper are based in part on data obtained from the Chandra Data Archive: 
ObsID 506, 507, 524, 552, 928, 1649, 1657, 1665, 2320, 2321, 3257, 3258, 3271, 3275, 3280, 3282, 3582, 3592, 4195, 4974, 5250, 6107, 6108, 6110, 7225, 7718, 7719, 7721, 9370, 9375, 9381, 9382, 13516, 13999, 14009, and 14407.
Our numerical simulations were run on the National Center for High-Performance Computing in Taiwan.
This work is supported in part by the Ministry of Science and Technology of Taiwan (grant MOST 106-2628-M-001-003-MY3) and by Academia Sinica (grant AS-IA-107-M01). 
This work is also supported by the Grants-in-Aid for Scientific Research by the Japan Society for the Promotion of Science with KAKENHI Grant Numbers JP18H05458 (YI) and JP18K03704 (TK).

\clearpage

\appendix
\section{Information of individual galaxy clusters}

We summarized the brief information of individual galaxy cluster about previous X-ray observations and related results. Since \cite{Donahue14} and \cite{Donahue16} have studied all of the sample, we do not mention their results in this section.

\subsection{A383}

This cluster is analyzed in X-rays by \cite{Allen08} and \cite{Newman11}, while no detailed study is reported thus far. This cluster seems to experience line-of-sight gas sloshing. The pattern in the X-ray residual image is similar to that found in Abell\,907, which is identified as the system of line-of-sight sloshing \citep{Ueda19}. In addition, the best-fit parameters of the ICM properties in both the positive and negative excess regions are consistent with those expected by line-of-sight gas sloshing.

\subsection{MACSJ\,0329.6-0211}

\cite{Giacintucci14} have studied this cluster using \Chandra ~and VLA. This cluster appears to be relaxed and a candidate of radio mini-halo is found in the cluster center.

\subsection{MACSJ\,0429.6-0253}

\cite{Hlavacek-Larrondo12} have reported that two clear, small ($r \sim 5$\,kpc) cavities are found in the cluster center. However, the power inferred from the size of the cavities is significantly smaller than the bolometric X-ray luminosity in the core.

\subsection{MACSJ\,1115.8+0129}

The X-ray analysis of this cluster is carried out by \cite{Allen08} and \cite{Repp18}, while no detailed study is found thus far.

\subsection{MACSJ\,1311.0-0310}

\cite{Hlavacek-Larrondo12} have reported that no cavity is found in this cluster.

\subsection{RXJ\,1347.5-1145}

This cluster is one of the most luminous X-ray clusters and hosts a strong cool core. In addition, this cluster is recognized as a major merging cluster. A substructure exists in the southeast region in X-rays, SZE, and SL \citep[e.g.,][]{Komatsu01, Allen02, Kitayama04, Kohlinger14, Kitayama16, Ueda18}. The dipolar pattern in the X-ray residual image indicates that gas sloshing takes place in the core \citep{Johnson12, Kreisch16, Ueda18}. The velocity of sloshing motion is inferred solving the equation of state for gas density perturbations using both X-ray and SZE data. Thus, the inferred velocity is smaller than the adiabatic sound speed, indicating that the gas motion is subsonic \citep{Ueda18}. This result is also shown by \cite{Di_Mascolo19}.

\subsection{MACSJ\,1423.8+2404}

\cite{Hlavacek-Larrondo12} have studied this cluster and reported the presence of cavities. The power stored in the cavities is estimated to be sufficient to suppress gas cooling within the cooling radius.

\subsection{RXJ\,1532.9+3021}

\cite{Hlavacek-Larrondo13} have studied this cluster in detail. This cluster has an apparent X-ray cavity in the west and relatively weak X-ray cavity in the east. They estimated the total mechanical power of the cavities and indicated that it is sufficient to offset runaway cooling in the cluster core. \cite{Giacintucci14} have confirmed the presence of a radio mini-halo in this cluster.

\subsection{MACS\,J1720.2+3536}

\cite{Hlavacek-Larrondo12} have shown two cavities in the core. The inferred power in the cavities using the buoyancy time-scale does not meet the requirement to prevent cooling.

\subsection{MACS\,J1931.8-2634}

A luminous X-ray AGN exists in the center of this cluster. \cite{Ehlert11} have studies this cluster deeply and have reported that an extended radio emission is observed and elongated in the east-west direction, which is spatially associated with the X-ray cavities. This cluster therefore has one of the most powerful AGN outbursts known. This cluster also hosts a strong cool core but the ICM temperature in the innermost region increases \citep{Ehlert11}. This feature is also reported by \cite{Hlavacek-Larrondo12}. \cite{Ehlert11} found a spiral-like feature composed of relatively cool, dense ICM in the core, which is likely to be associated with gas sloshing.

\subsection{RXJ\,2129.6+0005}

\cite{Giles17} have reported the radial profile of the ICM temperature. This cluster hosts a strong cool core.

\subsection{MS\,2137.3-2353}

\cite{Hlavacek-Larrondo12} have reported a cavity in the south direction from the center.

\section{Comparison of X-ray surface brightness to SL mass map}
\label{sec:X-SL}

We summarized the comparison of the X-ray surface brightness to the SL mass map in each galaxy cluster in this section.

\begin{figure*}
 \begin{center}
  \includegraphics[width=7.2cm]{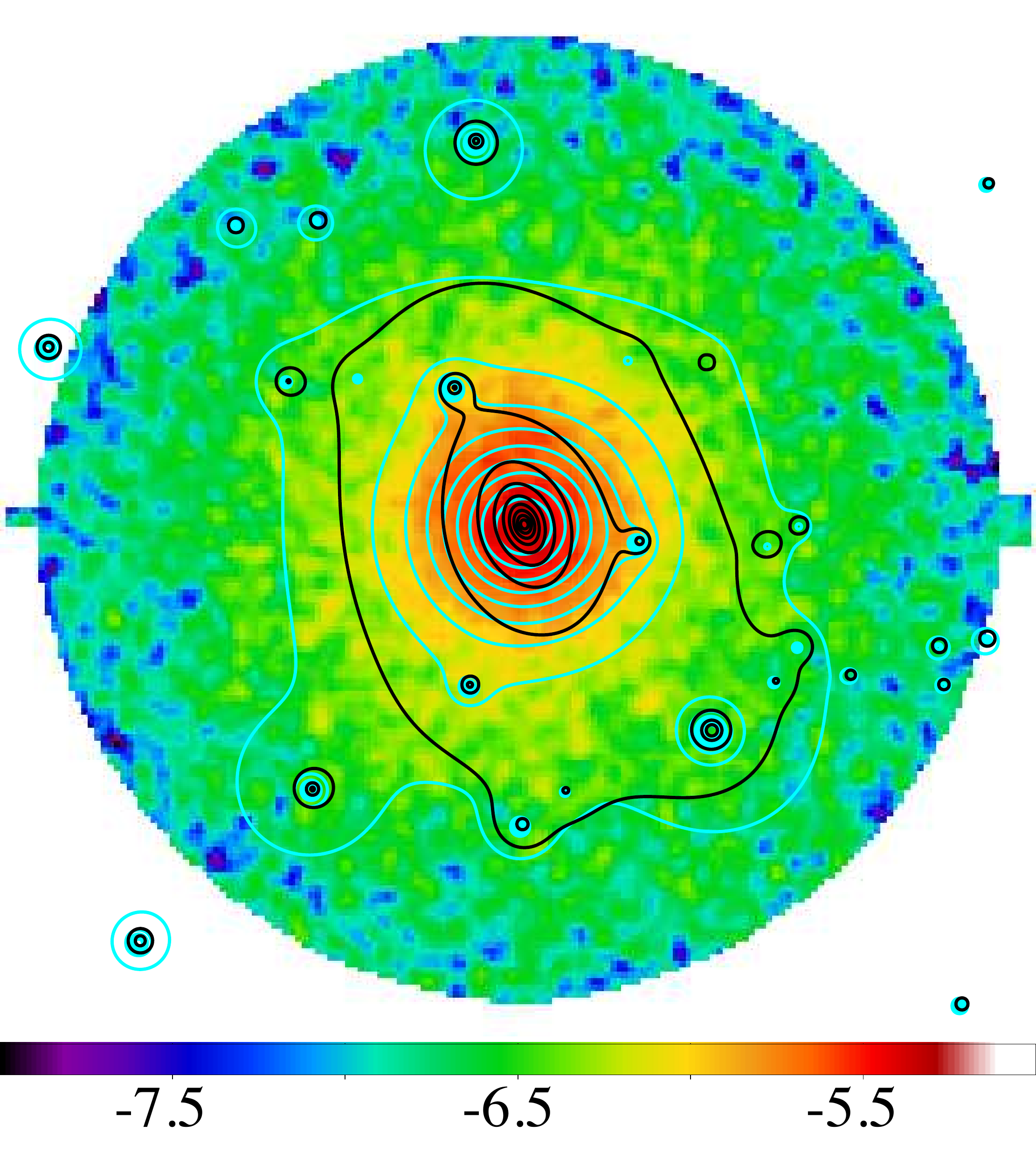}
  \includegraphics[width=7.2cm]{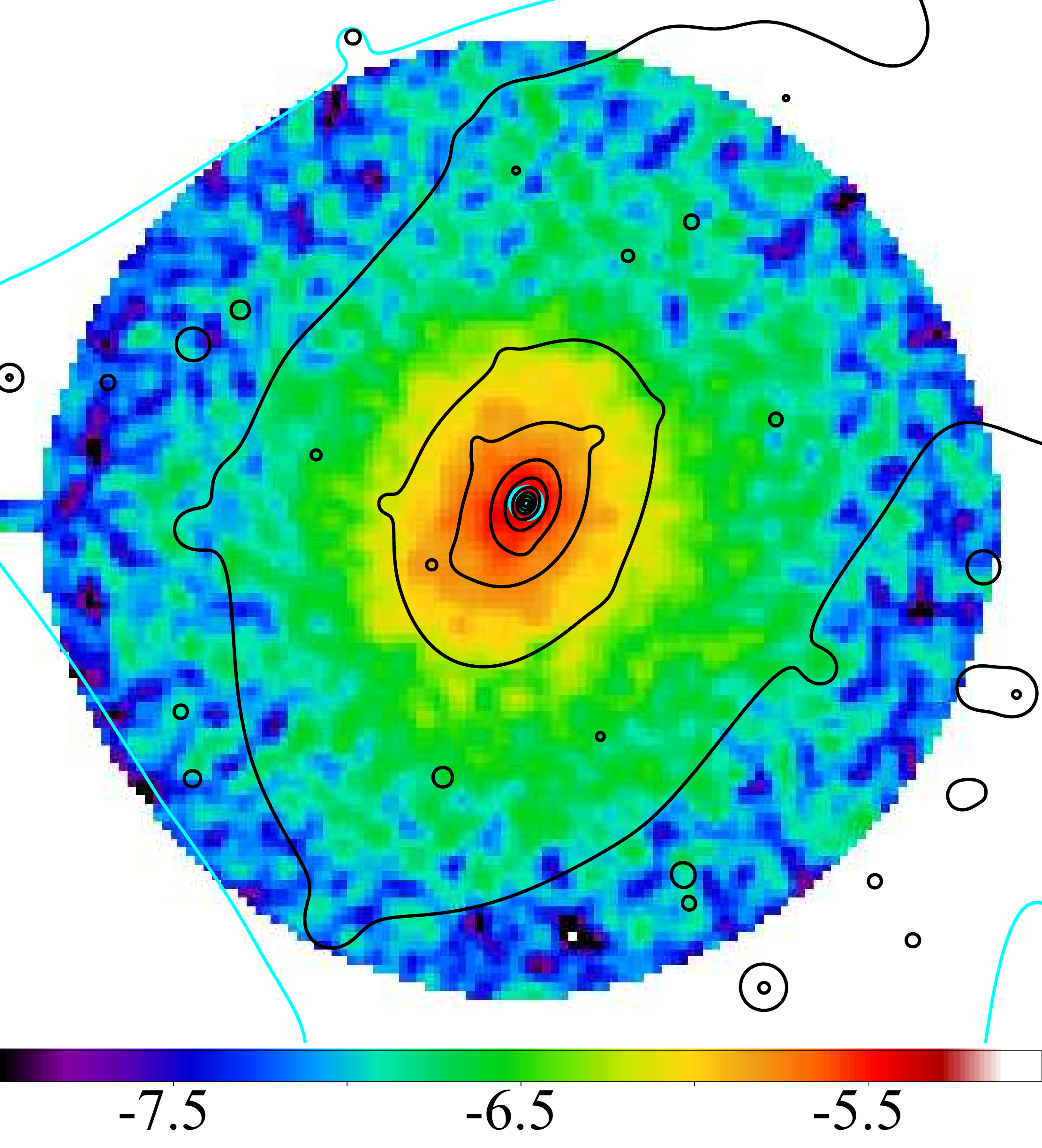} 
  \includegraphics[width=7.2cm]{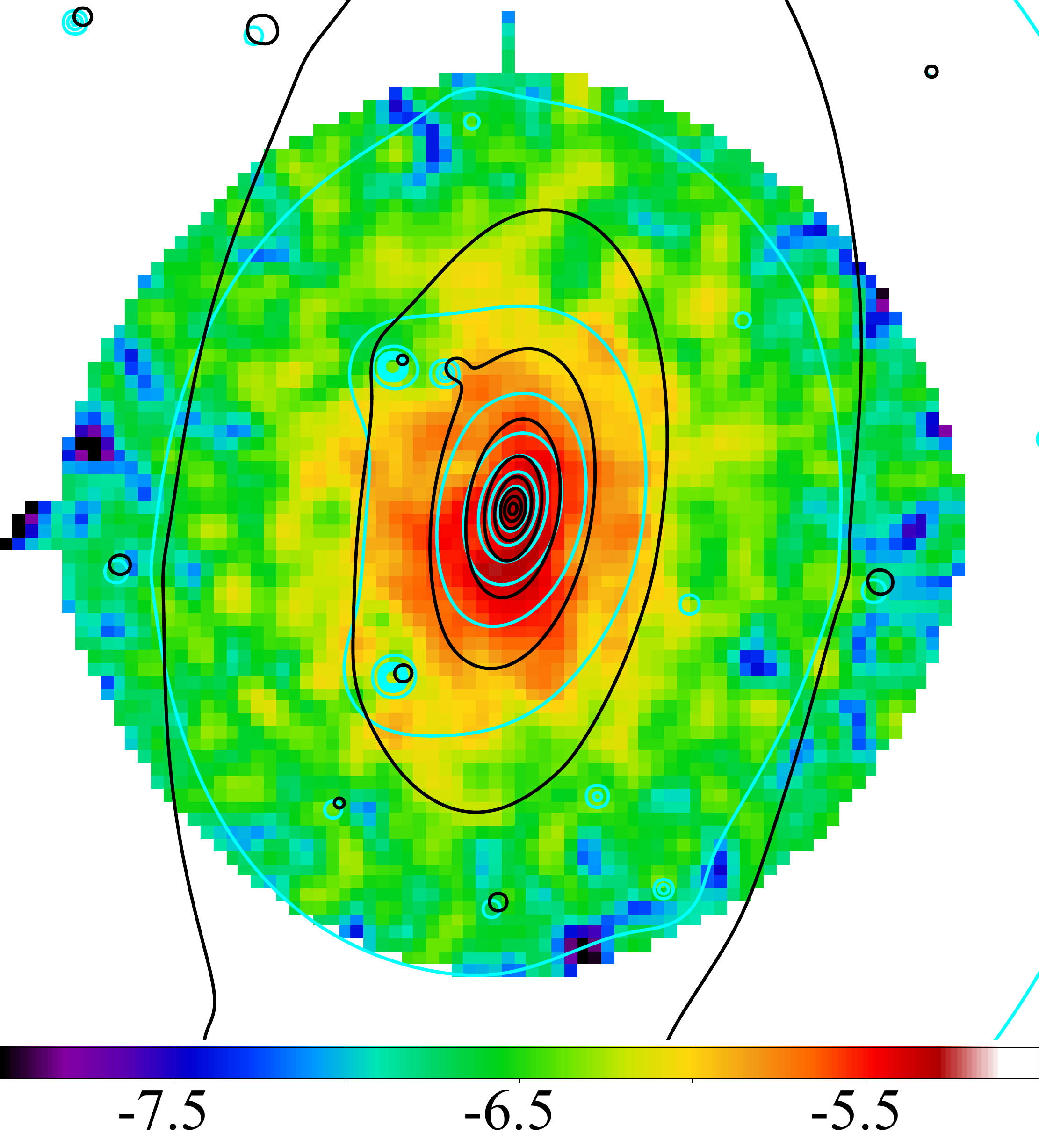}
  \includegraphics[width=7.2cm]{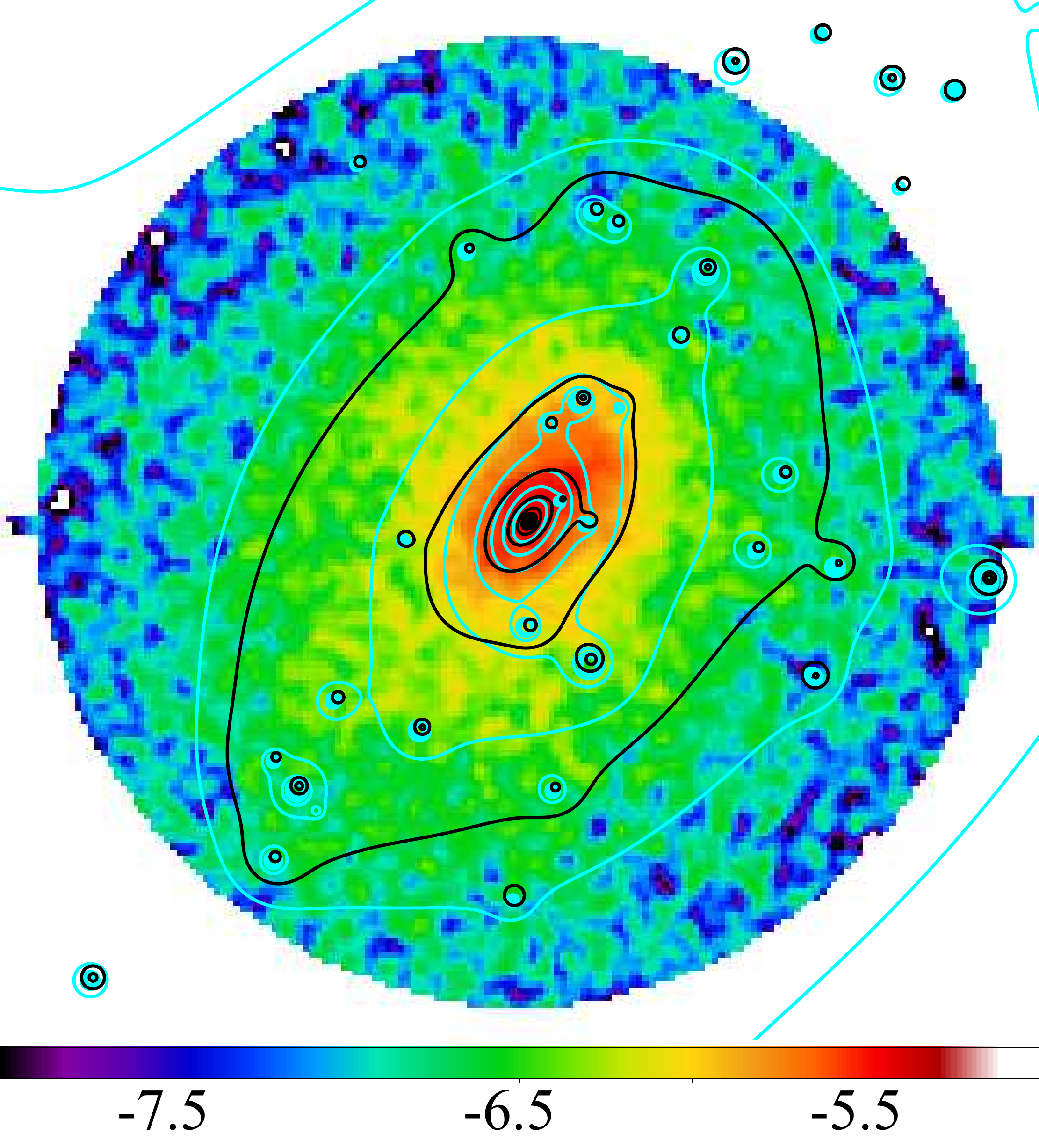} 
 \end{center}
\caption{Contours of the central mass distribution derived by the SL modeling with the NFW profile (black) and the LTM model (cyan) overlaid on the X-ray surface brightness, respectively. Contours correspond to the levels of $100\,\%, 90\,\%, ..., 10\,\%$ of the extreme value of each mass distribution. The panels from top to bottom are A383 (top left), MACSJ\,0329.6-0211 (top right), MACSJ\,0429.6-0253 (bottom left), and MACSJ\,1115.8+0129 (bottom right).
}
\label{fig:wSL}
\end{figure*}

\begin{figure*}
\addtocounter{figure}{-1}
 \begin{center}
  \includegraphics[width=7.2cm]{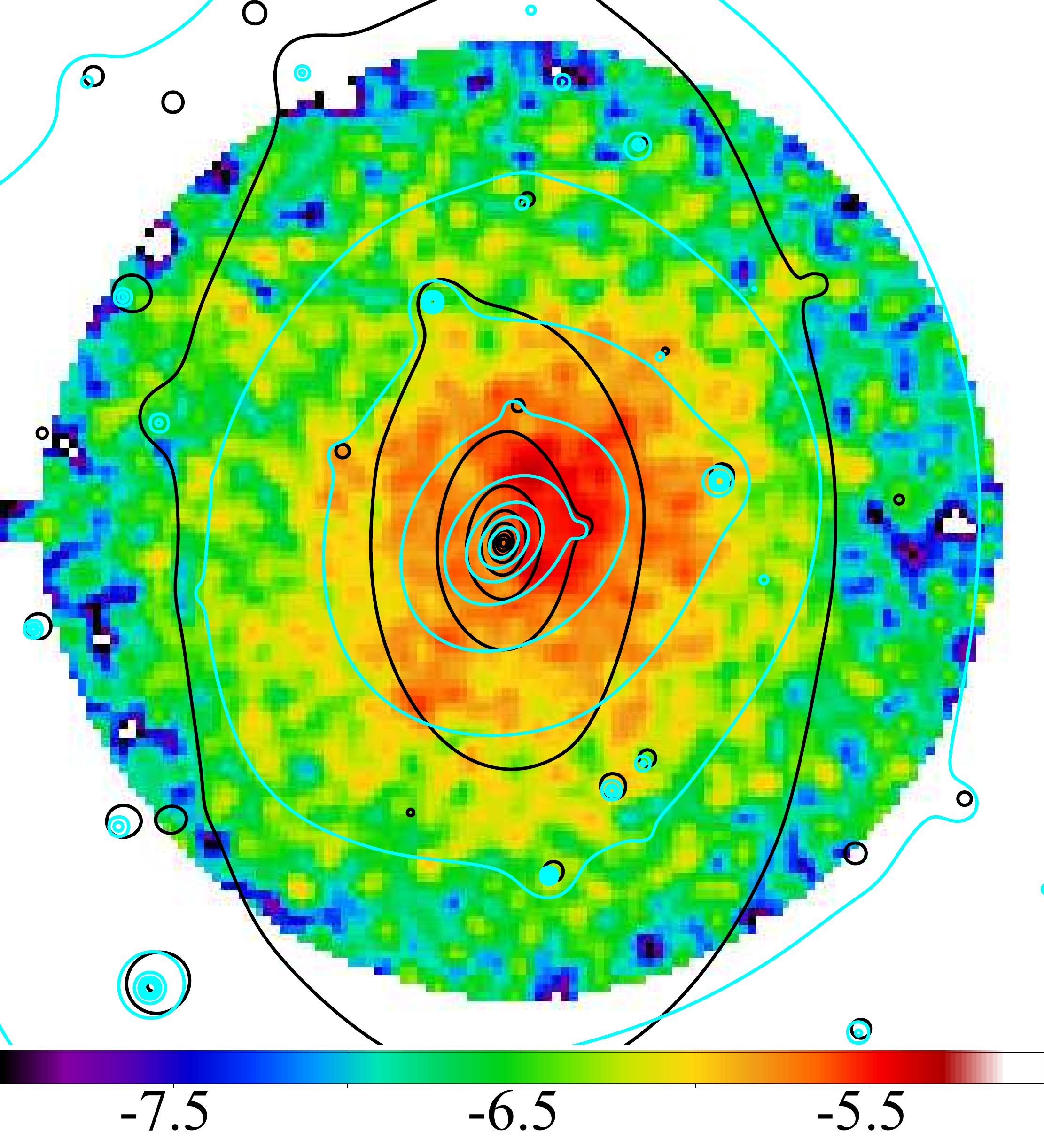}
  \includegraphics[width=7.2cm]{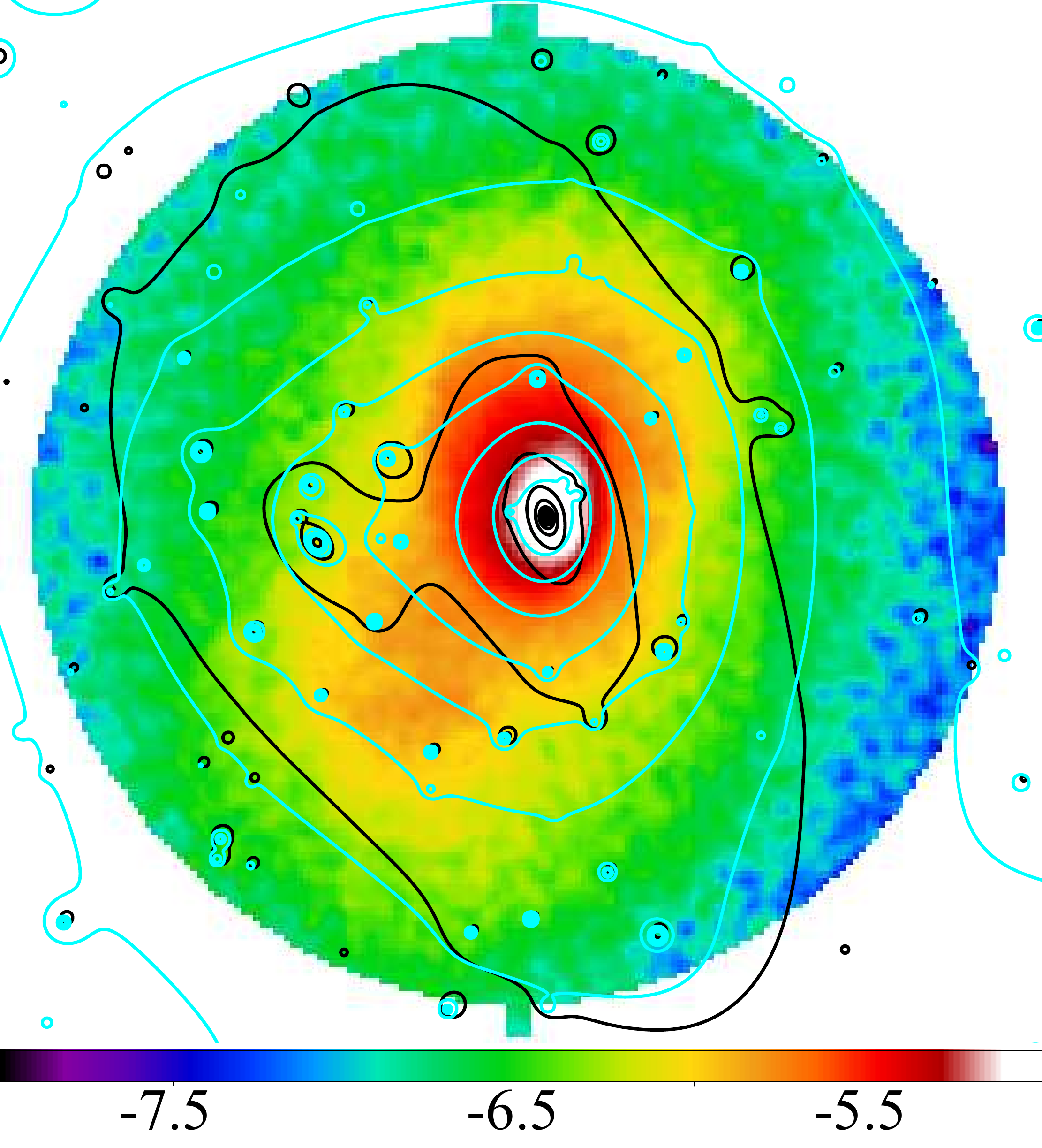}
  \includegraphics[width=7.2cm]{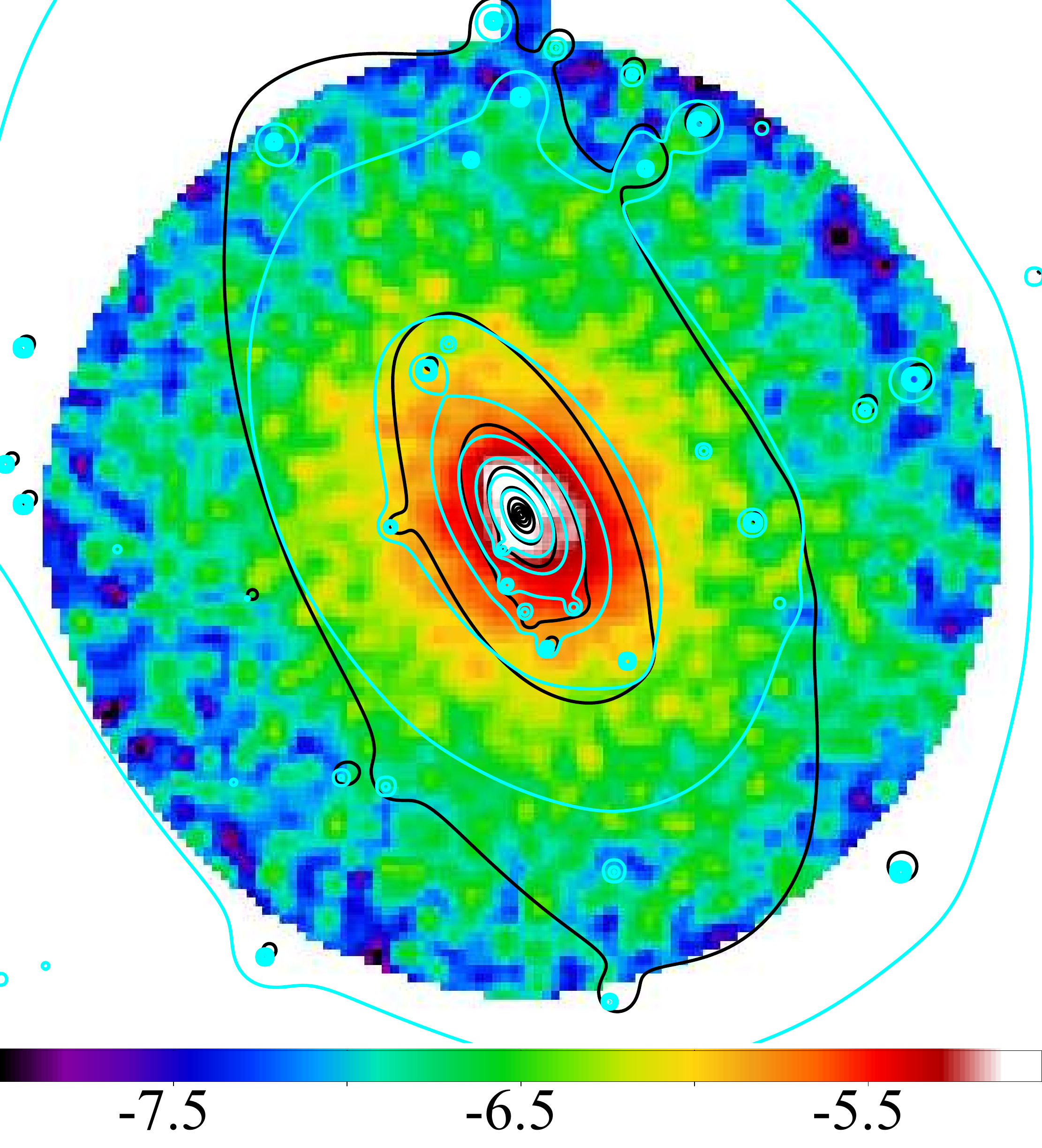}
  \includegraphics[width=7.2cm]{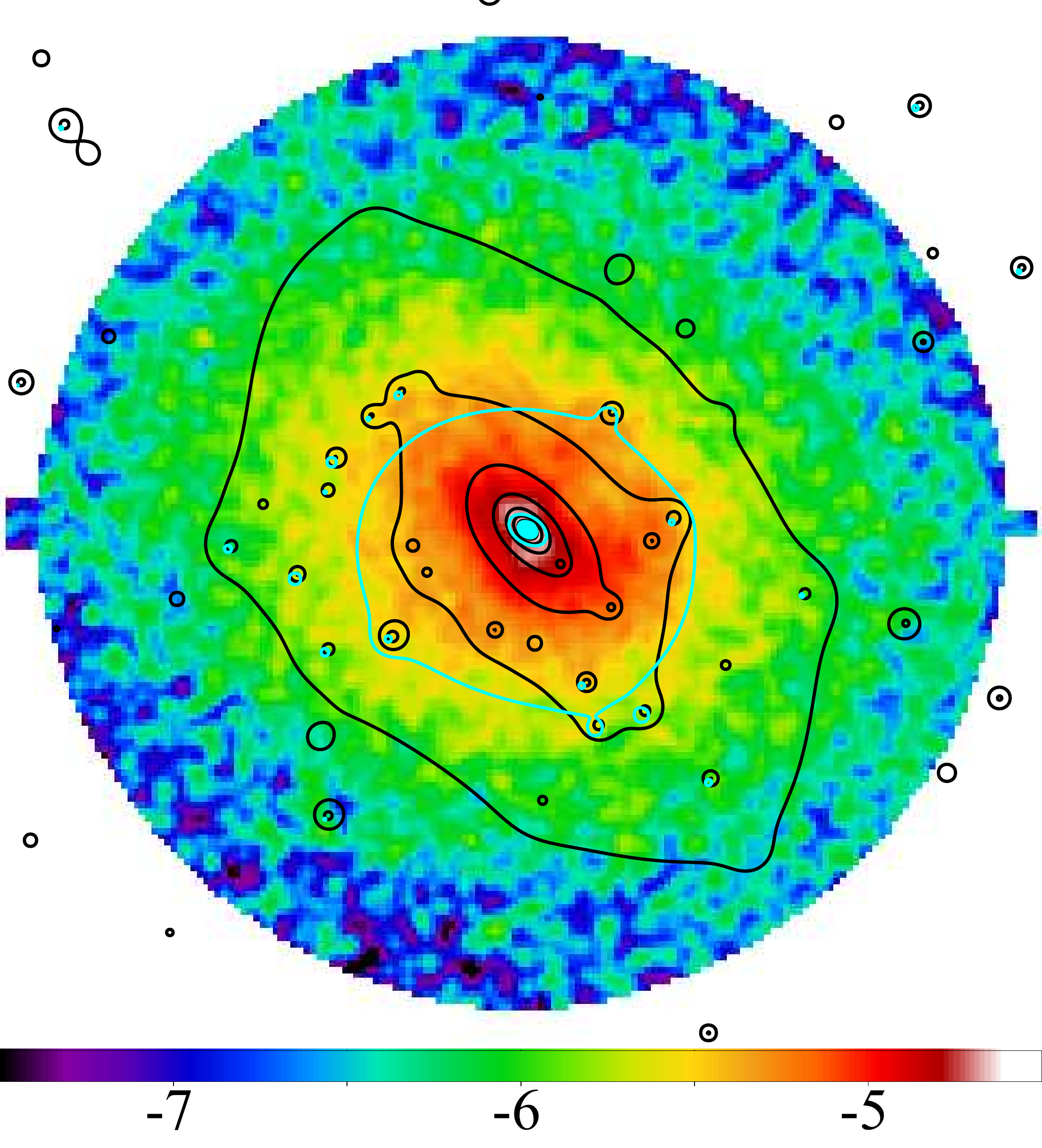}
  \includegraphics[width=7.2cm]{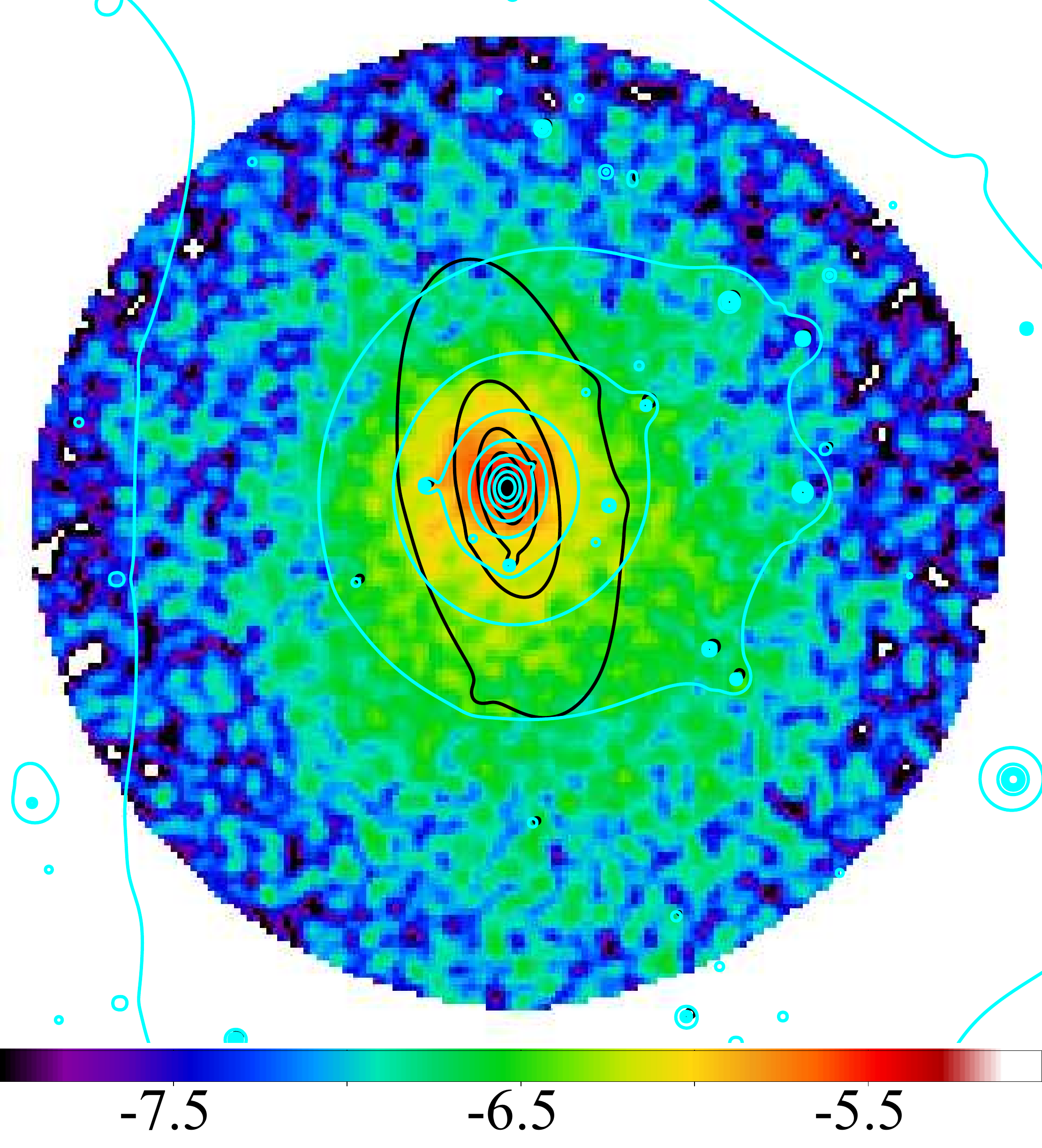} 
  \includegraphics[width=7.2cm]{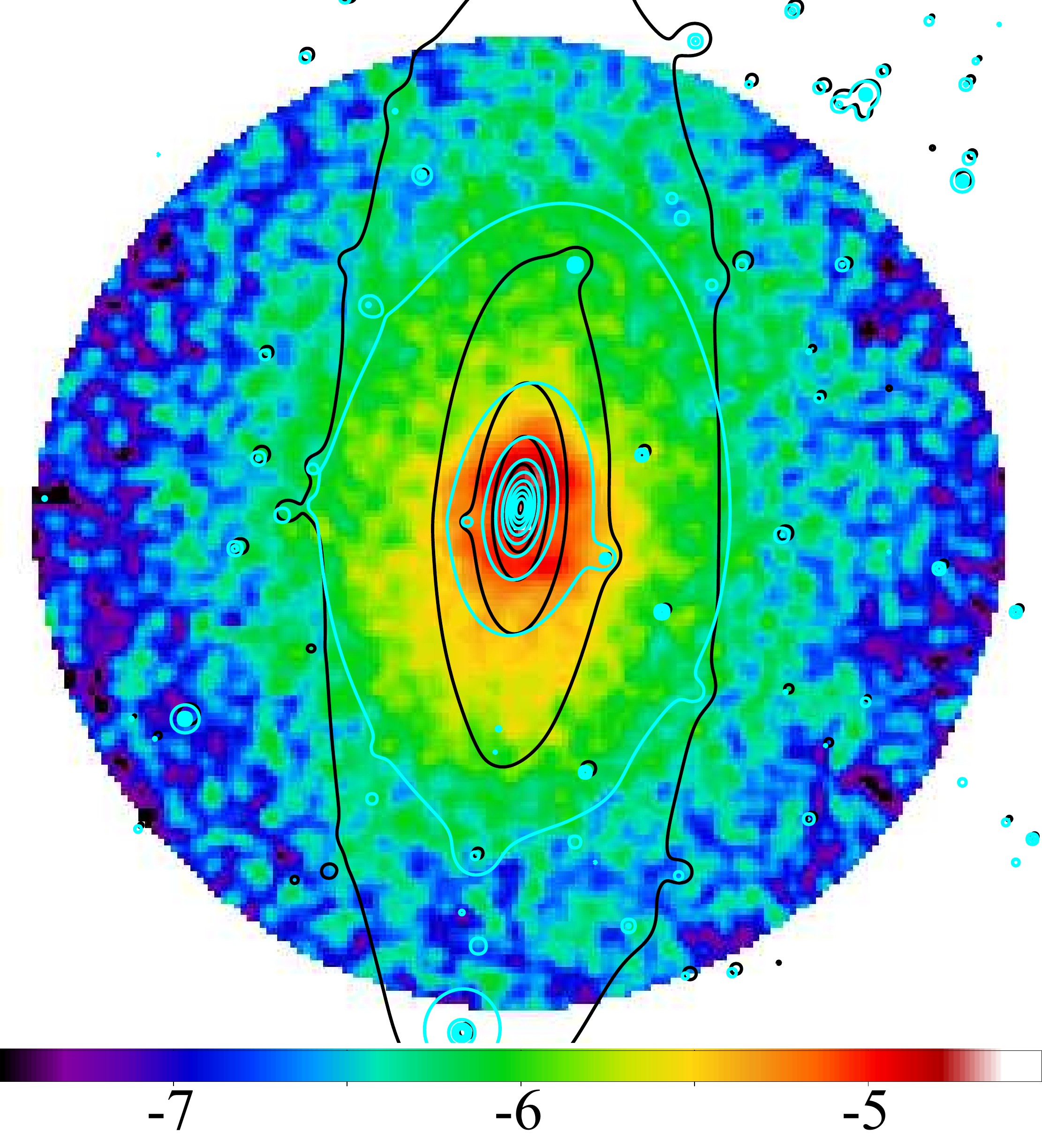}
 \end{center}
\caption{Continued. MACSJ\,1311.0-0310 (top left), RXJ\,1347.5-1145 (top right), MACSJ\,1423.8+2404 (middle left), RXJ\,1532.9+3021 (middle right), MACS\,J1720.2+3536 (bottom left), and MACS\,J1931.8-2634 (bottom right).
}
\end{figure*}

\begin{figure*}
\addtocounter{figure}{-1}
 \begin{center}
  \includegraphics[width=7.2cm]{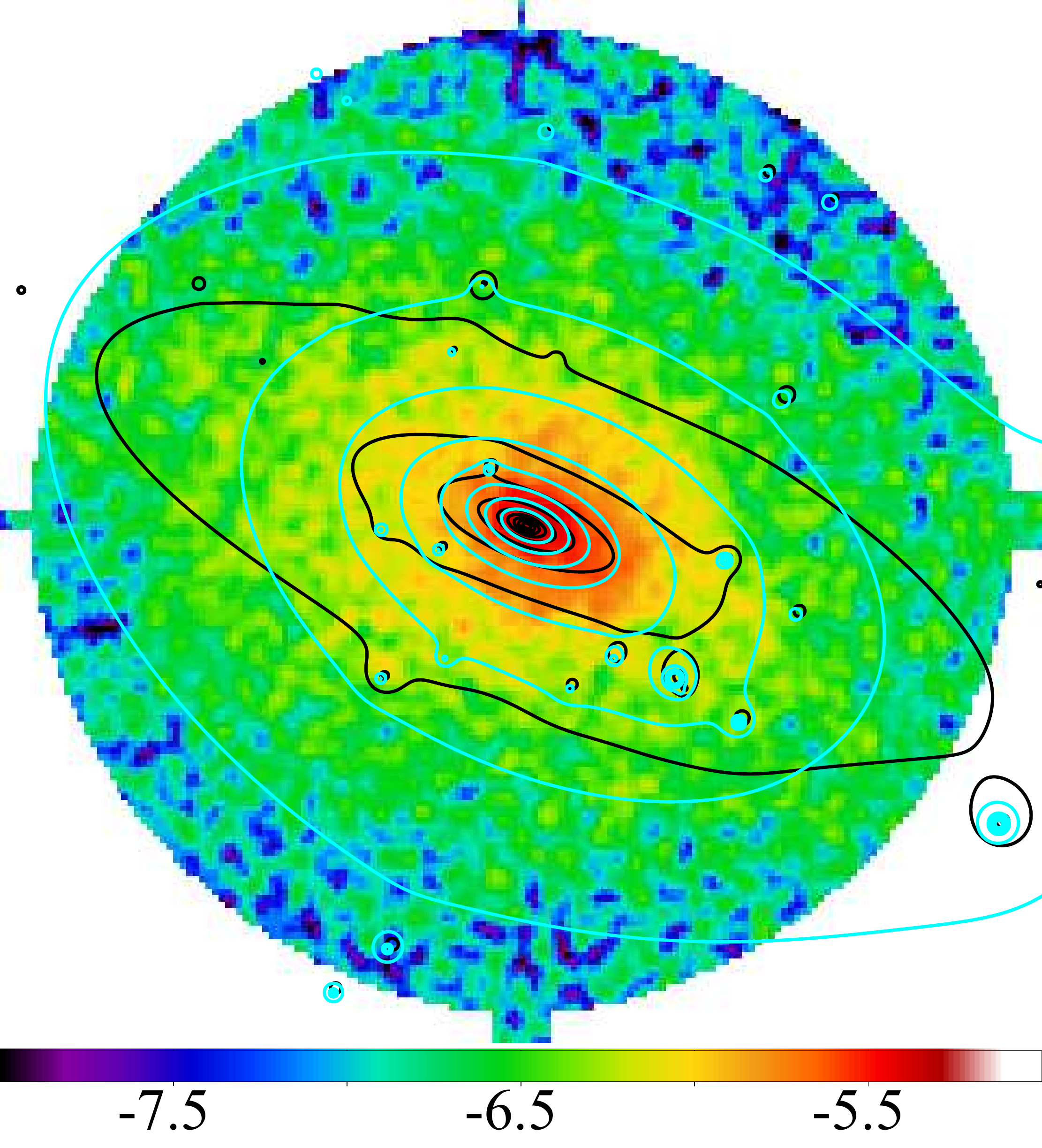}     
  \includegraphics[width=7.2cm]{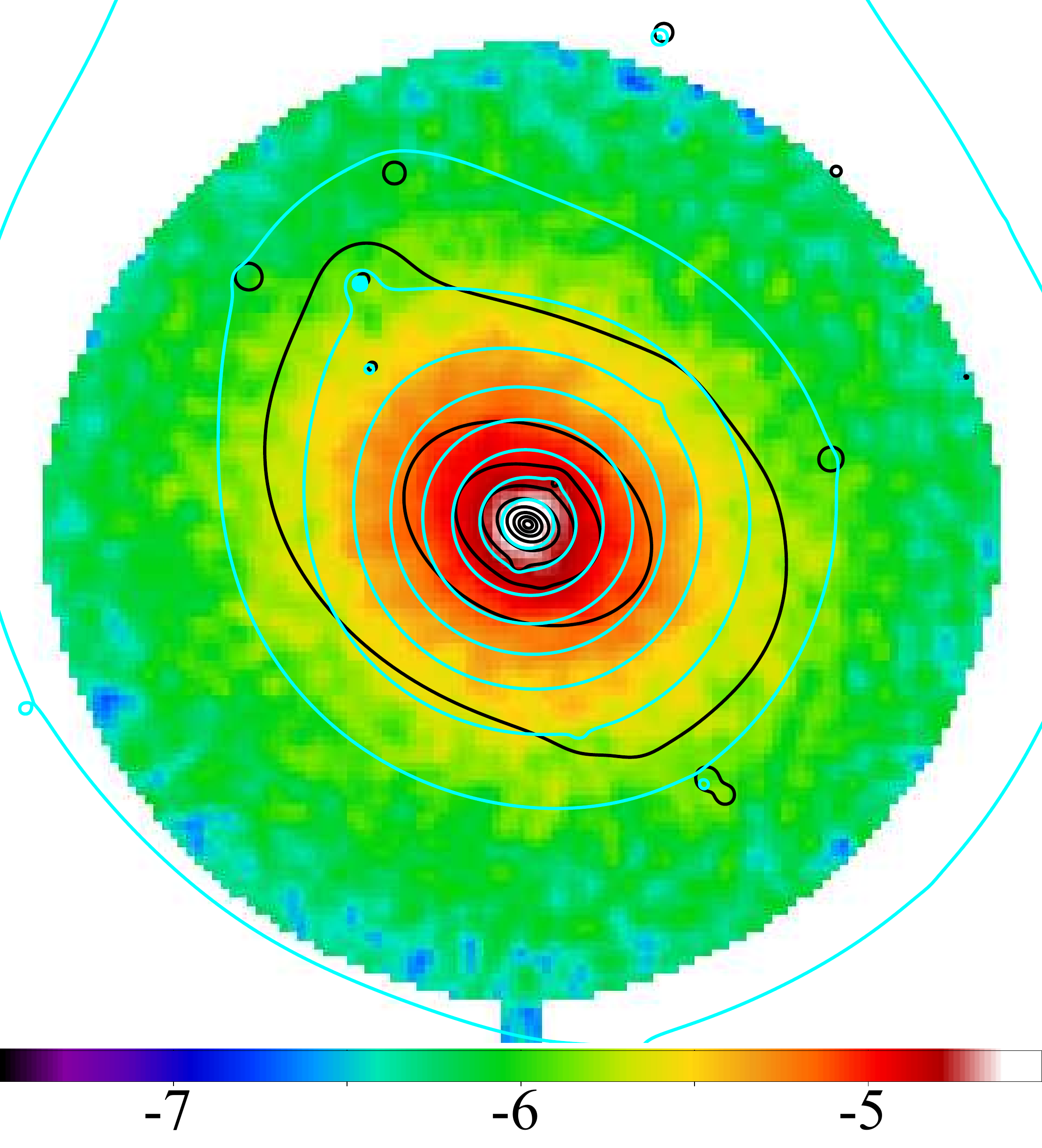}
 \end{center}
\caption{Continued. RXJ\,2129.6+0005 (left) and MS\,2137.3-2353 (right).
}
\end{figure*}

\section{Substructure in MACSJ\,0329.6-0211}
\label{sec:sub0329}

\begin{figure}
 \begin{center}
  \includegraphics[width=7.5cm]{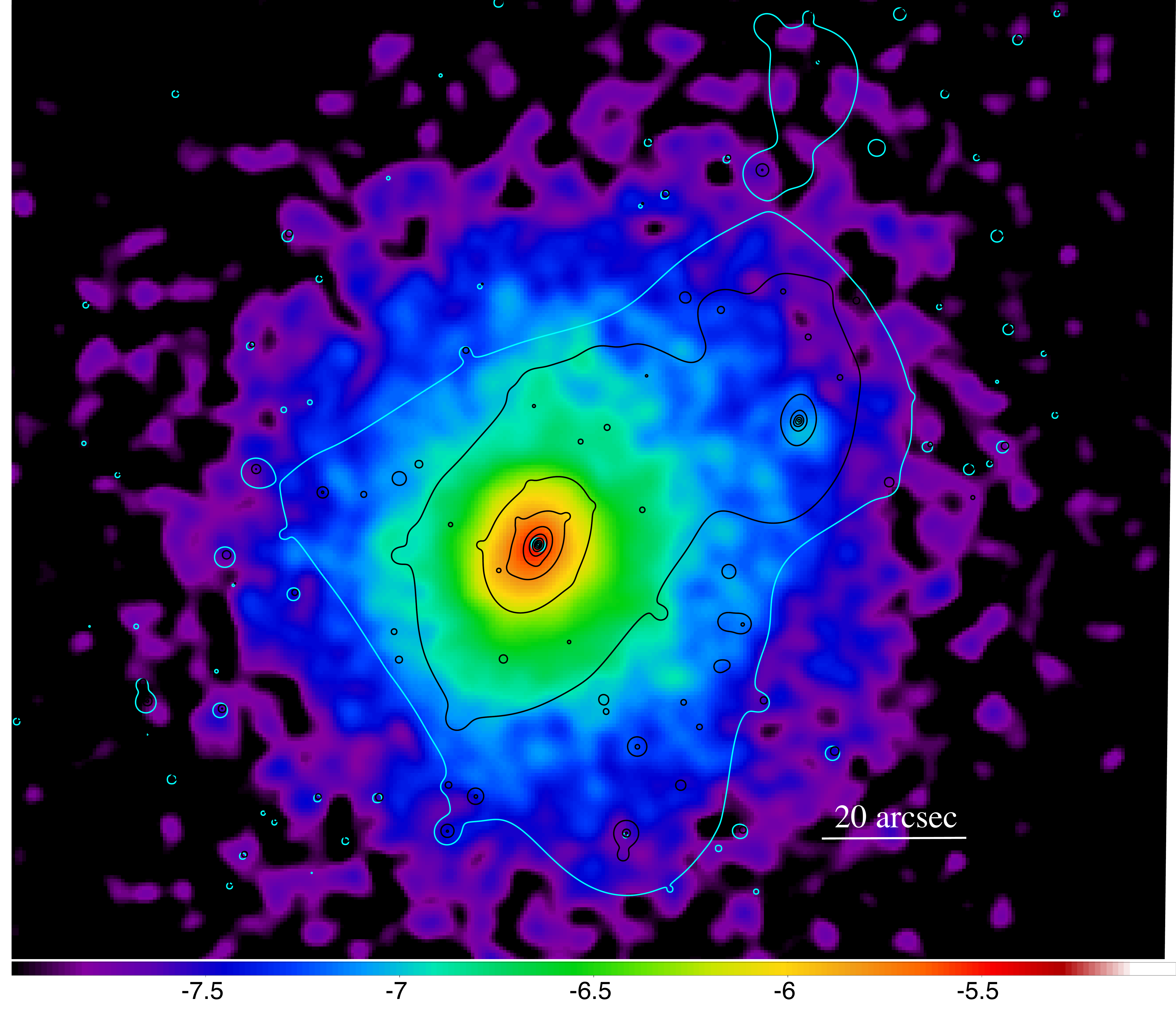}
 \end{center}
\caption{
Substructures in the central mass distribution of both the NFW and LTM models in MACSJ\,0329.6-0211 overlaid on its X-ray surface brightness. The data are the same as those shown in Figure~\ref{fig:wSL} but the plotting area is extended.
}
\label{fig:sub0329}
\end{figure}

As mentioned in Section~\ref{sec:mass}, we found the substructure in the central mass distribution in MACSJ\,0329.6-0211 as shown in Figure~\ref{fig:sub0329}. The secondary component is $\sim 40''$ (or $\sim 230$\,kpc) away from the primary peak toward the northwest direction. No apparent X-ray substructure is found around the secondary component.On the other hand, as shown in Figure~\ref{fig:image}, the X-ray residual image of this cluster shows a spiral-like feature. The positive excess region is located at the southeast region and the negative excess region is found in the northwest region. If the secondary component is associated with a past merger and the case to induce gas sloshing in its cool core, the morphology of this spiral pattern is consistent with this merger scenario. In addition, under this scenario, the trajectory of the secondary component is expected to be from the northeast to southwest direction. 

If the secondary component is in the first passage, the stripped gas originally in a subcluster could be found along with its trajectory such as RXJ\,1347.5-1145 \citep{Ueda18}. However, no apparent feature of gas stripping is found. In this sense, it is attracted that the subcluster is in the second passage or more. To find clear evidence of the X-ray substructure, deep X-ray data are required. 

\bibliographystyle{apj}
\bibliography{00_BibTeX_library}

\begin{thebibliography}{107}
\expandafter\ifx\csname natexlab\endcsname\relax\def\natexlab#1{#1}\fi

\bibitem[{{Allen} {et~al.}(2008){Allen}, {Rapetti}, {Schmidt}, {Ebeling},
  {Morris}, \& {Fabian}}]{Allen08}
{Allen}, S.~W., {Rapetti}, D.~A., {Schmidt}, R.~W., {et~al.} 2008, \mnras, 383,
  879

\bibitem[{{Allen} {et~al.}(2002){Allen}, {Schmidt}, \& {Fabian}}]{Allen02}
{Allen}, S.~W., {Schmidt}, R.~W., \& {Fabian}, A.~C. 2002, \mnras, 335, 256

\bibitem[{{Andrade-Santos} {et~al.}(2017){Andrade-Santos}, {Jones}, {Forman},
  {Lovisari}, {Vikhlinin}, {van Weeren}, {Murray}, {Arnaud}, {Pratt},
  {D{\'e}mocl{\`e}s}, {Kraft}, {Mazzotta}, {B{\"o}hringer}, {Chon},
  {Giacintucci}, {Clarke}, {Borgani}, {David}, {Douspis}, {Pointecouteau},
  {Dahle}, {Brown}, {Aghanim}, \& {Rasia}}]{Andrade-Santos17}
{Andrade-Santos}, F., {Jones}, C., {Forman}, W.~R., {et~al.} 2017, \apj, 843,
  76

\bibitem[{{Arnaud}(1996)}]{Arnaud96}
{Arnaud}, K.~A. 1996, in Astronomical Society of the Pacific Conference Series,
  Vol. 101, Astronomical Data Analysis Software and Systems V, ed. G.~H.
  {Jacoby} \& J.~{Barnes}, 17

\bibitem[{{Ascasibar} \& {Markevitch}(2006)}]{Ascasibar06}
{Ascasibar}, Y., \& {Markevitch}, M. 2006, \apj, 650, 102

\bibitem[{{Berger} \& {Colella}(1989)}]{Berger89}
{Berger}, M.~J., \& {Colella}, P. 1989, Journal of Computational Physics, 82,
  64

\bibitem[{{Berger} \& {Oliger}(1984)}]{Berger84}
{Berger}, M.~J., \& {Oliger}, J. 1984, Journal of Computational Physics, 53,
  484

\bibitem[{{Biviano} {et~al.}(2013){Biviano}, {Rosati}, {Balestra}, {Mercurio},
  {Girardi}, {Nonino}, {Grillo}, {Scodeggio}, {Lemze}, {Kelson}, {Umetsu},
  {Postman}, {Zitrin}, {Czoske}, {Ettori}, {Fritz}, {Lombardi}, {Maier},
  {Medezinski}, {Mei}, {Presotto}, {Strazzullo}, {Tozzi}, {Ziegler},
  {Annunziatella}, {Bartelmann}, {Benitez}, {Bradley}, {Brescia}, {Broadhurst},
  {Coe}, {Demarco}, {Donahue}, {Ford}, {Gobat}, {Graves}, {Koekemoer},
  {Kuchner}, {Melchior}, {Meneghetti}, {Merten}, {Moustakas}, {Munari},
  {Reg{\H{o}}s}, {Sartoris}, {Seitz}, \& {Zheng}}]{Biviano13}
{Biviano}, A., {Rosati}, P., {Balestra}, I., {et~al.} 2013, \aap, 558, A1

\bibitem[{{Blanton} {et~al.}(2011){Blanton}, {Randall}, {Clarke}, {Sarazin},
  {McNamara}, {Douglass}, \& {McDonald}}]{Blanton11}
{Blanton}, E.~L., {Randall}, S.~W., {Clarke}, T.~E., {et~al.} 2011, \apj, 737,
  99

\bibitem[{{Calzadilla} {et~al.}(2019){Calzadilla}, {Russell}, {McDonald},
  {Fabian}, {Baum}, {Combes}, {Donahue}, {Edge}, {McNamara}, \&
  {Nulsen}}]{Calzadilla19}
{Calzadilla}, M.~S., {Russell}, H.~R., {McDonald}, M.~A., {et~al.} 2019, \apj,
  875, 65

\bibitem[{{Canning} {et~al.}(2013){Canning}, {Sun}, {Sanders}, {Clarke},
  {Fabian}, {Giacintucci}, {Lal}, {Werner}, {Allen}, {Donahue}, {Edge},
  {Johnstone}, {Nulsen}, {Salom{\'e}}, \& {Sarazin}}]{Canning13}
{Canning}, R.~E.~A., {Sun}, M., {Sanders}, J.~S., {et~al.} 2013, \mnras, 435,
  1108

\bibitem[{{Chiu} {et~al.}(2018){Chiu}, {Umetsu}, {Sereno}, {Ettori},
  {Meneghetti}, {Merten}, {Sayers}, \& {Zitrin}}]{Chiu18}
{Chiu}, I.~N., {Umetsu}, K., {Sereno}, M., {et~al.} 2018, \apj, 860, 126

\bibitem[{{Cho} {et~al.}(2003){Cho}, {Lazarian}, {Honein}, {Knaepen},
  {Kassinos}, \& {Moin}}]{Cho03}
{Cho}, J., {Lazarian}, A., {Honein}, A., {et~al.} 2003, \apjl, 589, L77

\bibitem[{{Churazov} {et~al.}(2016){Churazov}, {Arevalo}, {Forman}, {Jones},
  {Schekochihin}, {Vikhlinin}, \& {Zhuravleva}}]{Churazov16}
{Churazov}, E., {Arevalo}, P., {Forman}, W., {et~al.} 2016, \mnras, 463, 1057

\bibitem[{{Churazov} {et~al.}(2003){Churazov}, {Forman}, {Jones}, \&
  {B{\"o}hringer}}]{Churazov03}
{Churazov}, E., {Forman}, W., {Jones}, C., \& {B{\"o}hringer}, H. 2003, \apj,
  590, 225

\bibitem[{{Clarke} {et~al.}(2004){Clarke}, {Blanton}, \& {Sarazin}}]{Clarke04}
{Clarke}, T.~E., {Blanton}, E.~L., \& {Sarazin}, C.~L. 2004, \apj, 616, 178

\bibitem[{{Di Mascolo} {et~al.}(2019){Di Mascolo}, {Churazov}, \&
  {Mroczkowski}}]{Di_Mascolo19}
{Di Mascolo}, L., {Churazov}, E., \& {Mroczkowski}, T. 2019, \mnras, 487, 4037

\bibitem[{{Donahue} {et~al.}(2014){Donahue}, {Voit}, {Mahdavi}, {Umetsu},
  {Ettori}, {Merten}, {Postman}, {Hoffer}, {Baldi}, {Coe}, {Czakon},
  {Bartelmann}, {Benitez}, {Bouwens}, {Bradley}, {Broadhurst}, {Ford},
  {Gastaldello}, {Grillo}, {Infante}, {Jouvel}, {Koekemoer}, {Kelson}, {Lahav},
  {Lemze}, {Medezinski}, {Melchior}, {Meneghetti}, {Molino}, {Moustakas},
  {Moustakas}, {Nonino}, {Rosati}, {Sayers}, {Seitz}, {Van der Wel}, {Zheng},
  \& {Zitrin}}]{Donahue14}
{Donahue}, M., {Voit}, G.~M., {Mahdavi}, A., {et~al.} 2014, \apj, 794, 136

\bibitem[{{Donahue} {et~al.}(2016){Donahue}, {Ettori}, {Rasia}, {Sayers},
  {Zitrin}, {Meneghetti}, {Voit}, {Golwala}, {Czakon}, {Yepes}, {Baldi},
  {Koekemoer}, \& {Postman}}]{Donahue16}
{Donahue}, M., {Ettori}, S., {Rasia}, E., {et~al.} 2016, \apj, 819, 36

\bibitem[{{Eddington}(1916)}]{Eddington16}
{Eddington}, A.~S. 1916, \mnras, 76, 572

\bibitem[{{Ehlert} {et~al.}(2011){Ehlert}, {Allen}, {von der Linden},
  {Simionescu}, {Werner}, {Taylor}, {Gentile}, {Ebeling}, {Allen}, {Applegate},
  {Dunn}, {Fabian}, {Kelly}, {Million}, {Morris}, {Sanders}, \&
  {Schmidt}}]{Ehlert11}
{Ehlert}, S., {Allen}, S.~W., {von der Linden}, A., {et~al.} 2011, \mnras, 411,
  1641

\bibitem[{{Fabian}(2012)}]{Fabian12}
{Fabian}, A.~C. 2012, \araa, 50, 455

\bibitem[{{Feretti} {et~al.}(2012){Feretti}, {Giovannini}, {Govoni}, \&
  {Murgia}}]{Feretti12}
{Feretti}, L., {Giovannini}, G., {Govoni}, F., \& {Murgia}, M. 2012, \aapr, 20,
  54

\bibitem[{{Fruscione} {et~al.}(2006){Fruscione}, {McDowell}, {Allen},
  {Brickhouse}, {Burke}, {Davis}, {Durham}, {Elvis}, {Galle}, {Harris},
  {Huenemoerder}, {Houck}, {Ishibashi}, {Karovska}, {Nicastro}, {Noble},
  {Nowak}, {Primini}, {Siemiginowska}, {Smith}, \& {Wise}}]{Fruscione06}
{Fruscione}, A., {McDowell}, J.~C., {Allen}, G.~E., {et~al.} 2006, in
  \procspie, Vol. 6270, Society of Photo-Optical Instrumentation Engineers
  (SPIE) Conference Series, 62701V

\bibitem[{{Fujita} {et~al.}(2019){Fujita}, {Cen}, \& {Zhuravleva}}]{Fujita19}
{Fujita}, Y., {Cen}, R., \& {Zhuravleva}, I. 2019, arXiv e-prints,
  arXiv:1912.01012

\bibitem[{{Fujita} {et~al.}(2004){Fujita}, {Matsumoto}, \& {Wada}}]{Fujita04b}
{Fujita}, Y., {Matsumoto}, T., \& {Wada}, K. 2004, \apjl, 612, L9

\bibitem[{{Fujita} {et~al.}(2018){Fujita}, {Umetsu}, {Rasia}, {Meneghetti},
  {Donahue}, {Medezinski}, {Okabe}, \& {Postman}}]{Fujita18}
{Fujita}, Y., {Umetsu}, K., {Rasia}, E., {et~al.} 2018, \apj, 857, 118

\bibitem[{{Garmire} {et~al.}(2003){Garmire}, {Bautz}, {Ford}, {Nousek}, \&
  {Ricker}}]{Garmire03}
{Garmire}, G.~P., {Bautz}, M.~W., {Ford}, P.~G., {Nousek}, J.~A., \& {Ricker},
  Jr., G.~R. 2003, in Society of Photo-Optical Instrumentation Engineers (SPIE)
  Conference Series, Vol. 4851, Society of Photo-Optical Instrumentation
  Engineers (SPIE) Conference Series, ed. J.~E. {Truemper} \& H.~D.
  {Tananbaum}, 28--44

\bibitem[{{Ghizzardi} {et~al.}(2014){Ghizzardi}, {De Grandi}, \&
  {Molendi}}]{Ghizzardi14}
{Ghizzardi}, S., {De Grandi}, S., \& {Molendi}, S. 2014, \aap, 570, A117

\bibitem[{{Giacintucci} {et~al.}(2017){Giacintucci}, {Markevitch}, {Cassano},
  {Venturi}, {Clarke}, \& {Brunetti}}]{Giacintucci17}
{Giacintucci}, S., {Markevitch}, M., {Cassano}, R., {et~al.} 2017, \apj, 841,
  71

\bibitem[{{Giacintucci} {et~al.}(2014){Giacintucci}, {Markevitch}, {Venturi},
  {Clarke}, {Cassano}, \& {Mazzotta}}]{Giacintucci14}
{Giacintucci}, S., {Markevitch}, M., {Venturi}, T., {et~al.} 2014, \apj, 781, 9

\bibitem[{{Giles} {et~al.}(2017){Giles}, {Maughan}, {Dahle}, {Bonamente},
  {Landry}, {Jones}, {Joy}, {Murray}, \& {van der Pyl}}]{Giles17}
{Giles}, P.~A., {Maughan}, B.~J., {Dahle}, H., {et~al.} 2017, \mnras, 465, 858

\bibitem[{{Gitti} {et~al.}(2007){Gitti}, {Ferrari}, {Domainko}, {Feretti}, \&
  {Schindler}}]{Gitti07}
{Gitti}, M., {Ferrari}, C., {Domainko}, W., {Feretti}, L., \& {Schindler}, S.
  2007, \aap, 470, L25

\bibitem[{{Hitomi Collaboration} {et~al.}(2016){Hitomi Collaboration},
  {Aharonian}, {Akamatsu}, {Akimoto}, {Allen}, {Anabuki}, {Angelini}, {Arnaud},
  {Audard}, {Awaki}, {Axelsson}, {Bamba}, {Bautz}, {Blandford}, {Brenneman},
  {Brown}, {Bulbul}, {Cackett}, {Chernyakova}, {Chiao}, {Coppi}, {Costantini},
  {de Plaa}, {den Herder}, {Done}, {Dotani}, {Ebisawa}, {Eckart}, {Enoto},
  {Ezoe}, {Fabian}, {Ferrigno}, {Foster}, {Fujimoto}, {Fukazawa}, {Furuzawa},
  {Galeazzi}, {Gallo}, {Gandhi}, {Giustini}, {Goldwurm}, {Gu}, {Guainazzi},
  {Haba}, {Hagino}, {Hamaguchi}, {Harrus}, {Hatsukade}, {Hayashi}, {Hayashi},
  {Hayashida}, {Hiraga}, {Hornschemeier}, {Hoshino}, {Hughes}, {Iizuka},
  {Inoue}, {Inoue}, {Ishibashi}, {Ishida}, {Ishikawa}, {Ishisaki}, {Itoh},
  {Iyomoto}, {Kaastra}, {Kallman}, {Kamae}, {Kara}, {Kataoka}, {Katsuda},
  {Katsuta}, {Kawaharada}, {Kawai}, {Kelley}, {Khangulyan}, {Kilbourne},
  {King}, {Kitaguchi}, {Kitamoto}, {Kitayama}, {Kohmura}, {Kokubun}, {Koyama},
  {Koyama}, {Kretschmar}, {Krimm}, {Kubota}, {Kunieda}, {Laurent}, {Lebrun},
  {Lee}, {Leutenegger}, {Limousin}, {Loewenstein}, {Long}, {Lumb}, {Madejski},
  {Maeda}, {Maier}, {Makishima}, {Markevitch}, {Matsumoto}, {Matsushita},
  {McCammon}, {McNamara}, {Mehdipour}, {Miller}, {Miller}, {Mineshige},
  {Mitsuda}, {Mitsuishi}, {Miyazawa}, {Mizuno}, {Mori}, {Mori}, {Moseley},
  {Mukai}, {Murakami}, {Murakami}, {Mushotzky}, {Nagino}, {Nakagawa},
  {Nakajima}, {Nakamori}, {Nakano}, {Nakashima}, {Nakazawa}, {Nobukawa},
  {Noda}, {Nomachi}, {O'Dell}, {Odaka}, {Ohashi}, {Ohno}, {Okajima}, {Ota},
  {Ozaki}, {Paerels}, {Paltani}, {Parmar}, {Petre}, {Pinto}, {Pohl}, {Porter},
  {Pottschmidt}, {Ramsey}, {Reynolds}, {Russell}, {Safi-Harb}, {Saito},
  {Sakai}, {Sameshima}, {Sato}, {Sato}, {Sato}, {Sawada}, {Schartel},
  {Serlemitsos}, {Seta}, {Shidatsu}, {Simionescu}, {Smith}, {Soong}, {Stawarz},
  {Sugawara}, {Sugita}, {Szymkowiak}, {Tajima}, {Takahashi}, {Takahashi},
  {Takeda}, {Takei}, {Tamagawa}, {Tamura}, {Tamura}, {Tanaka}, {Tanaka},
  {Tanaka}, {Tashiro}, {Tawara}, {Terada}, {Terashima}, {Tombesi}, {Tomida},
  {Tsuboi}, {Tsujimoto}, {Tsunemi}, {Tsuru}, {Uchida}, {Uchiyama}, {Uchiyama},
  {Ueda}, {Ueda}, {Ueno}, {Uno}, {Urry}, {Ursino}, {de Vries}, {Watanabe},
  {Werner}, {Wik}, {Wilkins}, {Williams}, {Yamada}, {Yamaguchi}, {Yamaoka},
  {Yamasaki}, {Yamauchi}, {Yamauchi}, {Yaqoob}, {Yatsu}, {Yonetoku}, {Yoshida},
  {Yuasa}, {Zhuravleva}, \& {Zoghbi}}]{Hitomi16}
{Hitomi Collaboration}, {Aharonian}, F., {Akamatsu}, H., {et~al.} 2016, \nat,
  535, 117

\bibitem[{{Hitomi Collaboration} {et~al.}(2018){Hitomi Collaboration},
  {Aharonian}, {Akamatsu}, {Akimoto}, {Allen}, {Angelini}, {Audard}, {Awaki},
  {Axelsson}, {Bamba}, {Bautz}, {Blandford}, {Brenneman}, {Brown}, {Bulbul},
  {Cackett}, {Canning}, {Chernyakova}, {Chiao}, {Coppi}, {Costantini}, {de
  Plaa}, {de Vries}, {den Herder}, {Done}, {Dotani}, {Ebisawa}, {Eckart},
  {Enoto}, {Ezoe}, {Fabian}, {Ferrigno}, {Foster}, {Fujimoto}, {Fukazawa},
  {Furuzawa}, {Galeazzi}, {Gallo}, {Gandhi}, {Giustini}, {Goldwurm}, {Gu},
  {Guainazzi}, {Haba}, {Hagino}, {Hamaguchi}, {Harrus}, {Hatsukade}, {Hayashi},
  {Hayashi}, {Hayashi}, {Hayashida}, {Hiraga}, {Hornschemeier}, {Hoshino},
  {Hughes}, {Ichinohe}, {Iizuka}, {Inoue}, {Inoue}, {Inoue}, {Ishida},
  {Ishikawa}, {Ishisaki}, {Iwai}, {Kaastra}, {Kallman}, {Kamae}, {Kataoka},
  {Katsuda}, {Kawai}, {Kelley}, {Kilbourne}, {Kitaguchi}, {Kitamoto},
  {Kitayama}, {Kohmura}, {Kokubun}, {Koyama}, {Koyama}, {Kretschmar}, {Krimm},
  {Kubota}, {Kunieda}, {Laurent}, {Lee}, {Leutenegger}, {Limousin},
  {Loewenstein}, {Long}, {Lumb}, {Madejski}, {Maeda}, {Maier}, {Makishima},
  {Markevitch}, {Matsumoto}, {Matsushita}, {McCammon}, {McNamara}, {Mehdipour},
  {Miller}, {Miller}, {Mineshige}, {Mitsuda}, {Mitsuishi}, {Miyazawa},
  {Mizuno}, {Mori}, {Mori}, {Mukai}, {Murakami}, {Mushotzky}, {Nakagawa},
  {Nakajima}, {Nakamori}, {Nakashima}, {Nakazawa}, {Nobukawa}, {Nobukawa},
  {Noda}, {Odaka}, {Ohashi}, {Ohno}, {Okajima}, {Ota}, {Ozaki}, {Paerels},
  {Paltani}, {Petre}, {Pinto}, {Porter}, {Pottschmidt}, {Reynolds},
  {Safi-Harb}, {Saito}, {Sakai}, {Sasaki}, {Sato}, {Sato}, {Sato}, {Sawada},
  {Schartel}, {Serlemtsos}, {Seta}, {Shidatsu}, {Simionescu}, {Smith}, {Soong},
  {Stawarz}, {Sugawara}, {Sugita}, {Szymkowiak}, {Tajima}, {Takahashi},
  {Takahashi}, {Takeda}, {Takei}, {Tamagawa}, {Tamura}, {Tanaka}, {Tanaka},
  {Tanaka}, {Tanaka}, {Tashiro}, {Tawara}, {Terada}, {Terashima}, {Tombesi},
  {Tomida}, {Tsuboi}, {Tsujimoto}, {Tsunemi}, {Tsuru}, {Uchida}, {Uchiyama},
  {Uchiyama}, {Ueda}, {Ueda}, {Uno}, {Urry}, {Ursino}, {Wang}, {Watanabe},
  {Werner}, {Wilkins}, {Williams}, {Yamada}, {Yamaguchi}, {Yamaoka},
  {Yamasaki}, {Yamauchi}, {Yamauchi}, {Yaqoob}, {Yatsu}, {Yonetoku},
  {Zhuravleva}, \& {Zoghbi}}]{Hitomi18d}
---. 2018, \pasj, 70, 9

\bibitem[{{Hlavacek-Larrondo} {et~al.}(2013{\natexlab{a}}){Hlavacek-Larrondo},
  {Fabian}, {Edge}, {Ebeling}, {Allen}, {Sanders}, \&
  {Taylor}}]{Hlavacek-Larrondo13b}
{Hlavacek-Larrondo}, J., {Fabian}, A.~C., {Edge}, A.~C., {et~al.}
  2013{\natexlab{a}}, \mnras, 431, 1638

\bibitem[{{Hlavacek-Larrondo} {et~al.}(2012){Hlavacek-Larrondo}, {Fabian},
  {Edge}, {Ebeling}, {Sanders}, {Hogan}, \& {Taylor}}]{Hlavacek-Larrondo12}
---. 2012, \mnras, 421, 1360

\bibitem[{{Hlavacek-Larrondo} {et~al.}(2013{\natexlab{b}}){Hlavacek-Larrondo},
  {Allen}, {Taylor}, {Fabian}, {Canning}, {Werner}, {Sand ers}, {Grimes},
  {Ehlert}, \& {von der Linden}}]{Hlavacek-Larrondo13}
{Hlavacek-Larrondo}, J., {Allen}, S.~W., {Taylor}, G.~B., {et~al.}
  2013{\natexlab{b}}, \apj, 777, 163

\bibitem[{{Hlavacek-Larrondo} {et~al.}(2015){Hlavacek-Larrondo}, {McDonald},
  {Benson}, {Forman}, {Allen}, {Bleem}, {Ashby}, {Bocquet}, {Brodwin},
  {Dietrich}, {Jones}, {Liu}, {Reichardt}, {Saliwanchik}, {Saro}, {Schrabback},
  {Song}, {Stalder}, {Vikhlinin}, \& {Zenteno}}]{Hlavacek-Larrondo15}
{Hlavacek-Larrondo}, J., {McDonald}, M., {Benson}, B.~A., {et~al.} 2015, \apj,
  805, 35

\bibitem[{{Ichinohe} {et~al.}(2019){Ichinohe}, {Simionescu}, {Werner},
  {Fabian}, \& {Takahashi}}]{Ichinohe19}
{Ichinohe}, Y., {Simionescu}, A., {Werner}, N., {Fabian}, A.~C., \&
  {Takahashi}, T. 2019, \mnras, 483, 1744

\bibitem[{{Ichinohe} {et~al.}(2015){Ichinohe}, {Werner}, {Simionescu}, {Allen},
  {Canning}, {Ehlert}, {Mernier}, \& {Takahashi}}]{Ichinohe15}
{Ichinohe}, Y., {Werner}, N., {Simionescu}, A., {et~al.} 2015, \mnras, 448,
  2971

\bibitem[{{Johnson} {et~al.}(2012){Johnson}, {Zuhone}, {Jones}, {Forman}, \&
  {Markevitch}}]{Johnson12}
{Johnson}, R.~E., {Zuhone}, J., {Jones}, C., {Forman}, W.~R., \& {Markevitch},
  M. 2012, \apj, 751, 95

\bibitem[{{Kalberla} {et~al.}(2005){Kalberla}, {Burton}, {Hartmann}, {Arnal},
  {Bajaja}, {Morras}, \& {P{\"o}ppel}}]{Kalberla05}
{Kalberla}, P.~M.~W., {Burton}, W.~B., {Hartmann}, D., {et~al.} 2005, \aap,
  440, 775

\bibitem[{{Kale} {et~al.}(2015){Kale}, {Venturi}, {Cassano}, {Giacintucci},
  {Bardelli}, {Dallacasa}, \& {Zucca}}]{Kale15}
{Kale}, R., {Venturi}, T., {Cassano}, R., {et~al.} 2015, \aap, 581, A23

\bibitem[{{Kale} {et~al.}(2013){Kale}, {Venturi}, {Giacintucci}, {Dallacasa},
  {Cassano}, {Brunetti}, {Macario}, \& {Athreya}}]{Kale13}
{Kale}, R., {Venturi}, T., {Giacintucci}, S., {et~al.} 2013, \aap, 557, A99

\bibitem[{{Khochfar} \& {Ostriker}(2008)}]{Khochfar08}
{Khochfar}, S., \& {Ostriker}, J.~P. 2008, \apj, 680, 54

\bibitem[{{Kim} \& {Narayan}(2003)}]{Kim03}
{Kim}, W.-T., \& {Narayan}, R. 2003, \apjl, 596, L139

\bibitem[{{Kitayama} {et~al.}(2004){Kitayama}, {Komatsu}, {Ota}, {Kuwabara},
  {Suto}, {Yoshikawa}, {Hattori}, \& {Matsuo}}]{Kitayama04}
{Kitayama}, T., {Komatsu}, E., {Ota}, N., {et~al.} 2004, \pasj, 56, 17

\bibitem[{{Kitayama} {et~al.}(2016){Kitayama}, {Ueda}, {Takakuwa}, {Tsutsumi},
  {Komatsu}, {Akahori}, {Iono}, {Izumi}, {Kawabe}, {Kohno}, {Matsuo}, {Ota},
  {Suto}, {Takizawa}, \& {Yoshikawa}}]{Kitayama16}
{Kitayama}, T., {Ueda}, S., {Takakuwa}, S., {et~al.} 2016, \pasj, 68, 88

\bibitem[{{K{\"o}hlinger} \& {Schmidt}(2014)}]{Kohlinger14}
{K{\"o}hlinger}, F., \& {Schmidt}, R.~W. 2014, \mnras, 437, 1858

\bibitem[{{Komatsu} {et~al.}(2001){Komatsu}, {Matsuo}, {Kitayama}, {Hattori},
  {Kawabe}, {Kohno}, {Kuno}, {Schindler}, {Suto}, \& {Yoshikawa}}]{Komatsu01}
{Komatsu}, E., {Matsuo}, H., {Kitayama}, T., {et~al.} 2001, \pasj, 53, 57

\bibitem[{{Kreisch} {et~al.}(2016){Kreisch}, {Machacek}, {Jones}, \&
  {Randall}}]{Kreisch16}
{Kreisch}, C.~D., {Machacek}, M.~E., {Jones}, C., \& {Randall}, S.~W. 2016,
  \apj, 830, 39

\bibitem[{{Liu} {et~al.}(2018){Liu}, {Yu}, {Diaferio}, {Tozzi}, {Hwang},
  {Umetsu}, {Okabe}, \& {Yang}}]{Liu18}
{Liu}, A., {Yu}, H., {Diaferio}, A., {et~al.} 2018, \apj, 863, 102

\bibitem[{{Lodders} \& {Palme}(2009)}]{Lodders09}
{Lodders}, K., \& {Palme}, H. 2009, Meteoritics and Planetary Science
  Supplement, 72, 5154

\bibitem[{{Lyskova} {et~al.}(2019){Lyskova}, {Churazov}, {Zhang}, {Forman},
  {Jones}, {Dolag}, {Roediger}, \& {Sheardown}}]{Lyskova19}
{Lyskova}, N., {Churazov}, E., {Zhang}, C., {et~al.} 2019, \mnras, 485, 2922

\bibitem[{{Markevitch} \& {Vikhlinin}(2007)}]{Markevitch07}
{Markevitch}, M., \& {Vikhlinin}, A. 2007, \physrep, 443, 1

\bibitem[{{Mazzotta} \& {Giacintucci}(2008)}]{Mazzotta08}
{Mazzotta}, P., \& {Giacintucci}, S. 2008, \apjl, 675, L9

\bibitem[{{McDonald} {et~al.}(2018){McDonald}, {Gaspari}, {McNamara}, \&
  {Tremblay}}]{McDonald18}
{McDonald}, M., {Gaspari}, M., {McNamara}, B.~R., \& {Tremblay}, G.~R. 2018,
  \apj, 858, 45

\bibitem[{{McDonald} {et~al.}(2011){McDonald}, {Veilleux}, {Rupke},
  {Mushotzky}, \& {Reynolds}}]{McDonald11}
{McDonald}, M., {Veilleux}, S., {Rupke}, D.~S.~N., {Mushotzky}, R., \&
  {Reynolds}, C. 2011, \apj, 734, 95

\bibitem[{{McNamara} \& {Nulsen}(2007)}]{McNamara07}
{McNamara}, B.~R., \& {Nulsen}, P.~E.~J. 2007, \araa, 45, 117

\bibitem[{{McNamara} \& {Nulsen}(2012)}]{McNamara12}
---. 2012, New Journal of Physics, 14, 055023

\bibitem[{{McNamara} {et~al.}(2005){McNamara}, {Nulsen}, {Wise}, {Rafferty},
  {Carilli}, {Sarazin}, \& {Blanton}}]{McNamara05}
{McNamara}, B.~R., {Nulsen}, P.~E.~J., {Wise}, M.~W., {et~al.} 2005, \nat, 433,
  45

\bibitem[{{Meneghetti} {et~al.}(2014){Meneghetti}, {Rasia}, {Vega}, {Merten},
  {Postman}, {Yepes}, {Sembolini}, {Donahue}, {Ettori}, {Umetsu}, {Balestra},
  {Bartelmann}, {Ben{\'\i}tez}, {Biviano}, {Bouwens}, {Bradley}, {Broadhurst},
  {Coe}, {Czakon}, {De Petris}, {Ford}, {Giocoli}, {Gottl{\"o}ber}, {Grillo},
  {Infante}, {Jouvel}, {Kelson}, {Koekemoer}, {Lahav}, {Lemze}, {Medezinski},
  {Melchior}, {Mercurio}, {Molino}, {Moscardini}, {Monna}, {Moustakas},
  {Moustakas}, {Nonino}, {Rhodes}, {Rosati}, {Sayers}, {Seitz}, {Zheng}, \&
  {Zitrin}}]{Meneghetti14}
{Meneghetti}, M., {Rasia}, E., {Vega}, J., {et~al.} 2014, \apj, 797, 34

\bibitem[{{Merten} {et~al.}(2015){Merten}, {Meneghetti}, {Postman}, {Umetsu},
  {Zitrin}, {Medezinski}, {Nonino}, {Koekemoer}, {Melchior}, {Gruen},
  {Moustakas}, {Bartelmann}, {Host}, {Donahue}, {Coe}, {Molino}, {Jouvel},
  {Monna}, {Seitz}, {Czakon}, {Lemze}, {Sayers}, {Balestra}, {Rosati},
  {Ben{\'{\i}}tez}, {Biviano}, {Bouwens}, {Bradley}, {Broadhurst}, {Carrasco},
  {Ford}, {Grillo}, {Infante}, {Kelson}, {Lahav}, {Massey}, {Moustakas},
  {Rasia}, {Rhodes}, {Vega}, \& {Zheng}}]{Merten15}
{Merten}, J., {Meneghetti}, M., {Postman}, M., {et~al.} 2015, \apj, 806, 4

\bibitem[{{Molino} {et~al.}(2017){Molino}, {Ben{\'\i}tez}, {Ascaso}, {Coe},
  {Postman}, {Jouvel}, {Host}, {Lahav}, {Seitz}, {Medezinski}, {Rosati},
  {Schoenell}, {Koekemoer}, {Jimenez-Teja}, {Broadhurst}, {Melchior},
  {Balestra}, {Bartelmann}, {Bouwens}, {Bradley}, {Czakon}, {Donahue}, {Ford},
  {Graur}, {Graves}, {Grillo}, {Infante}, {Jha}, {Kelson}, {Lazkoz}, {Lemze},
  {Maoz}, {Mercurio}, {Meneghetti}, {Merten}, {Moustakas}, {Nonino}, {Orgaz},
  {Riess}, {Rodney}, {Sayers}, {Umetsu}, {Zheng}, \& {Zitrin}}]{Molino17}
{Molino}, A., {Ben{\'\i}tez}, N., {Ascaso}, B., {et~al.} 2017, \mnras, 470, 95

\bibitem[{{Nandra} {et~al.}(2013){Nandra}, {Barret}, {Barcons}, {Fabian}, {den
  Herder}, {Piro}, {Watson}, {Adami}, {Aird}, \& {Afonso}}]{Nandra13}
{Nandra}, K., {Barret}, D., {Barcons}, X., {et~al.} 2013, arXiv e-prints,
  arXiv:1306.2307

\bibitem[{{Navarro} {et~al.}(1997){Navarro}, {Frenk}, \& {White}}]{Navarro97}
{Navarro}, J.~F., {Frenk}, C.~S., \& {White}, S.~D.~M. 1997, \apj, 490, 493

\bibitem[{{Newman} {et~al.}(2011){Newman}, {Treu}, {Ellis}, \& {Sand
  }}]{Newman11}
{Newman}, A.~B., {Treu}, T., {Ellis}, R.~S., \& {Sand }, D.~J. 2011, \apj, 728,
  L39

\bibitem[{{O'Dea} {et~al.}(2008){O'Dea}, {Baum}, {Privon}, {Noel-Storr},
  {Quillen}, {Zufelt}, {Park}, {Edge}, {Russell}, {Fabian}, {Donahue},
  {Sarazin}, {McNamara}, {Bregman}, \& {Egami}}]{ODea08}
{O'Dea}, C.~P., {Baum}, S.~A., {Privon}, G., {et~al.} 2008, \apj, 681, 1035

\bibitem[{{O'Sullivan} {et~al.}(2012){O'Sullivan}, {Giacintucci}, {Babul},
  {Raychaudhury}, {Venturi}, {Bildfell}, {Mahdavi}, {Oonk}, {Murray},
  {Hoekstra}, \& {Donahue}}]{OSullivan12}
{O'Sullivan}, E., {Giacintucci}, S., {Babul}, A., {et~al.} 2012, \mnras, 424,
  2971

\bibitem[{{Owers} {et~al.}(2011){Owers}, {Nulsen}, \& {Couch}}]{Owers11}
{Owers}, M.~S., {Nulsen}, P.~E.~J., \& {Couch}, W.~J. 2011, \apj, 741, 122

\bibitem[{{Pandey-Pommier} {et~al.}(2016){Pandey-Pommier}, {Richard}, {Combes},
  {Edge}, {Guiderdoni}, {Narasimha}, {Bagchi}, \& {Jacob}}]{Pandey-Pommier16}
{Pandey-Pommier}, M., {Richard}, J., {Combes}, F., {et~al.} 2016, in SF2A-2016:
  Proceedings of the Annual meeting of the French Society of Astronomy and
  Astrophysics, 367--372

\bibitem[{{Peterson} \& {Fabian}(2006)}]{Peterson06}
{Peterson}, J.~R., \& {Fabian}, A.~C. 2006, \physrep, 427, 1

\bibitem[{{Peterson} {et~al.}(2001){Peterson}, {Paerels}, {Kaastra}, {Arnaud},
  {Reiprich}, {Fabian}, {Mushotzky}, {Jernigan}, \& {Sakelliou}}]{Peterson01}
{Peterson}, J.~R., {Paerels}, F.~B.~S., {Kaastra}, J.~S., {et~al.} 2001, \aap,
  365, L104

\bibitem[{{Postman} {et~al.}(2012){Postman}, {Coe}, {Ben{\'{\i}}tez},
  {Bradley}, {Broadhurst}, {Donahue}, {Ford}, {Graur}, {Graves}, {Jouvel},
  {Koekemoer}, {Lemze}, {Medezinski}, {Molino}, {Moustakas}, {Ogaz}, {Riess},
  {Rodney}, {Rosati}, {Umetsu}, {Zheng}, {Zitrin}, {Bartelmann}, {Bouwens},
  {Czakon}, {Golwala}, {Host}, {Infante}, {Jha}, {Jimenez-Teja}, {Kelson},
  {Lahav}, {Lazkoz}, {Maoz}, {McCully}, {Melchior}, {Meneghetti}, {Merten},
  {Moustakas}, {Nonino}, {Patel}, {Reg{\"o}s}, {Sayers}, {Seitz}, \& {Van der
  Wel}}]{Postman12}
{Postman}, M., {Coe}, D., {Ben{\'{\i}}tez}, N., {et~al.} 2012, \apjs, 199, 25

\bibitem[{{Repp} \& {Ebeling}(2018)}]{Repp18}
{Repp}, A., \& {Ebeling}, H. 2018, \mnras, 479, 844

\bibitem[{{Ricker} \& {Sarazin}(2001)}]{Ricker01}
{Ricker}, P.~M., \& {Sarazin}, C.~L. 2001, \apj, 561, 621

\bibitem[{{Roediger} {et~al.}(2011){Roediger}, {Br{\"u}ggen}, {Simionescu},
  {B{\"o}hringer}, {Churazov}, \& {Forman}}]{Roediger11}
{Roediger}, E., {Br{\"u}ggen}, M., {Simionescu}, A., {et~al.} 2011, \mnras,
  413, 2057

\bibitem[{{Roediger} {et~al.}(2012){Roediger}, {Lovisari}, {Dupke},
  {Ghizzardi}, {Br{\"u}ggen}, {Kraft}, \& {Machacek}}]{Roediger12}
{Roediger}, E., {Lovisari}, L., {Dupke}, R., {et~al.} 2012, \mnras, 420, 3632

\bibitem[{{Rossetti} {et~al.}(2013){Rossetti}, {Eckert}, {De Grandi},
  {Gastaldello}, {Ghizzardi}, {Roediger}, \& {Molendi}}]{Rossetti13}
{Rossetti}, M., {Eckert}, D., {De Grandi}, S., {et~al.} 2013, \aap, 556, A44

\bibitem[{{Sanders} {et~al.}(2014){Sanders}, {Fabian}, {Hlavacek-Larrondo},
  {Russell}, {Taylor}, {Hofmann}, {Tremblay}, \& {Walker}}]{Sanders14}
{Sanders}, J.~S., {Fabian}, A.~C., {Hlavacek-Larrondo}, J., {et~al.} 2014,
  \mnras, 444, 1497

\bibitem[{{Schive} {et~al.}(2018){Schive}, {ZuHone}, {Goldbaum}, {Turk},
  {Gaspari}, \& {Cheng}}]{Schive18}
{Schive}, H.-Y., {ZuHone}, J.~A., {Goldbaum}, N.~J., {et~al.} 2018, \mnras,
  481, 4815

\bibitem[{{Sereno} {et~al.}(2018){Sereno}, {Umetsu}, {Ettori}, {Sayers},
  {Chiu}, {Meneghetti}, {Vega-Ferrero}, \& {Zitrin}}]{Sereno18}
{Sereno}, M., {Umetsu}, K., {Ettori}, S., {et~al.} 2018, \apj, 860, L4

\bibitem[{{Shin} {et~al.}(2016){Shin}, {Woo}, \& {Mulchaey}}]{Shin16}
{Shin}, J., {Woo}, J.-H., \& {Mulchaey}, J.~S. 2016, \apjs, 227, 31

\bibitem[{{Smith} {et~al.}(2001){Smith}, {Brickhouse}, {Liedahl}, \&
  {Raymond}}]{Smith01}
{Smith}, R.~K., {Brickhouse}, N.~S., {Liedahl}, D.~A., \& {Raymond}, J.~C.
  2001, \apjl, 556, L91

\bibitem[{{Su} {et~al.}(2017){Su}, {Nulsen}, {Kraft}, {Roediger}, {ZuHone},
  {Jones}, {Forman}, {Sheardown}, {Irwin}, \& {Randall}}]{Su17}
{Su}, Y., {Nulsen}, P.~E.~J., {Kraft}, R.~P., {et~al.} 2017, \apj, 851, 69

\bibitem[{{Tamura} {et~al.}(2001){Tamura}, {Kaastra}, {Peterson}, {Paerels},
  {Mittaz}, {Trudolyubov}, {Stewart}, {Fabian}, {Mushotzky}, {Lumb}, \&
  {Ikebe}}]{Tamura01}
{Tamura}, T., {Kaastra}, J.~S., {Peterson}, J.~R., {et~al.} 2001, \aap, 365,
  L87

\bibitem[{{The Lynx Team}(2018)}]{Lynx18}
{The Lynx Team}. 2018, arXiv e-prints, arXiv:1809.09642

\bibitem[{{Ueda} {et~al.}(2019){Ueda}, {Ichinohe}, {Kitayama}, \&
  {Umetsu}}]{Ueda19}
{Ueda}, S., {Ichinohe}, Y., {Kitayama}, T., \& {Umetsu}, K. 2019, \apj, 871,
  207

\bibitem[{{Ueda} {et~al.}(2017){Ueda}, {Kitayama}, \& {Dotani}}]{Ueda17}
{Ueda}, S., {Kitayama}, T., \& {Dotani}, T. 2017, \apj, 837, 34

\bibitem[{{Ueda} {et~al.}(2018){Ueda}, {Kitayama}, {Oguri}, {Komatsu},
  {Akahori}, {Iono}, {Izumi}, {Kawabe}, {Kohno}, {Matsuo}, {Ota}, {Suto},
  {Takakuwa}, {Takizawa}, {Tsutsumi}, \& {Yoshikawa}}]{Ueda18}
{Ueda}, S., {Kitayama}, T., {Oguri}, M., {et~al.} 2018, \apj, 866, 48

\bibitem[{{Umetsu} \& {Diemer}(2017)}]{Umetsu17}
{Umetsu}, K., \& {Diemer}, B. 2017, \apj, 836, 231

\bibitem[{{Umetsu} {et~al.}(2016){Umetsu}, {Zitrin}, {Gruen}, {Merten},
  {Donahue}, \& {Postman}}]{Umetsu16}
{Umetsu}, K., {Zitrin}, A., {Gruen}, D., {et~al.} 2016, \apj, 821, 116

\bibitem[{{Umetsu} {et~al.}(2014){Umetsu}, {Medezinski}, {Nonino}, {Merten},
  {Postman}, {Meneghetti}, {Donahue}, {Czakon}, {Molino}, {Seitz}, {Gruen},
  {Lemze}, {Balestra}, {Ben{\'{\i}}tez}, {Biviano}, {Broadhurst}, {Ford},
  {Grillo}, {Koekemoer}, {Melchior}, {Mercurio}, {Moustakas}, {Rosati}, \&
  {Zitrin}}]{Umetsu14}
{Umetsu}, K., {Medezinski}, E., {Nonino}, M., {et~al.} 2014, \apj, 795, 163

\bibitem[{{Umetsu} {et~al.}(2018){Umetsu}, {Sereno}, {Tam}, {Chiu}, {Fan},
  {Ettori}, {Gruen}, {Okumura}, {Medezinski}, {Donahue}, {Meneghetti}, {Frye},
  {Koekemoer}, {Broadhurst}, {Zitrin}, {Balestra}, {Ben{\'\i}tez}, {Higuchi},
  {Melchior}, {Mercurio}, {Merten}, {Molino}, {Nonino}, {Postman}, {Rosati},
  {Sayers}, \& {Seitz}}]{Umetsu18}
{Umetsu}, K., {Sereno}, M., {Tam}, S.-I., {et~al.} 2018, \apj, 860, 104

\bibitem[{{van Weeren} {et~al.}(2019){van Weeren}, {de Gasperin}, {Akamatsu},
  {Br{\"u}ggen}, {Feretti}, {Kang}, {Stroe}, \& {Zandanel}}]{van_Weeren19}
{van Weeren}, R.~J., {de Gasperin}, F., {Akamatsu}, H., {et~al.} 2019, \ssr,
  215, 16

\bibitem[{{Voigt} \& {Fabian}(2004)}]{Voigt04}
{Voigt}, L.~M., \& {Fabian}, A.~C. 2004, \mnras, 347, 1130

\bibitem[{{Walker} {et~al.}(2017){Walker}, {Hlavacek-Larrondo},
  {Gendron-Marsolais}, {Fabian}, {Intema}, {Sanders}, {Bamford}, \& {van
  Weeren}}]{Walker17}
{Walker}, S.~A., {Hlavacek-Larrondo}, J., {Gendron-Marsolais}, M., {et~al.}
  2017, \mnras, 468, 2506

\bibitem[{{Yu} {et~al.}(2018){Yu}, {Tozzi}, {van Weeren}, {Liuzzo},
  {Giovannini}, {Donahue}, {Balestra}, {Rosati}, \& {Aravena}}]{Yu18}
{Yu}, H., {Tozzi}, P., {van Weeren}, R., {et~al.} 2018, \apj, 853, 100

\bibitem[{{Zhuravleva} {et~al.}(2018){Zhuravleva}, {Allen}, {Mantz}, \&
  {Werner}}]{Zhuravleva18}
{Zhuravleva}, I., {Allen}, S.~W., {Mantz}, A., \& {Werner}, N. 2018, \apj, 865,
  53

\bibitem[{{Zitrin} {et~al.}(2015){Zitrin}, {Fabris}, {Merten}, {Melchior},
  {Meneghetti}, {Koekemoer}, {Coe}, {Maturi}, {Bartelmann}, {Postman},
  {Umetsu}, {Seidel}, {Sendra}, {Broadhurst}, {Balestra}, {Biviano}, {Grillo},
  {Mercurio}, {Nonino}, {Rosati}, {Bradley}, {Carrasco}, {Donahue}, {Ford},
  {Frye}, \& {Moustakas}}]{Zitrin15}
{Zitrin}, A., {Fabris}, A., {Merten}, J., {et~al.} 2015, \apj, 801, 44

\bibitem[{{ZuHone}(2011)}]{ZuHone11b}
{ZuHone}, J.~A. 2011, \apj, 728, 54

\bibitem[{{ZuHone} {et~al.}(2013{\natexlab{a}}){ZuHone}, {Markevitch},
  {Brunetti}, \& {Giacintucci}}]{ZuHone13}
{ZuHone}, J.~A., {Markevitch}, M., {Brunetti}, G., \& {Giacintucci}, S.
  2013{\natexlab{a}}, \apj, 762, 78

\bibitem[{{ZuHone} {et~al.}(2010){ZuHone}, {Markevitch}, \&
  {Johnson}}]{ZuHone10}
{ZuHone}, J.~A., {Markevitch}, M., \& {Johnson}, R.~E. 2010, \apj, 717, 908

\bibitem[{{ZuHone} {et~al.}(2011){ZuHone}, {Markevitch}, \& {Lee}}]{ZuHone11}
{ZuHone}, J.~A., {Markevitch}, M., \& {Lee}, D. 2011, \apj, 743, 16

\bibitem[{{ZuHone} {et~al.}(2013{\natexlab{b}}){ZuHone}, {Markevitch},
  {Ruszkowski}, \& {Lee}}]{ZuHone13b}
{ZuHone}, J.~A., {Markevitch}, M., {Ruszkowski}, M., \& {Lee}, D.
  2013{\natexlab{b}}, \apj, 762, 69

\bibitem[{{ZuHone} \& {Roediger}(2016)}]{ZuHone16b}
{ZuHone}, J.~A., \& {Roediger}, E. 2016, Journal of Plasma Physics, 82,
  535820301

\end{thebibliography}

\end{document}